\documentclass[]{pasj02} 

\usepackage[switch,mathlines]{lineno} 
\usepackage{url}
\usepackage{lscape}
\usepackage{siunitx}
\usepackage{tabularx}
\usepackage{multirow}
\usepackage{tabularx}
\usepackage{amsmath}
\usepackage{booktabs}

\usepackage{rotating}

\jyear{2026}
\Received{2025/09/11}
\Accepted{2025/12/18}
\Published{2026 February 23}

\begin{document} 

\title{Spectral and photometric variability of SS 433 observed with XRISM and simultaneous optical and near-infrared telescopes}



\author{
Yusuke \textsc{Sakai}\altaffilmark{1}\orcid{0000-0002-5809-3516}\email{sakai.yusuke.d@rikkyo.ac.jp}\altemailmark,
Shinya \textsc{Yamada}\altaffilmark{1}\orcid{0000-0003-4808-893X}\email{syamada@rikkyo.ac.jp}\altemailmark,
Yuta \textsc{Okada}\altaffilmark{2}\email{okada.kusastro@gmail.com}\altemailmark,
Toshihiro \textsc{Takagi}\altaffilmark{3}\email{takagi.toshihiro.bb@ehime-u.ac.jp}\altemailmark,
Tomoya \textsc{Usuki}\altaffilmark{3}\email{n809002b@mails.cc.ehime-u.ac.jp}\altemailmark,
Megumi \textsc{Shidatsu}\altaffilmark{3}\orcid{0000-0001-8195-6546},
Shogo \textsc{B. Kobayashi}\altaffilmark{4}\orcid{0000-0001-7773-9266},
Robert \textsc{Petre}\altaffilmark{5}\orcid{0000-0003-3850-2041},
Yoshihiro \textsc{Ueda}\altaffilmark{2}\orcid{0000-0001-7821-6715},
Hideki \textsc{Uchiyama}\altaffilmark{6}\orcid{0000-0003-4580-4021},
Miho \textsc{Tan}\altaffilmark{7,8}\orcid{0009-0000-2006-2688},
Taro \textsc{Kotani}\altaffilmark{9},
Taichi \textsc{Igarashi}\altaffilmark{8,1}\orcid{0000-0003-4369-7314},
Mami \textsc{Machida}\altaffilmark{8,7}\orcid{0000-0001-6353-7639},
Haruka \textsc{Sakemi}\altaffilmark{10}\orcid{0000-0002-4037-1346},
Nobuyuki \textsc{Kawai}\altaffilmark{11},
Daiki \textsc{Miura}\altaffilmark{12,13}\orcid{0009-0009-0439-1866},
Hiroya \textsc{Yamaguchi}\altaffilmark{13,12,14}\orcid{0000-0002-5092-6085},
Kanta \textsc{Fujiwara}\altaffilmark{2}\orcid{0009-0006-4377-4219},
Daichi \textsc{Hiramatsu}\altaffilmark{15,16,17}\orcid{0000-0002-1125-9187},
Keisuke \textsc{Isogai}\altaffilmark{18,19}\orcid{0000-0002-6480-3799},
Chulsoo \textsc{Kang}\altaffilmark{20},
Mariko \textsc{Kimura}\altaffilmark{21}\orcid{0009-0002-1729-8416},
Katsuhiro \textsc{L. Murata}\altaffilmark{18}\orcid{0000-0002-9579-731X},
Takahiro \textsc{Nagayama}\altaffilmark{22},
Taichi \textsc{Nakamoto}\altaffilmark{20},
Kosuke \textsc{Namekata}\altaffilmark{23,24,25,26}\orcid{0000-0002-1297-9485},
Yuki \textsc{Niida}\altaffilmark{20},
Yuu \textsc{Niino}\altaffilmark{27}\orcid{0000-0001-5322-5076},
Masafumi \textsc{Niwano}\altaffilmark{8}\orcid{0000-0003-3102-7452},
Kyuseok \textsc{Oh}\altaffilmark{28}\orcid{0000-0002-5037-951X},
Shigeyuki \textsc{Sako}\altaffilmark{29,30,31}\orcid{0000-0002-8792-2205},
Mahito \textsc{Sasada}\altaffilmark{32}\orcid{0000-0001-5946-9960},
Hiromasa \textsc{Suzuki}\altaffilmark{33}\orcid{0000-0002-8152-6172},
Kenta \textsc{Taguchi}\altaffilmark{18,2}\orcid{0000-0002-8482-8993},
Ichiro \textsc{Takahashi}\altaffilmark{34}\orcid{0000-0003-2691-4444},
Miyu \textsc{Uenishi}\altaffilmark{3},
Yoichi \textsc{Yatsu}\altaffilmark{34}\orcid{0000-0003-1890-3913},
Marina \textsc{Yoshimoto}\altaffilmark{3}\orcid{0009-0005-0819-0819}
}

\altaffiltext{1}{Department of Physics, Rikkyo University, 3-34-1 Nishi Ikebukuro, Toshima-ku, Tokyo 171-8501, Japan}
\altaffiltext{2}{Department of Astronomy, Kyoto University, Kitashirakawa-Oiwake-cho, Sakyo-ku, Kyoto 606-8502, Japan}
\altaffiltext{3}{Department of Physics, Ehime University, 2-5 Bunkyo-cho, Matsuyama, Ehime 790-8577, Japan}
\altaffiltext{4}{Faculty of Physics, Tokyo University of Science, 1-3 Kagurazaka, Shinjuku-ku, Tokyo 162-8601, Japan}
\altaffiltext{5}{NASA/Goddard Space Flight Center, Greenbelt, MD 20771, USA}
\altaffiltext{6}{Faculty of Education, Shizuoka University, 836 Ohya, Suruga-ku, Shizuoka, Shizuoka 422-8529, Japan}
\altaffiltext{7}{Astronomical Science Program, The Graduate University for Advanced Studies, SOKENDAI, 2-21-1 Osawa, Mitaka, Tokyo 181-8588, Japan}
\altaffiltext{8}{Division of Science, National Astronomical Observatory of Japan, 2-21-1 Osawa, Mitaka, Tokyo 181-8588, Japan}
\altaffiltext{9}{Kanagawa University, Faculty of Engineering, 3-27-1 Rokkakubashi, Kanagawa-ku, Yokohama, Kanagawa 221-8686, Japan}
\altaffiltext{10}{Graduate School of Sciences and Technology for Innovation, Yamaguchi University, 1677-1 Yoshida, Yamaguchi 753-0841, Japan}
\altaffiltext{11}{Department of Physics, School of Science, Tokyo Institute of Technology, 2-12-1 Ookayama, Meguro-ku, Tokyo 152-8550, Japan}
\altaffiltext{12}{Department of Physics, Graduate School of Science, The University of Tokyo, 7-3-1 Hongo, Bunkyo-ku, Tokyo 113-0033, Japan}
\altaffiltext{13}{Institute of Space and Astronautical Science (ISAS), Japan Aerospace Exploration Agency (JAXA), 3-1-1 Yoshinodai, Chuo-ku, Sagamihara, Kanagawa 252-5210, Japan}
\altaffiltext{14}{Department of Science and Engineering, Graduate School of Science and Engineering, Aoyama Gakuin University, 5-10-1 Fuchinobe, Sagamihara 252-5258, Japan}
\altaffiltext{15}{Center for Astrophysics \textbar{} Harvard \& Smithsonian, 60 Garden Street, Cambridge, MA 02138-1516, USA}
\altaffiltext{16}{The NSF AI Institute for Artificial Intelligence and Fundamental Interactions, USA}
\altaffiltext{17}{Department of Astronomy, University of Florida, 211 Bryant Space Science Center, Gainesville, FL 32611-2055, USA}
\altaffiltext{18}{Okayama Observatory, Kyoto University, 3037-5 Honjo, Kamogata-cho, Asakuchi, Okayama 719-0232, Japan}
\altaffiltext{19}{Department of Multi-Disciplinary Sciences, Graduate School of Arts and Sciences, The University of Tokyo, 3-8-1 Komaba, Meguro, Tokyo 153-8902, Japan}
\altaffiltext{20}{Graduate School of Science and Engineering, Ehime University, 2-5, Bunkyo-cho, Matsuyama, Ehime 790-8577, Japan}
\altaffiltext{21}{Advanced Research Center for Space Science and Technology, Institute of Science and Engineering, Kanazawa University, Kakuma, Kanazawa, Ishikawa 920-1192, Japan}
\altaffiltext{22}{Graduate School of Science and Engineering, Kagoshima University, 1-21-35 Korimoto, Kagoshima, Kagoshima 890-0065, Japan}
\altaffiltext{23}{Heliophysics Science Division, NASA Goddard Space Flight Center, 8800 Greenbelt Road, Greenbelt, MD 20771, USA}
\altaffiltext{24}{The Catholic University of America, 620 Michigan Avenue, N.E. Washington, DC 20064, USA}
\altaffiltext{25}{The Hakubi Center for Advanced Research, Kyoto University, Yoshida-Honmachi, Sakyo-ku, Kyoto 606-8501, Japan}
\altaffiltext{26}{Department of Physics, Kyoto University, Kitashirakawa-Oiwake-cho, Sakyo-ku, Kyoto 606-8502, Japan}
\altaffiltext{27}{Kiso Observatory, Institute of Astronomy, Graduate School of Science, The University of Tokyo, 10762-30 Mitake, Kiso, Nagano 397-0101, Japan}
\altaffiltext{28}{Korea Astronomy and Space Science Institute, Daedeokdae-ro 776, Yuseong-gu, Daejeon 34055, Republic of Korea}
\altaffiltext{29}{Institute of Astronomy, Graduate School of Science, The University of Tokyo, 2-21-1 Osawa, Mitaka, Tokyo 181-0015, Japan}
\altaffiltext{30}{UTokyo Organization for Planetary Space Science, The University of Tokyo, 7-3-1 Hongo, Bunkyo-ku, Tokyo 113-0033, Japan}
\altaffiltext{31}{Collaborative Research Organization for Space Science and Technology, The University of Tokyo, 7-3-1 Hongo, Bunkyo-ku, Tokyo 113-0033, Japan}
\altaffiltext{32}{Institute of Integrated Research, Institute of Science Tokyo, 2-12-1 Ookayama, Meguro-ku, Tokyo 152-8550, Japan}
\altaffiltext{33}{Faculty of Engineering, University of Miyazaki, Miyazaki 889-2192, Japan}
\altaffiltext{34}{Department of Physics, Institute of Science Tokyo, 2-12-1 Ookayama, Meguro-ku, Tokyo 152-8550, Japan}

\KeyWords{X-rays: binaries --- stars: jets --- stars: individual (SS 433) --- accretion, accretion disks}

\maketitle

\begin{abstract}
We present results from coordinated multiwavelength observations of the microquasar SS~433, obtained with XRISM, optical telescopes (Seimei, LCO, Tomo-e Gozen, MITSuME), and the kSIRIUS near-infrared camera during April 2024 and March 2025. The XRISM exposures amounted to $\sim$200~ks in 2024 and $\sim$100~ks in 2025, corresponding to different combinations of orbital and precessional phases. With XRISM/Resolve's high spectral resolution and large effective area, we clearly resolved numerous emission lines even in short time segments, achieving improved accuracy in Doppler-shift measurements relative to earlier observations. The simultaneously obtained X-ray and optical Doppler shifts suggest a possible tendency for the optical emission to lag slightly behind the X-rays. In the Resolve data, the Doppler shifts of the two jet components exhibited apparent asymmetries, with jet speeds fluctuating around $\sim0.26 \pm 0.01c$ in 2024 and $\sim0.30 \pm 0.01c$ in 2025. The velocity variations indicated modulations on a timescale of $\sim$6.3~d, with a phase offset of about $-90^{\circ}$ relative to the nutation cycle. The observed line widths and flux of the approaching and receding jets appear consistent with the expected geometrical effects, indicating systematically larger line widths in the inner regions of the jets, as proposed by \citet{Shidatsu_2025}. Optical light curves show flares of $\sim$400~s in 2024 and $\sim$1600~s in 2025, with amplitudes up to $\sim$15\% during out-of-eclipse intervals, while the XRISM/Xtend light curves show no significant variability within the overlapping intervals and given the statistical uncertainties. Near-infrared photometry in 2024, obtained during an out-of-eclipse interval at a different epoch from the optical observations, showed no flare-like variability, and the X-ray band also remained constant within uncertainties. These coordinated observations provide a foundation for future XRISM studies aimed at probing the dynamical properties of the relativistic jets in SS~433.
\end{abstract}


\section{Introduction}\label{sec:Introduction}

\begin{figure*}[ht!]
 \includegraphics[width=1\linewidth]{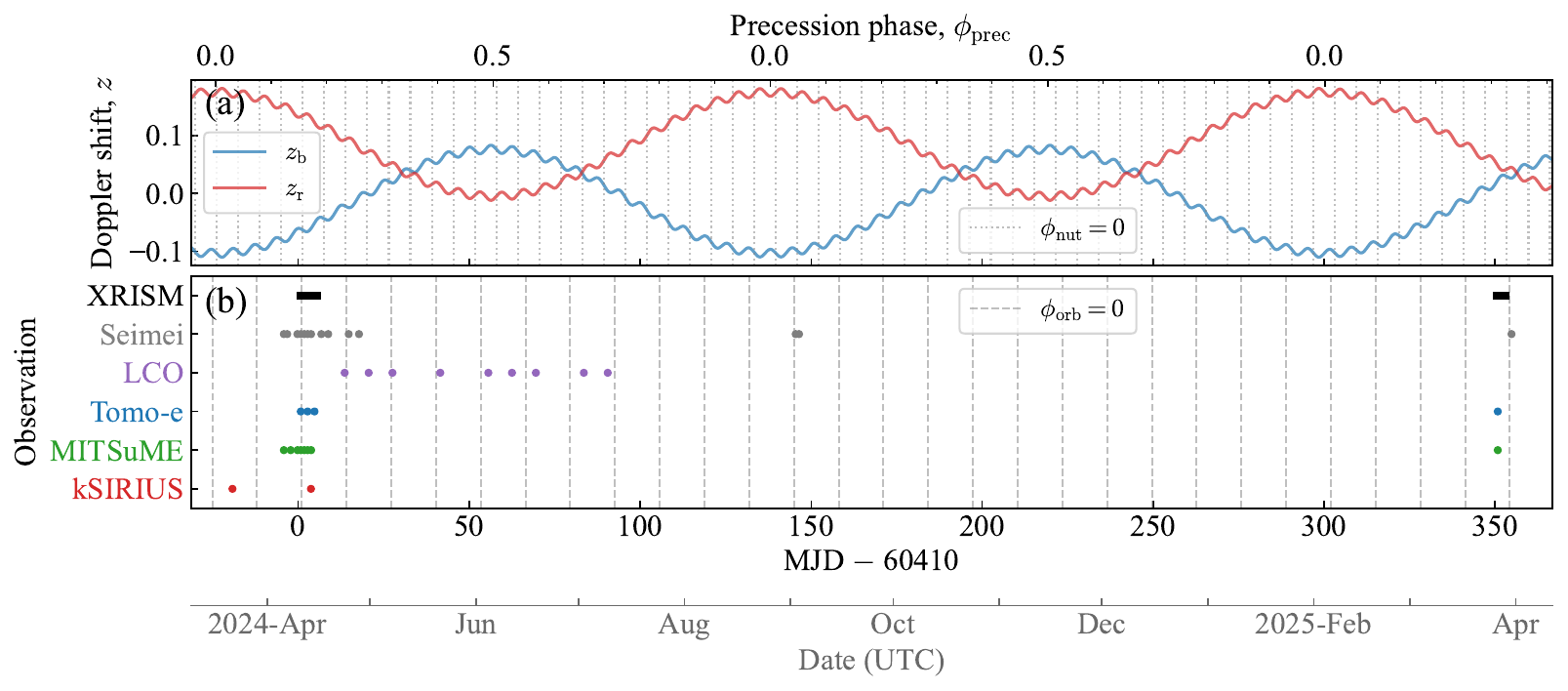}
\caption{
Overview of the multiwavelength observation campaign of SS~433 conducted in 2024--2025. 
(a) Doppler shifts of the approaching and receding jets ($z_\mathrm{b}$ and $z_\mathrm{r}$).
The curves are calculated from the precession and nutation models based on the ephemerides and parameters of \citet{Gies_2002} and \citet{Davydov_2008}, with the phase offset adjusted to match the observed Doppler shifts (see text). The dotted line indicates the nutation phase zero ($\phi_\mathrm{nut} = 0$).
(b) Timeline of observations. Observing epochs of XRISM (black), Seimei (gray), LCO (purple), Tomo-e Gozen (blue), MITSuME (green), and kSIRIUS (red) are shown. 
The dashed line indicates the orbital phase zero ($\phi_\mathrm{orb} = 0$), based on the ephemeris of \citet{Cherepashchuk_2023}.
}

 \label{observed_overview}
\end{figure*}

The microquasar SS~433 is one of the most intensively studied jet sources in our Galaxy, with over four decades of multiwavelength observations (see reviews by \cite{Margon_1984, Fabrika_2004, Cherepashchuk_2025}). It is an X-ray binary consisting of an A-type supergiant and a compact object---either a black hole or a neutron star---in a 13.1-d orbit \citep{Crampton_1980, Hillwig_2004}. 
A defining feature of the system is its baryonic jets, which show prominent Doppler-shifted emission lines in the X-ray, optical, and near-infrared bands. These jets precess with a period of 162.5~d, with an average bulk velocity of $\sim 0.26c$ (e.g., \cite{Abell_1979, Kotani_1996, Marshall_2002, Waisberg_2019}).
The system is located at a distance of $\sim$5.5~kpc \citep{Blundell_2004}. On arcsecond scales ($\sim 10^{17}$~cm), the jets trace helical paths in radio and X-ray images \citep{Hjellming_1981, Sakai_2025}, while on degree scales ($\sim 100$~pc) they inflate the surrounding W50 nebula through interactions with the interstellar medium (e.g., \cite{Geldzahler_1980, Dubner_1998, Brinkmann_1996, Sakemi_2023, Kayama_2025}).

In addition to precession and orbital motion, SS~433 also shows a 6.3-d nutation cycle, first proposed by \citet{Katz_1982}.
These three periodicities together define the system's long-term kinematic behavior.
Although generally stable, long-term optical \citep{Cherepashchuk_2022} and X-ray \citep{Medvedev_2019} monitoring has revealed subtle variations in jet properties, including phase shifts and small changes in the precession or nutation periods, on timescales of years. 
The physical origin of the nutation remains uncertain, with various models proposed to explain its nature and driving mechanisms (e.g., \cite{Collins_1988, Eikenberry_2001, Collins_2002}).

The interpretation of nutation signatures depends on the spatial origin of the jet emission. In the optical band, emission is thought to arise from extended regions along the jet ($\sim10^{14\text{--}15}$~cm; \cite{Fabrika_2004}), leading to light-travel delays of up to $\sim$0.5~d for the receding component. These delays complicate the isolation of intrinsic kinematic changes. X-ray emission lines, by contrast, are known to originate much closer to the launch point ($\sim 10^{12}$~cm; \cite{Brinkmann_1991, Marshall_2002}), reducing geometric delays and providing a more direct view of the jet dynamics, although the emission may still be affected by obscuration from the companion star and/or accretion disk, whose dimensions can be comparable to those of the emission region (e.g., \cite{Fabrika_2004, Lopez_2006}).

Simultaneous optical and X-ray studies have provided important constraints on the spatial and dynamical structure of the jets. For example, \citet{Kubota_2010b} found a $\sim$0.5~d delay between Suzaku X-ray and optical jet Doppler shifts, with short-term X-ray variations not fully explained by precession and nutation alone. \citet{Marshall_2013} confirmed this delay and reported variations in the jet speed when compared with the symmetric jet model, 
which assumes that the two jets are ejected in opposite directions with the same velocity. Such deviations may arise from changes in jet speed, asymmetries between the jets, or motions of the launching region.

Photometric studies have also linked optical and X-ray variability in SS~433. \citet{Revnivtsev_2004} and \citet{Atapin_2015} detected rapid optical fluctuations with timescales similar to X-rays but delayed by $\sim$100~s, which were interpreted as reprocessing of X-rays on the outer funnel walls into variable UV--optical emission. \citet{Burenin_2011} found that the optical variability amplitude decreases during eclipses, implying a compact origin near the disk, and identified a break in the power spectral density at $f \approx 2.4 \times 10^{-3}$~Hz, corresponding to a timescale of $\sim$70~s.

In this paper, we present a simultaneous multiwavelength campaign combining observations from the X-ray Imaging and Spectroscopy Mission (XRISM; \cite{Tashiro_2022}) conducted in 2024 and 2025, alongside optical spectroscopy and minute-cadence optical/near-infrared photometry. 
XRISM carries two instruments: the Resolve X-ray microcalorimeter \citep{Ishisaki_2022} and the Xtend X-ray CCD camera \citep{Mori_2022, Noda_2025, Uchida_2025}, which we employed for high-resolution spectroscopy and X-ray light-curve measurements, respectively. XRISM/Resolve's high spectral resolution and large effective area, as demonstrated in its first SS~433 analysis \citep{Shidatsu_2025}, enable precise measurements of jet line widths in emission lines such as Si and Fe, allowing time-resolved spectroscopy with unprecedented sensitivity.

This paper is structured as follows. Section~\ref{sec:data_reduction} describes the data reduction procedures. Section~\ref{Analysis and results} presents time-resolved measurements of Doppler shifts in the jets, together with the analysis of light curves. Section~\ref{sec:Discussion} provides a discussion of our results. Finally, section~\ref{sec:Conclusion} summarizes our conclusions. Unless otherwise noted, all quoted uncertainties correspond to 1$\sigma$ confidence intervals.

\section{Observations and data reduction}\label{sec:data_reduction}

An overview of the X-ray, optical, and near-infrared observations of SS~433 is shown in figure~\ref{observed_overview}. 
All times were standardized to Universal Time (UTC) for consistency across instruments.

\subsection{XRISM} \label{XRISM}

XRISM observations were carried out from 2024 April 10 to 15 (OBSID = 300041010) 
and from 2025 March 26 to 28 (OBSID = 201014010). 
The corresponding exposure times were $\sim$204.3~ks and $\sim$98.4~ks, respectively. 
Spectroscopic analysis was performed with the Resolve microcalorimeter, 
and light-curve data were obtained with the Xtend CCD camera.
The data reduction was performed using \texttt{HEAsoft} version 6.34 and the XRISM Calibration Database available as of 2024 September, 
following the XRISM Quick Start Guide v2.3\footnote{\url{https://heasarc.gsfc.nasa.gov/docs/xrism/analysis/quickstart/}}.

We focused our spectral analysis on the Doppler-shifted Fe and Ni emission features from the jets using events in the 5.5--9~keV band. Although the Resolve bandpass extends down to $\sim$1.6~keV even with the gate valve closed, time-resolved spectroscopy of lower-energy features such as Si and S lines would be challenging due to limited photon statistics.
For the analysis, we used only the ``High-resolution primary'' (Hp) events, which are best calibrated in energy. The 6~$\times$~6 pixel array includes one pixel dedicated to energy calibration. The response matrix file (RMF) was created using \texttt{rslmkrmf} with the ``large'' configuration (\texttt{whichrmf=L}), and the ancillary response file (ARF) was generated using \texttt{xaarfgen}, assuming a point source at the aim point. Background subtraction was not applied, as the contributions from the non-X-ray background (NXB), cosmic X-ray background (CXB), and Galactic ridge X-ray emission (GRXE) were all estimated to be negligible. In particular, the NXB level was at most a few percent of the source flux in the 2--10 keV band, and the CXB and GRXE were even smaller.

We divided the Resolve data into time segments to investigate temporal variations in the emission lines. The number of segments was empirically chosen to balance temporal resolution and photon statistics: 30 for the brighter 2024 data and 5 for the fainter 2025 data. Each segment was assigned a sequential identification number (ID) starting from 0 (see table~\ref{tab:xray_doppler}). The resulting exposure per segment was $\sim$6.8~ks in 2024 and $\sim$19.7~ks in 2025.

Xtend consists of four CCD chips arranged in a 2~$\times$~2 configuration, providing a field of view of approximately 38$\farcm$5~$\times$~38$\farcm$5. It was operated in the full-window mode during both the 2024 and 2025 observations. Source events were extracted from a circular region with a 4$^\prime$ radius centered on SS~433, which was fully contained within a single CCD chip. We selected events in the 2--10~keV energy range to minimize contamination from the background. Background subtraction was not applied, as the background contribution was estimated to be $\lesssim$~5\% of the total count rate in this band.

\subsection{Seimei} \label{Seimei}

The Seimei telescope is a 3.8~m optical and near-infrared telescope located at Okayama Observatory, Kyoto University, Japan \citep{Kurita_2020}. In this work, we conducted spectroscopy using the Kyoto Okayama Optical Low-dispersion Spectrograph with optical-fiber integral field unit (KOOLS-IFU; \cite{Yoshida_2005, Matsubayashi_2019}). The observations were conducted at 14 nights in total: 11 nights between 2024 April 5 and April 27, 2 nights on 2024 September 2 and 3, and 1 night on 2025 March 30. The VPH-blue grism, covering 4100--8900~\AA~with a wavelength resolution of $R \sim 500$ was adopted for all epochs, except for the observation on 2025 March 30, in which the VPH-683 grism, covering 5800--8000~\AA~with $R \sim 2000$ was utilized. The integration time for each object frame was 60~s, and we acquired between 3 (on 2024 April 27) and 50 (on 2024 April 12) frames per night.

We performed standard data reduction for all frames, including bias subtraction, flat-fielding, wavelength calibration, sky background subtraction, flux calibration, and barycentric correction, using IRAF \citep{Tody_1986} and the KOOLS-IFU pipeline tool\footnote{\url{https://www.kusastro.kyoto-u.ac.jp/~iwamuro/KOOLS/}} to extract the spectra of the individual frames. The Hg, Ne and Xe comparison lamp data were used in the wavelength calibration. We adopted HR7596 as the standard star for the flux calibration, except for the nights of 2024 April 12 and 2025 March 30, for which HR5501 was adopted. In cases where bias, flat, comparison, or standard star frames were not available for a specific night, we utilized the corresponding data from the closest available night. We performed the above data reduction for the individual object frames, and adopted the spectra from a single object frame obtained at approximately the same time each night. The remaining spectra will be used in a separate paper to investigate intra-night variability.

\subsection{LCO} \label{LCO}

Las Cumbres Observatory (LCO) operates a global network of robotic optical telescopes, comprising two 2~m, nine 1~m, and nine 0.4~m instruments distributed across six sites worldwide \citep{Brown_2013}. In this work, we conducted spectroscopic observations using the FLOYDS, a cross-dispersed, low resolution spectrograph, mounted on the 2~m Faulkes Telescope North (FTN) at Haleakala (USA). The observations were conducted on 9 nights between April 23 and July 9, 2024. We used the spectra for the individual nights obtained through the FLOYDS pipeline\footnote{\url{https://lco.global/documentation/data/floyds-pipeline/}} \citep{Valenti_2014}.

\subsection{Tomo-e Gozen} \label{Tomo-e Gozen}

Tomo-e Gozen is a wide-field CMOS mosaic camera mounted on the 1.05~m Kiso Schmidt telescope \citep{Sako_2018}. 
It consists of 84 CMOS sensors, which cover approximately 30\% of the focal plane and correspond to a circular field of view about $9^\circ$ in diameter. 
Each sensor covers $39\farcm7 \times 22\farcm4$ with a pixel scale of $1\farcs19$, and all sensors are read simultaneously at 2~fps. The camera is sensitive to wavelengths from $\sim$370 to 730~nm, peaking at 500~nm \citep{Kojima_2018}, and is particularly suited for detecting short-timescale transients through continuous sky monitoring.

Tomo-e Gozen observed SS~433 on April 10, 12, and 14, 2024, and on March 26, 2025. We obtained 23, 24, and 24 data cubes on the respective 2024 nights, and 69 cubes in 2025. Each cube nominally consists of 240 frames at 2~fps. Note that several data cubes from March 26 contain fewer than 240 frames due to elevated background levels during morning twilight. Standard preprocessing steps---including bias subtraction, dark correction, and flat-fielding---were applied. To balance photometric stability and temporal resolution, we stacked every 10 frames, corresponding to the typical PSF under average seeing conditions (FWHM $\sim$3--4$^{\prime\prime}$). Aperture photometry was performed using a fixed 13-pixel radius ($\sim$15$^{\prime\prime}$), centered on the targets.

Photometric zero-points were derived using Gaia DR3 $G$-band magnitudes \citep{Gaia_2023}. Although the \mbox{Tomo-e Gozen} response differs from the Gaia $G$-band, their spectral shapes are broadly similar, and this calibration method has been adopted in previous studies (e.g., \cite{Beniyama_2022, Nishino_2022}). Since the same calibration stars were used across all frames, systematic offsets are expected to be uniform, ensuring reliable relative photometry. Stars with $G > 16.5$ were excluded to avoid sources near the detection limit. We also required no other cataloged object within $30^{\prime\prime}$ to minimize contamination and limited the calibration sample to stars with $9 < G < 14.5$. Instrumental fluxes were measured via aperture photometry, and the zero-point for each frame was computed by fitting the instrumental magnitudes to Gaia values. Frames with zero-point standard deviation exceeding 0.1~mag were discarded, and we retained only measurements with signal-to-noise ratio (S/N) > 10 to remove low-confidence outliers. A visual overview of the source environment around SS~433 is provided in appendix~\ref{appendix:tomoe_fov}.

\subsection{MITSuME} \label{MITSuME}

The Multicolor Imaging Telescopes for Survey and Monstrous Explosions (MITSuME) is a robotic optical telescope system consisting of 50-cm instruments \citep{Kotani_2005, Yatsu_2007, Shimokawabe_2009}.  
Two telescopes are currently in operation, located at the Akeno Observatory and the Okayama Astrophysical Observatory, and are managed by the Institute of Science Tokyo and Kyoto University, respectively.  
Each telescope is equipped with three CCD cameras, enabling simultaneous photometric observations in the $g'$, $R_\mathrm{C}$, and $I_\mathrm{C}$ bands.  
SS~433 was observed with MITSuME on April 5--13, 2024, and March 26, 2025. In 2024, both the Akeno and Okayama sites were used, while in 2025, observations were conducted only at Akeno.  
Exposure times of 10~s and 60~s per frame were adopted in 2024, whereas 10~s exposures were adopted in 2025.
The 2025 dataset was obtained contemporaneously with XRISM and Tomo-e Gozen.

The data reduction for both telescopes was performed using a common automated analysis pipeline \citep{Niwano_2021}.  
For the 10-second exposure data, every six consecutive frames (i.e., totaling 60 seconds of exposure time) were co-added to improve the S/N.  
The 60-second exposure data were analyzed on a frame-by-frame basis without co-adding.
For consistency with the quality of the Tomo-e Gozen data, only the MITSuME data with S/N > 10 were used in the 2025 analysis. We also removed a single anomalous $g'$-band point near MJD~60760.80 that was significantly ($\sim$1~mag) brighter than the surrounding data, likely due to a photometric error. For the color index analysis presented in appendix~\ref{appendix:color_index}, a more relaxed threshold of S/N > 3 was applied to the 2024 data.
In addition, images co-added over an entire night were also prepared. Among these, the 2024 data were primarily used to trace the orbital phase and identify the timing of eclipses in the accretion disk.
Aperture photometry was performed using relative photometry with reference to stars listed in the Pan-STARRS1 (PS1) catalog \citep{Chambers_2016} within the same field of view.  
For flux calibration, the catalog magnitudes in the PS1 $g$, $r$, and $i$ bands were converted to the magnitudes in $g'$, $R_\mathrm{C}$, and $I_\mathrm{C}$ bands following table~6 of \citet{Tonry_2012}. 
These magnitudes, initially in the AB system, were then converted to the Vega system using the offsets given in table~1 of \citet{Blanton_2007}.

\begin{figure}[ht!]
 \includegraphics[width=0.94\linewidth]{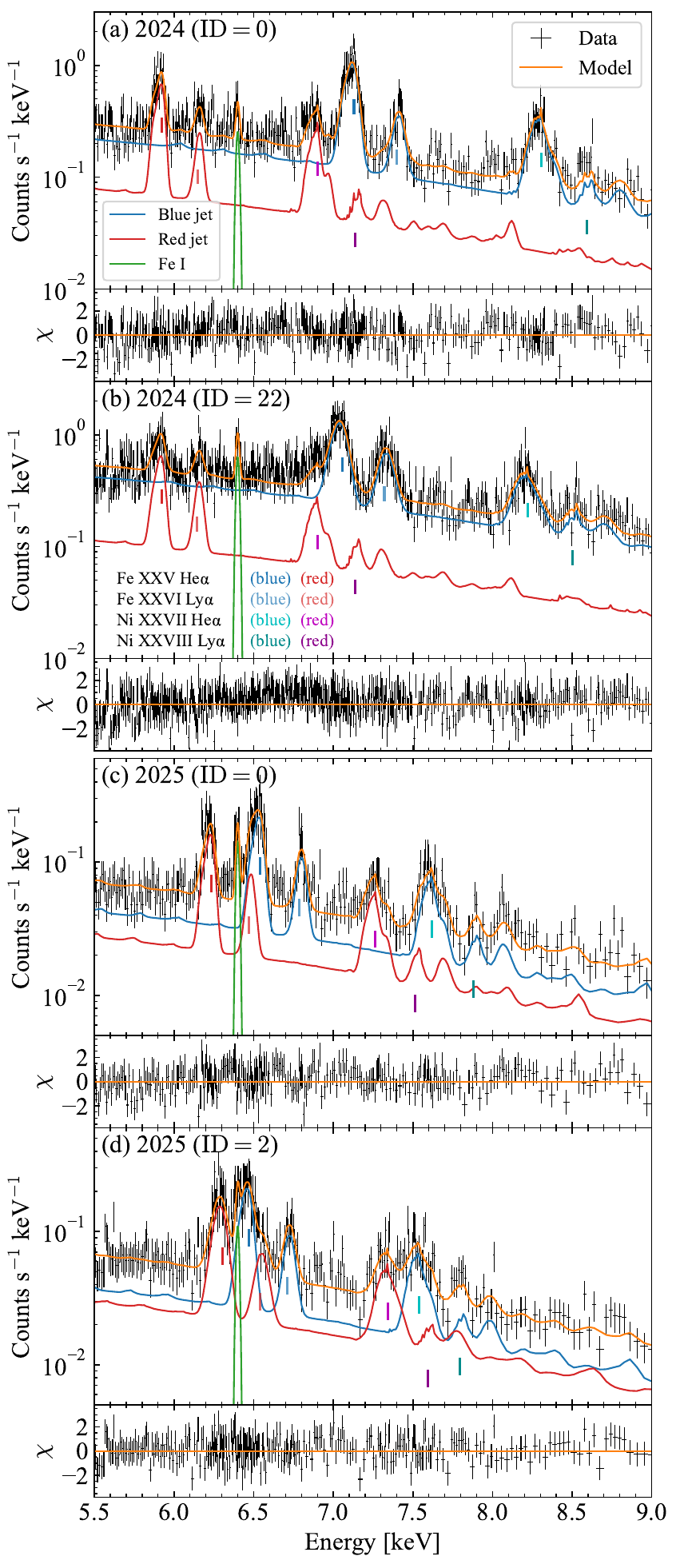}
\caption{XRISM/Resolve spectra from 2024 (a, b) and 2025 (c, d) represent time-resolved data, extracted from 30 intervals in 2024 and 5 in 2025. The panels show representative intervals: ID 0 and 22 (2024), and ID 0 and 2 (2025). The spectra are fitted with two thermal jet components and a stationary Fe line. Black points show the data; orange, the total model. Blue, red, and green lines represent the approaching jet, receding jet, and Fe~\textsc{i} K$\alpha$, respectively. Several representative jet emission lines are overplotted based on the best-fit Doppler shifts. Representative Fe and Ni lines (He$\alpha$, Ly$\alpha$) are shown according to the best-fit Doppler model. Each panel shows the spectra with residuals.
}
 \label{resolve_each_spectra}
\end{figure}
\begin{figure}[ht!]
 \includegraphics[width=0.96\linewidth]{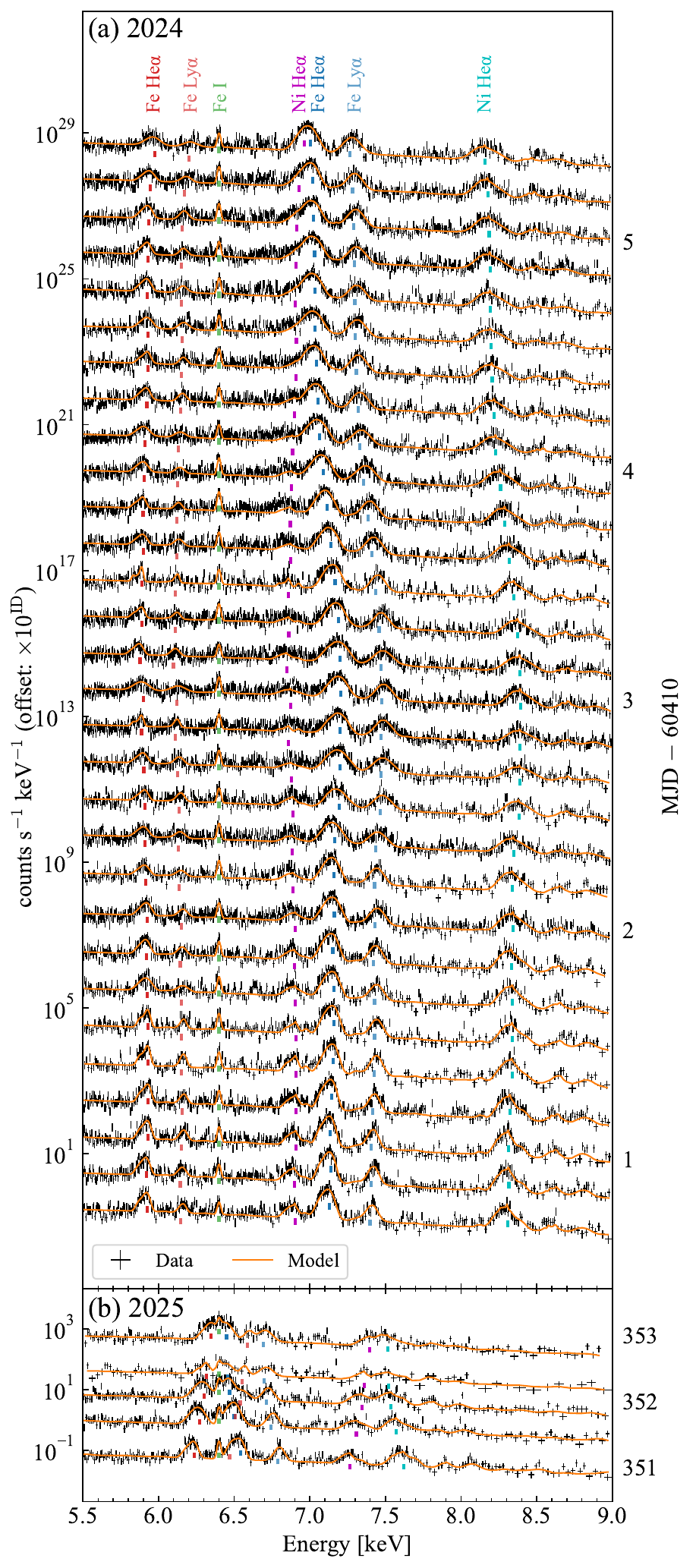}
\caption{
Time-resolved XRISM/Resolve spectra in the 5.5--9~keV band from 2024 (a) and 2025 (b). 
The data are divided into 30 and 5 segments for the 2024 and 2025 observations, respectively, with vertical offsets applied to the black points for clarity.  
Spectra are fitted with two thermal jet components and a stationary Fe line (orange).  
Major emission lines are shown with the following colors:  
Fe\,\textsc{xxv} He$\alpha$ (blue/red),  
Fe\,\textsc{xxvi} Ly$\alpha$ (light blue/light red),  
and Ni\,\textsc{xxvii} He$\alpha$ (cyan/magenta),  
corresponding to the blue and red jets.
The stationary Fe\,\textsc{i} K$\alpha$ line is shown in green.  
The right-hand axis shows the observation time in MJD. 
See table~\ref{tab:xray_doppler} for the Doppler shift values.
}
 \label{resolve_all_spectra}
\end{figure}

\subsection{kSIRIUS} \label{kSIRIUS}

kSIRIUS is a three-band ($JHK_{\mathrm{s}}$) simultaneous imager mounted on the 1.0-m telescope at Kagoshima University \citep{Nagayama_2024}, providing a field of view of $\sim3\farcm7\times2\farcm9$ with a pixel scale of $0\farcs69$ using dichroic beam splitters and InGaAs detectors. Monitoring observations were carried out on 2024 March 21 and April 13, employing a six-point dither pattern along the right ascension direction; the net exposure times were $\sim$40~s and $\sim$5600~s, respectively. Data reduction included dark subtraction, flat-fielding, sky subtraction from the dithered frames, and astrometric calibration using 2MASS \citep{Cutri_2003}.

Photometric quality degraded after MJD~60413.755 (April 13), likely because of deteriorating sky conditions. To ensure reliability, all data taken after this time were excluded, removing approximately 30\% of the frames and leaving a net usable exposure time of $\sim$3900~s for that night. Aperture photometry was performed on the reduced images to extract light curves, yielding measurements with S/N > 100. All magnitudes are reported in the Vega system.

\begin{figure}[ht!]
 \includegraphics[width=0.96\linewidth]{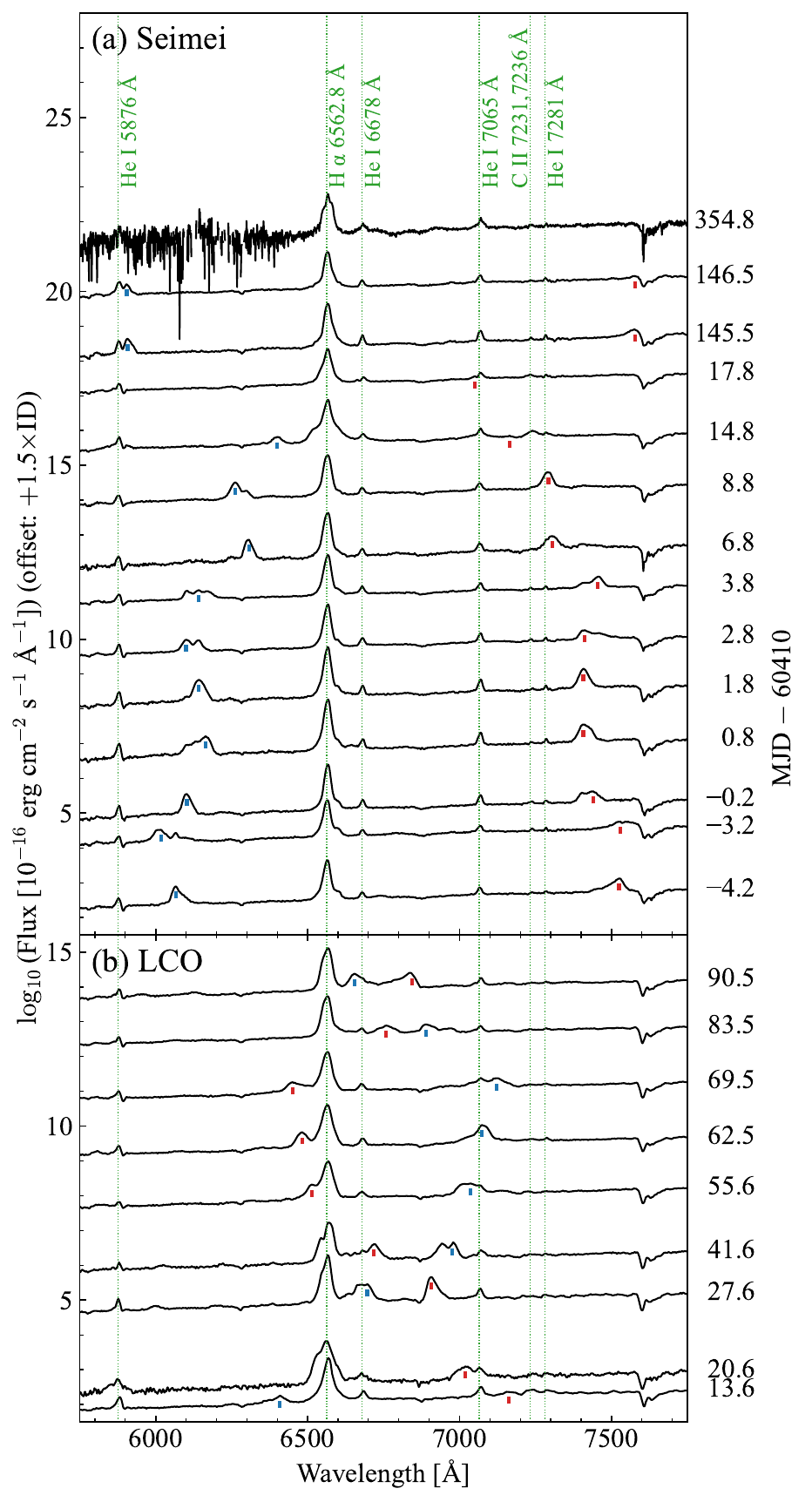} 
\caption{Optical spectra obtained with the Seimei and LCO telescopes. The blue- and red-shifted components of the H$\alpha$ line are shown in blue and red, respectively. Dotted green lines indicate stationary emission lines such as He\,\textsc{i}, C\,\textsc{ii}, and H$\alpha$, which are likely associated with the accretion disk wind or circumbinary material. 
The right-hand axis shows the observation time in MJD for each spectrum. 
The corresponding Doppler shift values are summarized in table~\ref{tab:optical_doppler}. 
}
 \label{optical_spectra_all}
\end{figure}

\section{Analysis and results} \label{Analysis and results}
\subsection{Spectral analysis of X-ray and optical data} \label{sec:spectral_analysis}
 
We measured the Doppler shifts of the approaching and receding jet components from time-resolved spectra in the X-ray and optical bands.
Spectral fitting of the X-ray data obtained with Resolve was performed following the method described in \citet{Shidatsu_2025}, employing the optically thin, collisionally ionized plasma model \texttt{APEC} \citep{Smith_2001}, supplemented by a stationary Fe\,\textsc{i} K$\alpha$ line.
The model was fitted using \texttt{XSPEC} \citep{xspec_tool} with the form \texttt{tbabs*(bvapec\_b + bvapec\_r + gaussian)}, where \texttt{bvapec\_b} and \texttt{bvapec\_r} represent the eastern (approaching) and western (receding) jet components, respectively.
The approaching and receding jets are hereafter referred to as the blue and red jets, respectively.
The hydrogen column density was fixed at $2.07 \times 10^{22}~\mathrm{cm^{-2}}$ \citep{Lopez_2006}. 
The Fe\,\textsc{i} K$\alpha$ line was modeled with a Gaussian component, which was used as a simple approximation for a stationary narrow line possibly originating from the outer accretion disk (\cite{Kotani_1996, Takagi_2025}).  
This line is known to have a FWHM $< 1000~\mathrm{km~s^{-1}}$ \citep{Marshall_2002}, and the Gaussian width was fixed to $\sigma_E = 10~\mathrm{eV}$ at 6.4~keV.
We adopted the solar abundance table of \citet{Lodders_2009} and the cross-section table of \citet{Verner_1996}.  
The Fe and Ni abundances in the plasma components were treated as free parameters.
In particular, Ni has been reported to be overabundant relative to the solar value in SS~433 (e.g., \cite{Marshall_2013, Shidatsu_2025}).
The free parameters were the plasma temperature $kT$, Fe and Ni abundances ($A_\mathrm{Fe}$, $A_\mathrm{Ni}$), plasma normalizations $N$, Gaussian normalization $N_\mathrm{gau}$, Doppler shift $z$, and the Gaussian sigma of the velocity broadening, $\sigma_v$. The corresponding FWHM is defined as $W_v = 2\sqrt{2\ln2}\,\sigma_v$, and is used throughout this paper.

All parameters were linked between the two jet components, except for $z$, $N$, and $W_v$.
More complex models, including multi-temperature plasma components or disk reflection, have been considered in previous studies (e.g., \cite{Medvedev_2018, Middleton_2021}). The model adopted in our analysis reproduces the spectral features sufficiently well for the purpose of Doppler shift measurements, although such more elaborate models will be explored in future work.

The fitting results are summarized in table~\ref{tab:xray_doppler}.
All parameters are generally consistent with those reported by \citep{Shidatsu_2025}.
The representative fitting results are presented in figures~\ref{resolve_each_spectra}a and \ref{resolve_each_spectra}b for 2024 and figures~\ref{resolve_each_spectra}c and \ref{resolve_each_spectra}d for 2025.  
The model components include the blue and red jets and the stationary Fe\,\textsc{i} K$\alpha$ line.
In the 2024 observations, intervals ID 0 (during eclipse) and ID 22 (out of eclipse) clearly show the blue and red jet lines separately.  
Notably, the out-of-eclipse spectrum also exhibits a broader Fe\,\textsc{xxv} He$\alpha$ line, consistent with the findings of \citet{Shidatsu_2025}.
In contrast, the 2025 spectra (IDs 0 and 2; out of eclipse) show only small Doppler shifts, indicating that the jets are oriented nearly perpendicular to the line of sight.
During these epochs, the blue and red jet lines appear at similar energies due to the jets being nearly perpendicular to the observer's direction.
The full set of fitting results is shown in figure~\ref{resolve_all_spectra}.  
The corresponding full-band spectra for these epochs are shown in figure~\ref{resolve_all_spectra_full} (Appendix~\ref{appendix:full_resolve_spectrum}), which illustrates the overall spectral features.

For the optical spectra obtained with Seimei and LCO,
the Doppler shifts of the blue and red jets were derived from the peak wavelengths of the corresponding H$\alpha$ line profiles,
using the rest-frame wavelength of H$\alpha$.
Absorption features and nearby stationary components were visually inspected and accounted for when determining the peak positions.
Three epochs were excluded from the analysis.
In one case, both the blue and red jet components were excluded because the jet was nearly perpendicular to the line of sight, resulting in very weak emission with no clear peaks.
In the other two cases, only the blue jet was excluded due to blending with stationary H$\alpha$ emission lines.
The measured peak positions and corresponding Doppler shifts are shown in figure~\ref{optical_spectra_all}.

\begin{table}[ht!]
\caption{Ephemeris and kinematic parameters adopted for SS~433}
\label{tab:ephemeris}
\begin{tabular}{lcl}
\toprule
\textbf{Parameter} & \textbf{Symbol} & \textbf{Value}\footnotemark[$*$] \\
\midrule
& \textbf{Precession}\footnotemark[$\ddag$] & \\
\midrule
Period (d) & $P_{\mathrm{prec}}$ & 162.15 \\
Reference epoch (MJD)\footnotemark[$\dag$] & $t_{0,\mathrm{prec}}$ & 51457.62 \\
\quad (updated, this work)\footnotemark[$\#$] & & $51467.67 \pm 0.01$ \\
Jet speed ($c$) & $\beta$ & 0.2602 \\
Inclination angle ($^\circ$) & $i$ & 78.83 \\
Half-opening angle ($^\circ$) & $\theta$ & 19.85 \\
\midrule
& \textbf{Nutation}\footnotemark[$\S$] & \\
\midrule
Period (d) & $P_{\mathrm{nut}}$ & 6.287599 \\
Reference epoch (MJD)\footnotemark[$\dag$] & $t_{0,\mathrm{nut}}$ & 43029.655 \\
\quad (updated, this work)\footnotemark[$\#$] & & $43032.423 \pm 0.007$ \\
Amplitude ($z$) & $A_{\mathrm{nut}}$ & 0.00689349 \\
\midrule
& \textbf{Orbit}\footnotemark[$\|$] & \\
\midrule
Period (d) & $P_{\mathrm{orb}}$ & 13.08250 \\
Period derivative (s s$^{-1}$) & $\dot{P}_{\mathrm{orb}}$ & $1.14 \times 10^{-7}$ \\
Reference epoch (MJD)\footnotemark[$\dag$] & $t_{0,\mathrm{orb}}$ & 51737.04 \\
\bottomrule
\end{tabular}

\begin{tabnote}
\footnotemark[$*$] Uncertainties reported in the original references are omitted here for clarity.\\
\footnotemark[$\dag$] Originally in JD, converted to MJD via $\mathrm{MJD} = \mathrm{JD} - 2400000.5$. \\
\footnotemark[$\ddag$] Ephemeris adopted from \citet{Gies_2002}. \\
\footnotemark[$\S$] Ephemeris adopted from \citet{Davydov_2008}. \\
\footnotemark[$\|$]  Ephemeris adopted from \citet{Cherepashchuk_2023}. \\
\footnotemark[$\#$] Updated values derived from Doppler-shift analysis 
of XRISM/Resolve data (see subsection~\ref{Doppler-shift variability in X-ray and optical}).
\end{tabnote}

\end{table}

\begin{figure*}[ht!]
 \includegraphics[width=1\linewidth]{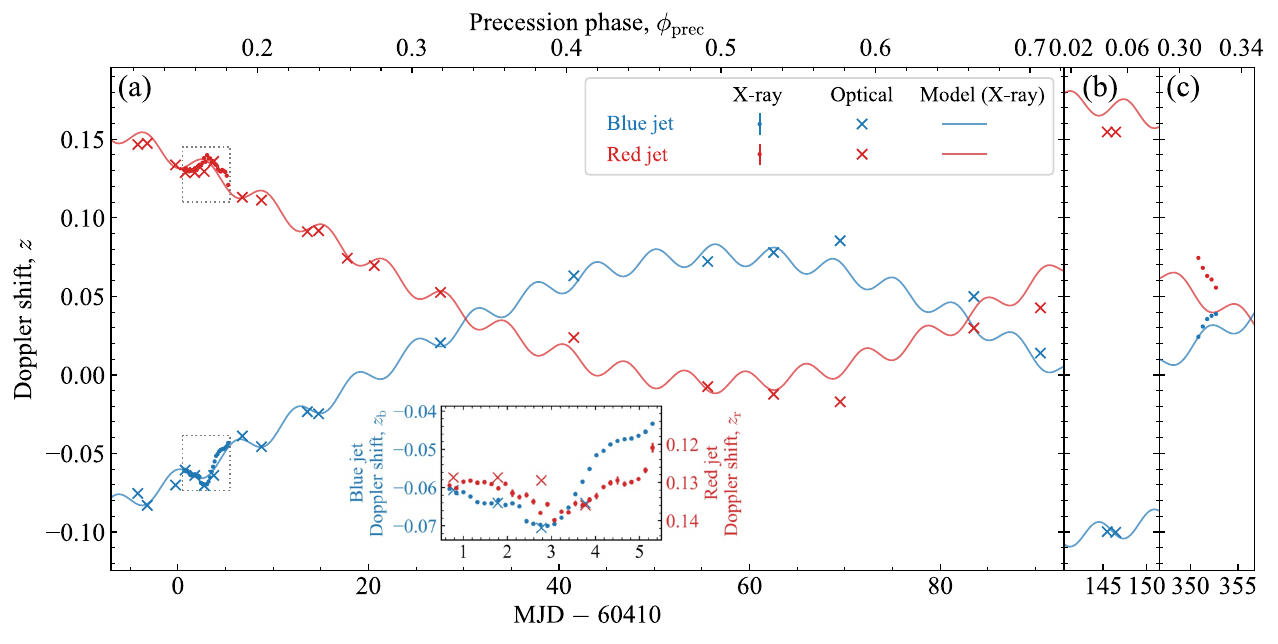}
\caption{Doppler shifts observed with XRISM/Resolve (X-ray) and Seimei/LCO (optical) from 2024 to 2025. (a--c) Blue and red data points indicate the blue and red jets, respectively. Panel (a) also includes an inset showing a zoom-in view of the blue and red jets within the highlighted rectangular region. The precession and nutation models are calculated based on the ephemeris and parameters from table~\ref{tab:ephemeris}, except for the reference epochs $t_{\mathrm{0,prec}}$ and $t_{\mathrm{nut}}$, which were fitted to the X-ray data as free parameters (see text).
}
 \label{doppler_shift_fit}
\end{figure*}

\subsection{Doppler-shift variability in X-ray and optical} \label{Doppler-shift variability in X-ray and optical}

The temporal variability of the Doppler shifts is illustrated in figure~\ref{doppler_shift_fit}. The blue jet shifts range from $z_\mathrm{b} \sim -0.10$ to $0.08$, and the red jet shifts from $z_\mathrm{r} \sim -0.02$ to $0.15$, exhibiting smooth periodic changes.
In the 2024 X-ray and optical data, the Doppler-shift variations appear to have different amplitudes for the blue and red jets (likely related to variations in the jet velocity, which are examined in detail in subsection~\ref{sec:jet_nutation_velocity}).

The Doppler shifts of SS~433 are often described by a commonly used kinematic model of jet precession \citep{Abell_1979}, with the nutation often represented phenomenologically by adding a sinusoidal term in redshift (e.g., \cite{Gies_2002, Davydov_2008}).
Such nutation is thought to arise from tidally driven torques from the companion star acting on the precessing disk \citep{Katz_1982}.
In practice, several modes may be present, but the $\sim$6.3-d component is dominant and is usually modeled by the sinusoidal term.
This combined precession and nutation model can be expressed as
\begin{equation}
\begin{split}
    z &= \gamma \left[1 \pm \beta (\cos i \cos \theta + \sin i \sin \theta \cos(2\pi \phi_{\mathrm{prec}})) \right] - 1,\\
    \delta z &= \pm A_{\mathrm{nut}} \sin(2\pi \phi_{\mathrm{nut}}),
    \label{prec_and_nut_eq}
\end{split}
\end{equation}
where $\beta = v/c$ is the intrinsic jet speed normalized to the speed of light, 
$\gamma = (1 - \beta^2)^{-1/2}$ is the Lorentz factor, 
$i$ is the inclination angle of the jet axis with respect to the line of sight, 
and $\theta$ is the half-opening angle of the precession cone. 
The precession and nutation phases are defined as 
$\phi_{\mathrm{prec}} = (t - t_{\mathrm{0,prec}})/P_{\mathrm{prec}}$ and 
$\phi_{\mathrm{nut}} = (t - t_{\mathrm{0,nut}})/P_{\mathrm{nut}}$, 
with $P_{\mathrm{prec}}$ and $P_{\mathrm{nut}}$ denoting their respective periods. 
The total Doppler shift is given by $z + \delta z$, 
with the minus and plus signs corresponding to the blue and red jets, respectively.

To enable direct comparison with earlier determinations of the ephemerides, we applied this model to our data. In the fitting, $\beta$, $i$, $\theta$, $A_{\mathrm{nut}}$, $P_{\mathrm{prec}}$, and $P_{\mathrm{nut}}$ were fixed to the tabulated values in table~\ref{tab:ephemeris}, as constrained by previous studies of the binary system, while $t_{\mathrm{0,prec}}$ and $t_{\mathrm{0,nut}}$ were treated as free parameters.
The blue and red jets were fitted simultaneously under the assumption of shared model parameters, using all X-ray data from both the 2024 and 2025 observations. 
The resulting model curves are shown in figure~\ref{doppler_shift_fit}. 
The updated reference epochs are summarized in 
table~\ref{tab:ephemeris}, showing offsets of about 
+10~d (precession; relative to \cite{Gies_2002}) and 
+2.8~d (nutation; relative to \cite{Davydov_2008}) 
with respect to the adopted ephemerides. 
Similar offsets have also been reported in previous studies 
(e.g., \cite{Kubota_2010b, Marshall_2013, Cherepashchuk_2018}).

\begin{figure}[ht!]
 \includegraphics[width=1\linewidth]{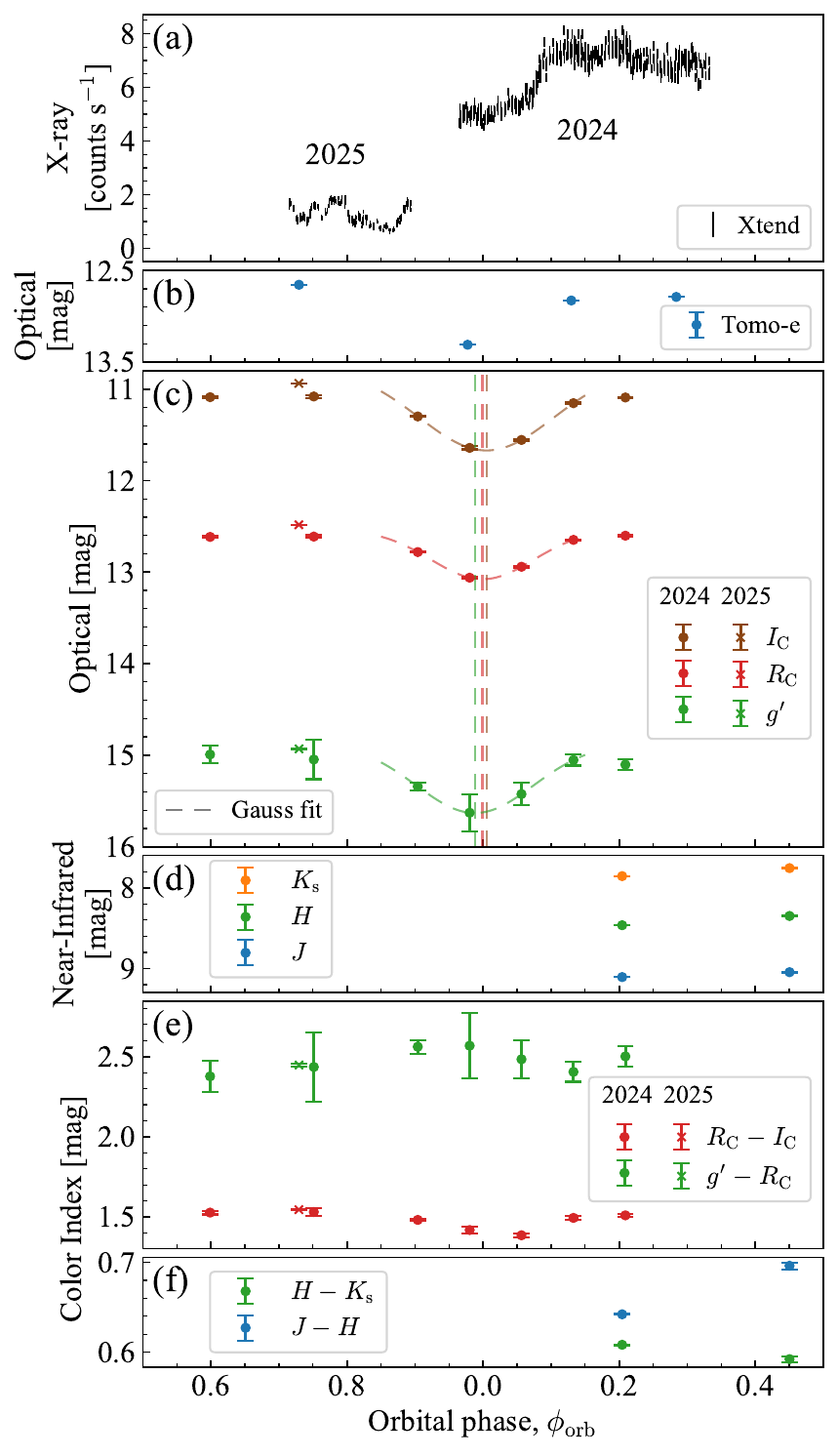}
\caption{Summary of simultaneous X-ray, optical and near-infrared observations conducted in 2024 and 2025. The orbital phase is calculated based on the ephemeris from \citet{Cherepashchuk_2023}.  
(a) XRISM/Xtend light curve in the 2--10~keV band.  
(b) Tomo-e Gozen optical light curve.  
(c) MITSuME optical light curves in the $I_\mathrm{C}$ (brown), $R_\mathrm{C}$ (red), and $g'$ (green) bands.  
Dashed curves represent Gaussian fits to the MITSuME data obtained in April 2024 within $|\phi_\mathrm{orb}| < 0.15$, and vertical dashed lines mark the fitted eclipse center times (see text).  
(d) kSIRIUS near-infrared light curves in the $K_\mathrm{s}$ (orange), $H$ (green), and $J$ (blue) bands.  
(e) Color indices derived from panel (c), calculated as $R_\mathrm{C} - I_\mathrm{C}$ (red) and $g' - R_\mathrm{C}$ (green).
(f) Color indices derived from panel (d), calculated as $H - K_\mathrm{s}$ (green) and $J - H$ (blue).
 }
 \label{light_curve_overview}
\end{figure}

\begin{figure*}[ht!]
 \includegraphics[width=1\linewidth]{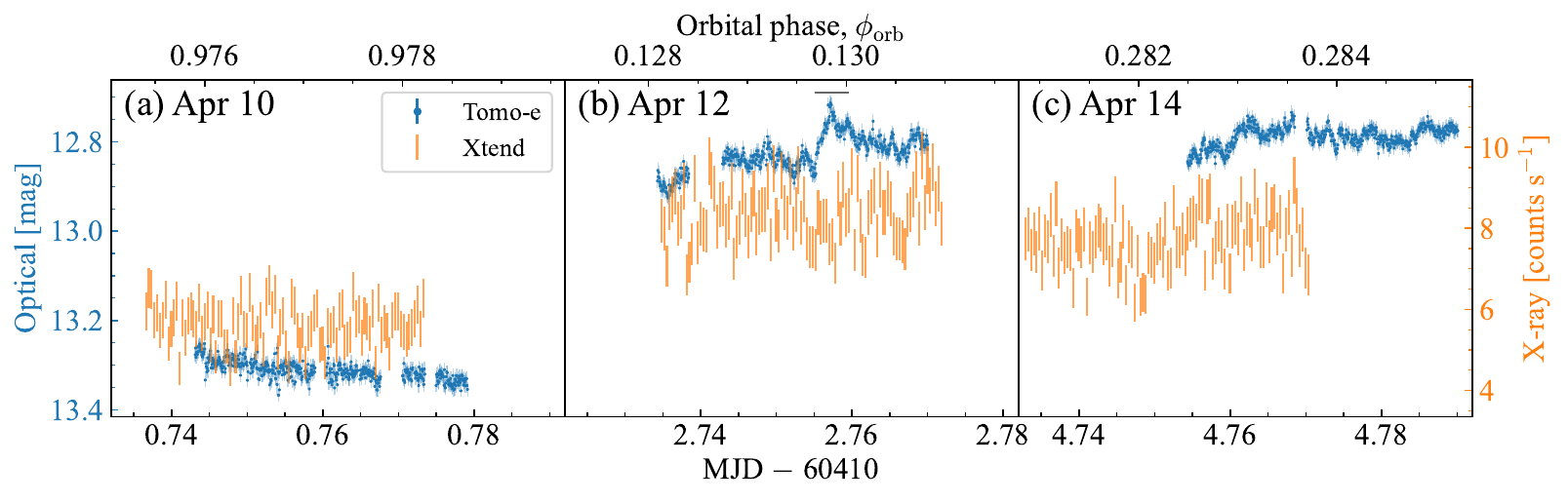}
\caption{Light curves obtained with Tomo-e Gozen and XRISM/Xtend in 2024.
(a)--(c) correspond to April 10, 12, and 14, 2024, respectively, showing optical data from Tomo-e Gozen (blue) and X-ray data from XRISM/Xtend (orange; 2--10 keV, 32 s bins).  
In panel (b), an optical flare is visible, with a 400~s gray bar denoting its approximate duration.
}
 \label{light_curve_2024}
\end{figure*}

\begin{figure}[ht!]
 \includegraphics[width=1\linewidth]{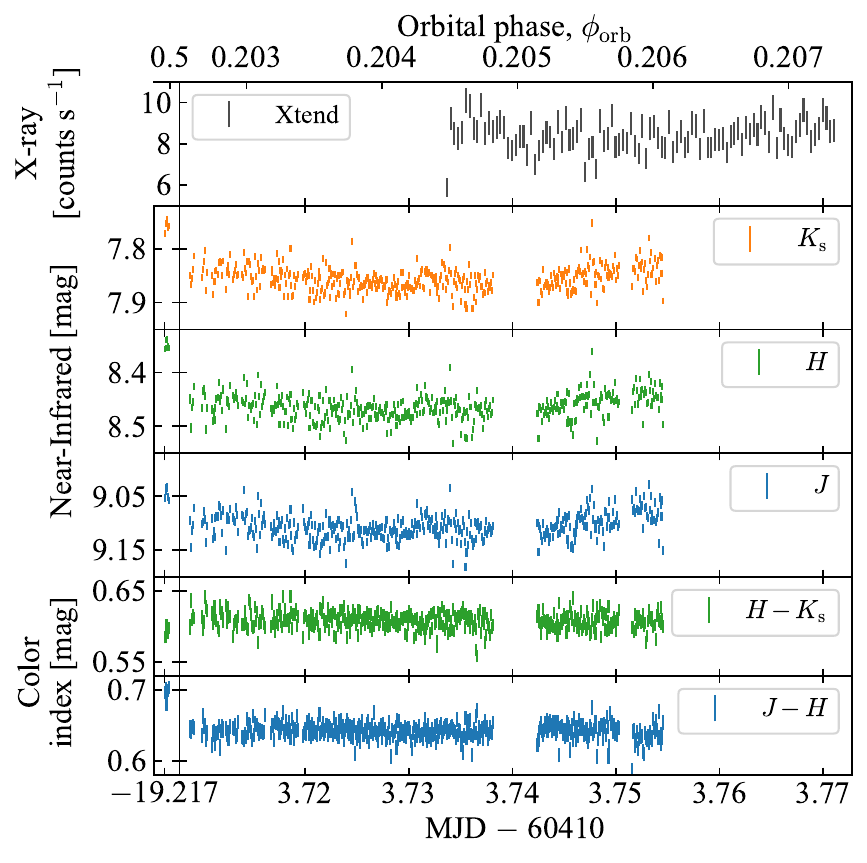}
\caption{kSIRIUS and XRISM/Xtend light curves in 2024.
From top to bottom, the panels show the XRISM/Xtend light curve (2--10 keV, 32~s bins), the $K_\mathrm{s}$, $H$, and $J$ bands, and the color indices $H - K_\mathrm{s}$ and $J - H$ from kSIRIUS. For the kSIRIUS panels, the left and right columns correspond to the observations on March 21 and April 14, respectively.
}
 \label{ksirius_light_curve_2024}
\end{figure}

\begin{figure}[ht!]
 \includegraphics[width=1\linewidth]{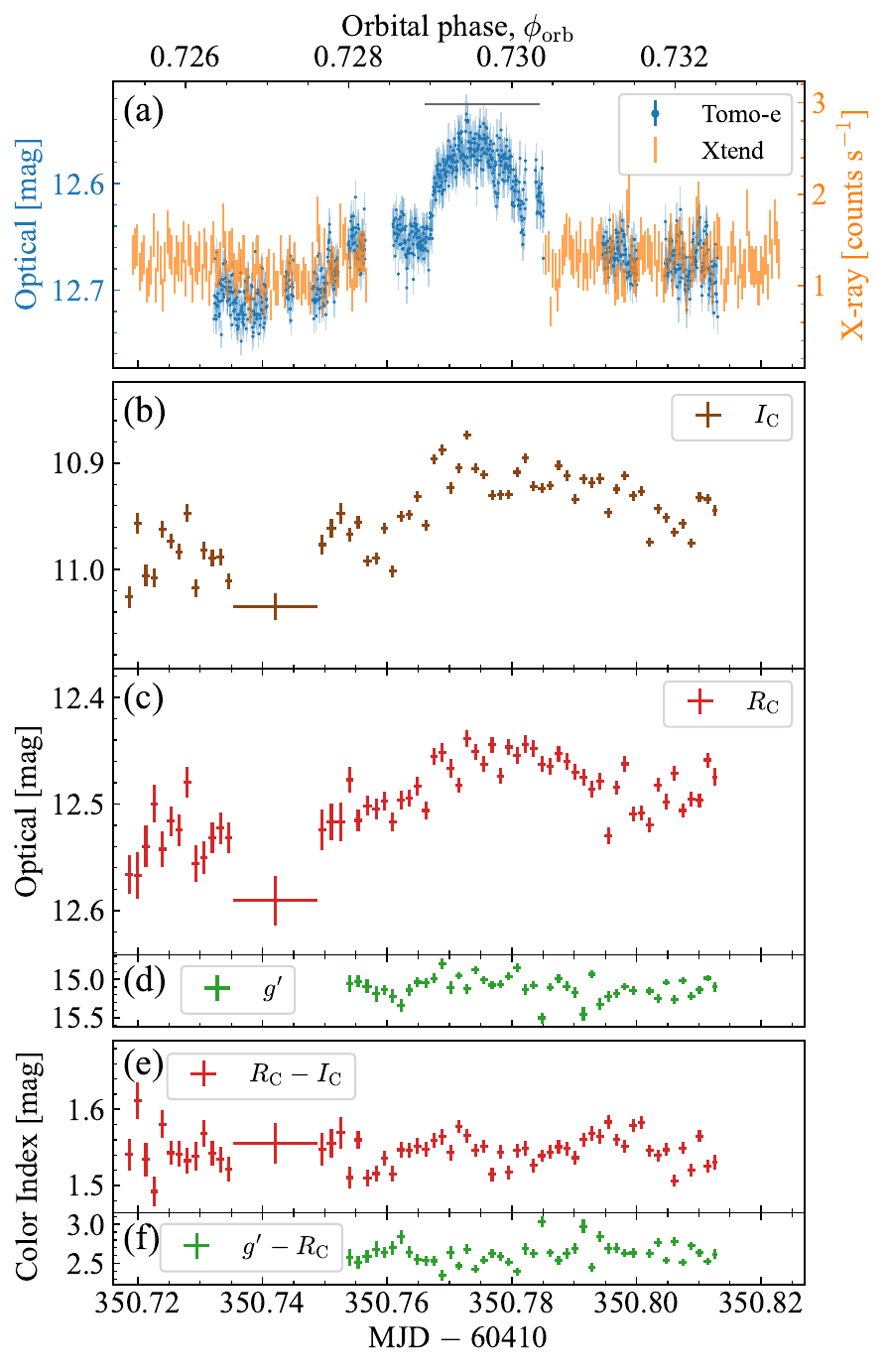}
\caption{ 
Light curves with Tomo-e Gozen, XRISM/Xtend, and MITSuME obtained in 2025.
(a) Light curve from March 26, showing optical data from Tomo-e Gozen (blue) and X-ray data from XRISM/Xtend (orange; 2--10~keV, 32 s bins). 
A gray bar with a width of 1600 s indicates the approximate duration of the optical flare.
(b)--(d) show MITSuME light curves in the $I_\mathrm{C}$, $R_\mathrm{C}$, and $g'$ bands.  
(e) and (f) show the color indices calculated as $R_\mathrm{C} - I_\mathrm{C}$ and $g' - R_\mathrm{C}$, respectively.
}
 \label{light_curve_2025}
\end{figure}

\subsection{Simultaneous multiwavelength light curves} \label{sec:light_curve}

Simultaneous multiwavelength light curves were obtained during the campaign.
The photometric results from all instruments are shown in figure~\ref{light_curve_overview}.
In the X-ray light curves (figure~\ref{light_curve_overview}a), the 2024 data ($\phi_{\mathrm{prec}} \sim 0.2$) show a higher count rate than the 2025 data ($\phi_{\mathrm{prec}} \sim 0.3$), consistent with previous reports that the X-ray brightness varies with precession phase (e.g., \cite{Gies_2002, Goranskij_2011}).

In the optical bands (figures~\ref{light_curve_overview}b and \ref{light_curve_overview}c), 
the light curves are primarily modulated by the orbital phase through eclipses, 
with additional variations in eclipse depth and overall brightness reported to depend on the precession phase 
(e.g., \cite{Cherepashchuk_2025}). 
Specifically, in the $V$ band, during phases $|\phi_{\mathrm{prec}}| < 0.2$, the brightness difference between eclipse and out-of-eclipse phases is typically $\Delta V \sim 0.6$~mag (e.g., \cite{Cherepashchuk_2021}), a trend also seen in the Tomo-e Gozen and MITSuME light curves. The photometric colors are consistent with previous studies (e.g., optical: \cite{Goranskij_2011, Cherepashchuk_2025}; near-infrared: \cite{Kodaira_1985, Cherepashchuk_2005}).
At similar orbital phases, the 2025 data are brighter by about $\sim0.1$~mag compared to the 2024 data, without accompanying changes in the color indices.

To check the consistency of the orbital phase assignment ($\phi_{\mathrm{orb}}$), we adopted the ephemeris of \citet{Cherepashchuk_2023}, which includes the orbital period derivative ($\dot{P}_{\mathrm{orb}}$). We compared the predicted eclipse timing with the 2024 MITSuME light curves and performed simple Gaussian fits near the eclipse phase ($|\phi_{\mathrm{orb}}| < 0.15$), following the procedure of \citet{Cherepashchuk_2023}. The fitted eclipse centers, shown in figure~\ref{light_curve_overview}c, are $\phi_{\mathrm{orb}} = 0.005 \pm 0.001$ in $I_\mathrm{C}$, $-0.001 \pm 0.001$ in $R_\mathrm{C}$, and $-0.01 \pm 0.01$ in $g'$, all consistent with $\phi_{\mathrm{orb}} = 0$ within $1\sigma$, except for the $I_\mathrm{C}$ band, which shows a slightly larger offset. However, given the limited number of data points near eclipse, this deviation is not considered significant. The 2024 X-ray light curve (figure~\ref{light_curve_overview}a) also shows egress transitions consistent with this phase reference, supporting the validity of the \citet{Cherepashchuk_2023} ephemeris for our dataset.

We now focus on the light curves obtained simultaneously with the X-ray data.
Figure~\ref{light_curve_2024} presents the optical and X-ray light curves obtained in 2024. 
Panel (a) corresponds to the eclipse phase, while panels (b) and (c) show out-of-eclipse intervals. 
In panel (b), an optical flare with a timescale of $\sim$400~s was observed, showing a flux enhancement of $\sim$15\% above the pre-flare level. Within the statistical uncertainties, no significant variability is seen in the X-ray light curve during the flare. The observed timescale and amplitude are similar to the $R_\mathrm{C}$-band optical variability reported by \citet{Burenin_2011}.
We also examined near-infrared light curves from kSIRIUS (figure~\ref{ksirius_light_curve_2024}) during the out-of-eclipse phase in 2024, finding that both the X-ray and near-infrared light curves showed no significant variability within the statistical uncertainties.

The optical and X-ray light curves obtained in 2025 are shown in figure~\ref{light_curve_2025}.
The Tomo-e Gozen light curve shows a clear brightening of $\sim$10\% over a timescale of $\sim$1600~s. 
This brightening is also seen in the $I_\mathrm{C}$ and $R_\mathrm{C}$ bands from MITSuME, and the $g'$ band shows no significant variation within the current signal-to-noise ratio.
The X-ray coverage does not reach the optical peak, but the X-ray light curve stays flat during the overlap, similar to the optical trend.
While the durations of the optical flares differ between 2024 and 2025, their rise and decay profiles are broadly similar. 
Given the limited number of detected events, interpretation of their physical origin is deferred to future investigations.

\section{Discussion}\label{sec:Discussion}

\begin{figure}[ht!]
 \includegraphics[width=1\linewidth]{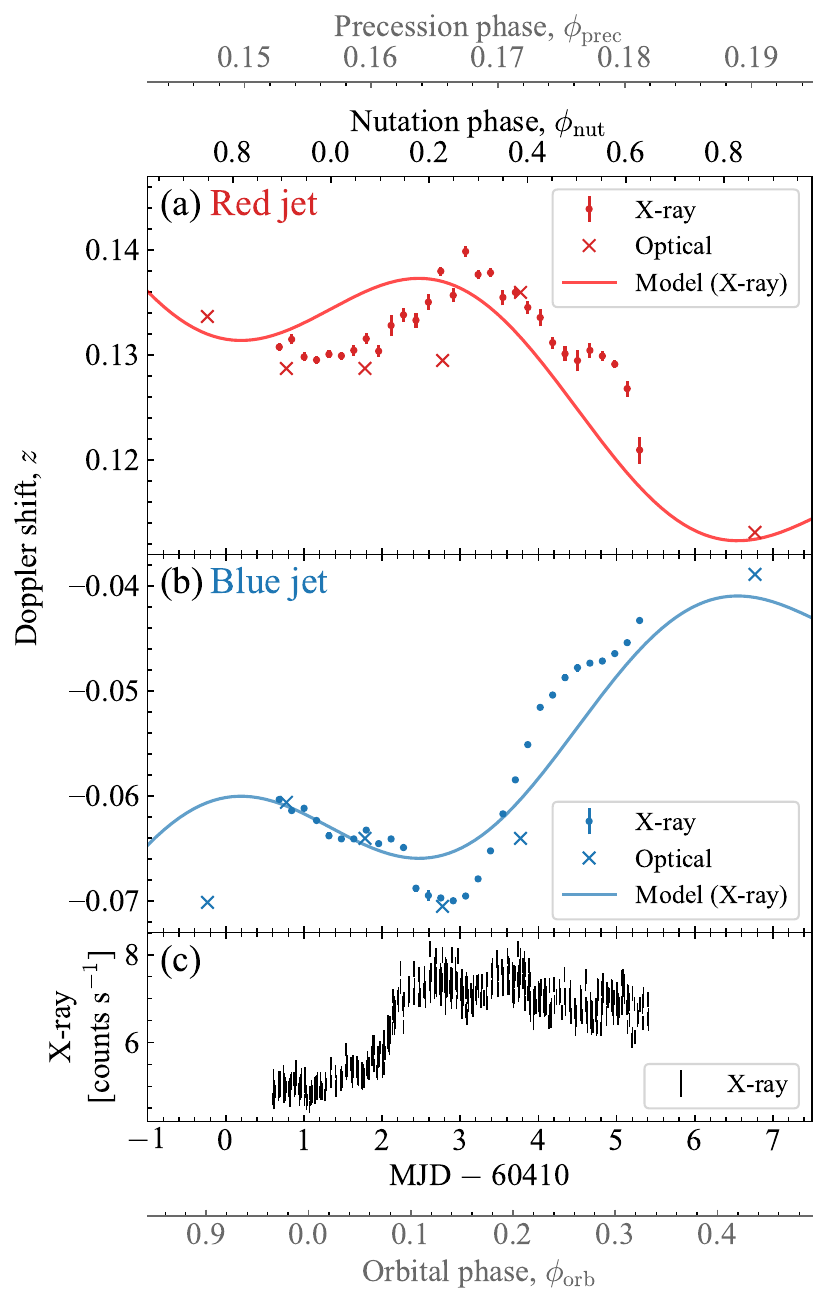}
\caption{
(a, b) Close-up views of figure~\ref{doppler_shift_fit}a, focusing on the 2024 data. The same symbols and line styles are used as in the original figure.
(c) XRISM/Xtend light curve in the 2--10~keV band. 
}
 \label{doppler_shift_zoom}
\end{figure}

\subsection{Summary of observational results}\label{sec:observation summary}

We summarize the main results from our coordinated X-ray, optical, and near-infrared campaign on SS~433. 
Time-resolved spectroscopy was conducted with XRISM/Resolve in X-rays and simultaneously with Seimei and LCO in the optical. 
The X-ray spectra were divided into multiple time segments with typical exposures on the order of 10 ks, from which the jet Doppler shifts were obtained by fitting thermal plasma emission models.
In the optical band, the shifts were measured from the peak positions of the H$\alpha$ emission lines. 
The statistical accuracy of the X-ray measurements reached $\Delta z \lesssim 3 \times 10^{-4}$ per segment, allowing variability on the nutation timescale to be probed. 
The X-ray and optical Doppler shifts showed closely aligned variations of both the blue and red jets. 
Applying the standard model of Doppler-shift variations, in which precession is supplemented by a sinusoidal nutation term (equation~\eqref{prec_and_nut_eq}), yielded phase offsets of a few days relative to the ephemerides of \citet{Gies_2002} and \citet{Davydov_2008}, consistent with previous results.

Photometric observations with Tomo-e Gozen, MITSuME, and kSIRIUS indicated that the optical and near-infrared colors, as well as the eclipse timings, were broadly consistent with earlier studies. 
The optical data also revealed short-timescale flares with durations of $\sim$400--1600~s and amplitudes up to $\sim$15\% in both 2024 and 2025. 
Although face-on configurations are generally associated with higher optical fluxes, the 2025 edge-on observations were brighter by $\sim$0.1~mag compared to the 2024 data at similar orbital phases, without significant changes in color indices. 
This behavior cannot be readily explained by a geometrically thin disk model, in which the projected emitting area would be expected to decrease toward edge-on phases.

\begin{figure}[ht!]
 \includegraphics[width=0.95\linewidth]{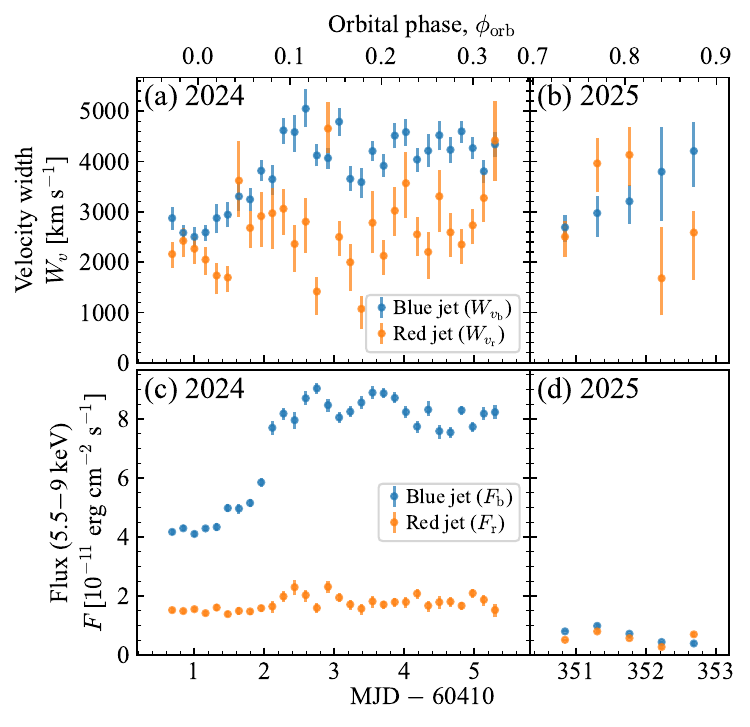}
\caption{
X-ray jet velocity widths and fluxes derived from Resolve spectral fitting.
(a, b) Velocity widths (FWHM; $W_v$) of the blue ($W_{v_\mathrm{b}}$) and red ($W_{v_\mathrm{r}}$) jets obtained from the Resolve spectra in the 5.5--9 keV band for 2024 and 2025, respectively (see also \citet{Shidatsu_2025} for details of the jet velocity width).
(c, d) Fluxes of the blue ($F_\mathrm{b}$) and red ($F_\mathrm{r}$) jets in the same energy band and epochs.
}
 \label{velocity_vs_flux}
\end{figure}

\subsection{X-ray vs. optical Doppler shifts and the impact of eclipse} \label{sec:jet_general}

A comparison between the temporal variations of Doppler shifts in the X-ray and optical bands provides important constraints on the spatial extent and physical conditions of the emitting regions. In previous studies, \citet{Kubota_2010b} analyzed Suzaku/XIS CCD spectra and, because of the limited energy resolution available at that time, modeled the data by superimposing Fe and Ni emission lines with Gaussian components on a continuum. They reported a delay of $\sim$0.5 d in the optical Doppler shifts relative to the X-rays. Furthermore, Chandra/HETG grating spectroscopy \citep{Marshall_2013} examined Mg\,\textsc{xii}, Si\,\textsc{xiv}, and Fe\,\textsc{xxvi} Ly$\alpha$ lines, as well as the Fe\,\textsc{xxv} He$\alpha$ line, and derived an upper limit of $\lesssim$~0.4 d for such a lag.

In the simultaneous XRISM/Resolve and optical observations in 2024, the superior spectral resolution and effective area of Resolve enabled direct application of physical plasma models to the time-resolved spectra. A magnified view of the Doppler-shift variations is shown in figures~\ref{doppler_shift_zoom}a and \ref{doppler_shift_zoom}b, showing the same data and model as presented in figure~\ref{doppler_shift_fit}. The high spectral resolution of Resolve allowed us to identify a large number of emission lines with high fidelity enabled a significant improvement in the accuracy of Doppler shift measurements compared with earlier studies. The results support the trend previously reported by \citet{Kubota_2010b} and \citet{Marshall_2013}, in which the optical Doppler shifts lag behind the X-rays. Future longer-duration monitoring covering different precessional phases will be required to establish the presence and possible phase dependence of such a delay with higher statistical significance.

The 2024 observation covered the egress, allowing us to test possible effects of partial occultation on the measured Doppler shifts.
If different velocity layers along the jet were selectively blocked by the companion, a sharp change in the Doppler shift could appear across the eclipse, but no significant variation is seen.
For a companion radius of order $\sim10^{12\text{--}13}$~cm (e.g.,~\cite{Burenin_2011}),
the expected light-travel delay is at most a few hundred seconds,
which is an order of magnitude shorter than the time resolution of our spectral analysis.
Therefore, the smooth evolution of the Doppler shift observed across the egress is consistent with the absence of detectable time-delay effects.

\subsection{Geometrical interpretation of jet flux and line broadening}\label{sec:jet_flux_line_beaming}

Following \citet{Shidatsu_2025}, who reported that the line widths of Fe and Ni lines (5.5--9~keV band) are larger than those of Si and S lines (2--4~keV band), and that the blue jet shows broader lines than the red jet in the 2024 data,
we investigate how these line-broadening properties are related to the jet fluxes derived from the Resolve spectra (figure~\ref{resolve_all_spectra}, table~\ref{tab:xray_doppler}).

Figures~\ref{velocity_vs_flux}a and \ref{velocity_vs_flux}b show the measured line widths of the blue ($W_{v_\mathrm{b}}$) and red ($W_{v_\mathrm{r}}$) jets. 
In 2024, the blue jet exhibited a slightly larger velocity width than the red jet, consistent with the tendency reported by \citet{Shidatsu_2025}, whereas in 2025 the velocity widths of the two jets were nearly identical.
The corresponding 5.5--9~keV fluxes ($F_\mathrm{b}$ and $F_\mathrm{r}$) are shown in figures~\ref{velocity_vs_flux}c and \ref{velocity_vs_flux}d. 
In 2024, the blue jet showed a stronger flux contrast between the in-eclipse and out-of-eclipse phases than the red jet, 
whereas in 2025 both jets exhibited similar brightness and variability throughout the observation.

Since the observed jet fluxes are affected by Doppler beaming,
we examined whether the measured flux variations can be explained by this effect.
In this framework, each jet flux scales as $F \propto D^{n+\Gamma}$,
where $D = 1/(1+z)$ is the Doppler factor calculated from the Doppler shifts of the blue and red jets measured in our spectral analysis.
Here, $n$ represents the jet emission state ($n = 2$ for a continuous flow and $n = 3$ for discrete blobs; e.g.,~\cite{Sikora_1997}).
We considered both cases of $n = 2$ and $n = 3$.
Although various interpretations of the appropriate value of $n$ have been discussed in the literature (e.g.,~\cite{Miller_2008}),
the difference is negligible for the present discussion, and we do not attempt to distinguish between them.
$\Gamma$ is the photon index, roughly 1.7 for thermal plasma emission in this band.

To isolate geometrical effects such as the emitting area,
we focus on the flux ratio under the assumption that both jets are intrinsically symmetric,
so that any intrinsic power variations cancel out in the ratio.
In 2024, the blue-to-red flux ratio expected from Doppler beaming was $\sim$2,
which can reasonably explain the in-eclipse data but not the out-of-eclipse phase.
In contrast, in 2025 the predicted ratio was close to unity, consistent with the observed values.
For a companion size on the order of the disk scale (e.g.,~\cite{Burenin_2011}),
such partial obscuration of the inner jet regions is plausible.
These results support a geometrical interpretation in which the larger line widths of the Fe and Ni emission
arise from inner jet regions that are partially obscured by the accretion disk and the companion star.

\begin{figure}[ht!]
 \includegraphics[width=1\linewidth]{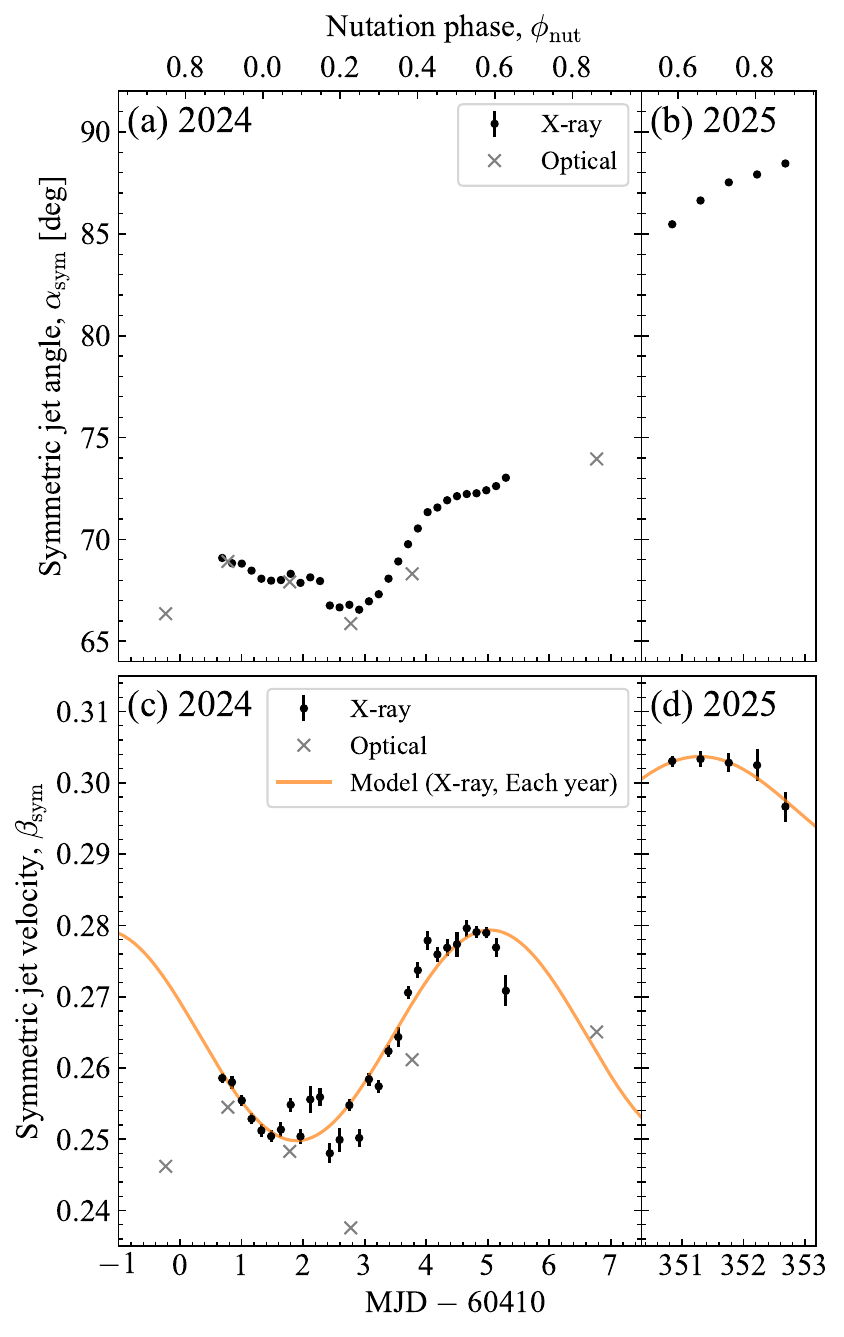}
\caption{
Jet angles ($\alpha_{\mathrm{sym}}$; panels a and b) and jet velocities ($\beta_{\mathrm{sym}}$; panels c and d) derived under the symmetric assumption (equation~\eqref{eq:jet_app}) for the years 2024 and 2025.
Black points and gray crosses indicate the observed blueshifted and redshifted components obtained with Resolve and Seimei, respectively.  
Solid lines in panels c and d show the best-fit sinusoidal models described by equation~\eqref{eq:beta_sym_fit}, with all parameters fitted independently for each year as listed in table~\ref{beta_sym_model}.
}
 \label{alpha_beta_nutation}
\end{figure}

\subsection{Observed jet angles and velocities}\label{sec:jet_nutation_velocity}

The observed Doppler shifts (figure~\ref{doppler_shift_fit}) can be used to investigate the kinematic properties of the jets, such as their ejection direction and velocity.
We assume that the blue and red jets are launched simultaneously in opposite directions, at symmetric angles, and share the same intrinsic speed \citep{Marshall_2002}.

Under this assumption, the Doppler shifts of the blue and red jets are expressed as
\begin{equation}
\begin{split}
z_{\mathrm{b}} &= \gamma_{\mathrm{sym}} \left(1 - \beta_{\mathrm{sym}} \cos\alpha_{\mathrm{sym}}\right) - 1, \\
z_{\mathrm{r}} &= \gamma_{\mathrm{sym}} \left(1 + \beta_{\mathrm{sym}} \cos\alpha_{\mathrm{sym}}\right) - 1,
\end{split}
\label{z_app}
\end{equation}
where $\alpha_{\mathrm{sym}}$ and $\beta_{\mathrm{sym}}$ denote the jet angle and speed under the symmetric assumption, and $\gamma_{\mathrm{sym}} = (1 - \beta_{\mathrm{sym}}^2)^{-1/2}$.
From equation~\eqref{z_app}, the apparent jet angle and velocity can then be derived as
\begin{equation}
\begin{split}
\cos \alpha_{\mathrm{sym}} &= \frac{z_\mathrm{r} - z_\mathrm{b}}{2 \left( 1+ \frac{z_\mathrm{b} + z_\mathrm{r}}{2} \right) \sqrt{ 1 - \left( 1 + \frac{z_{\mathrm{b}} + z_{\mathrm{r}}}{2} \right)^{-2}}}, \\
\beta_{\mathrm{sym}} &= \sqrt{1 - \left( 1 + \frac{z_{\mathrm{b}} + z_{\mathrm{r}}}{2} \right)^{-2}}.
\end{split}
\label{eq:jet_app}
\end{equation}

Figure~\ref{alpha_beta_nutation} shows the derived $\alpha_{\mathrm{sym}}$ and $\beta_{\mathrm{sym}}$, calculated from the Doppler shifts measured in the Resolve and optical data (figure~\ref{doppler_shift_fit}).
The Resolve measurements display a clear periodic modulation.
The jet angle $ \alpha_{\mathrm{sym}} $ follows the variation expected from the precession phase, 
with values of $\sim70^{\circ}$ in 2024 (near face-on) and $\sim90^{\circ}$ in 2025 (near edge-on), 
with additional variations attributable to nutation also being visible.
The jet velocity $\beta_{\mathrm{sym}}$ shows pronounced variations, ranging from $0.25c$--$0.28c$ in 2024 and from $0.29c$--$0.30c$ in 2025. These variations seem periodic, with a timescale comparable to that of nutation.

To quantify this apparent periodicity in $\beta_{\mathrm{sym}}$, 
we fit a simple sinusoid with a $-\pi/2$ phase offset, 
motivated by the observation that the modulation of $\beta_{\mathrm{sym}}$ 
appears shifted by about $-90^{\circ}$ relative to the nutation-phase variation of $\alpha_{\mathrm{sym}}$:
\begin{equation}
\beta_{\mathrm{sym,model}} = C_{\mathrm{sym}} + \delta \beta_{\mathrm{sym}}
\sin \left(2\pi \phi_{\mathrm{nut}} - \frac{\pi}{2} \right),
\label{eq:beta_sym_fit}
\end{equation}
where $C_\mathrm{sym}$ is the mean velocity and $\delta \beta_\mathrm{sym}$ the modulation amplitude. The nutation period was fixed to the value from \citet{Davydov_2008}, and fits were performed separately for 2024 and 2025 due to the large difference in $C_\mathrm{sym}$. The best-fit parameters from the Resolve data are summarized in table~\ref{beta_sym_model} and shown as solid lines in figures~\ref{alpha_beta_nutation}c and \ref{alpha_beta_nutation}d.

\begin{table}[htbp]
\caption{Fit results for the apparent jet velocity model in equation~\eqref{eq:beta_sym_fit}, corresponding to the curves shown in Figure~\ref{alpha_beta_nutation}.}
\label{beta_sym_model}
\begin{tabular}{lcc}
\toprule
\textbf{Parameter} & \textbf{2024} & \textbf{2025} \\
\midrule
$t_{0, \mathrm{nut}}$  &  $43032.97 \pm 0.02$ &  $43033.41 \pm 0.20$  \\
$C_\mathrm{sym}$    &  $0.2646 \pm 0.0002$ &       $0.2962 \pm 0.0031$  \\
$\delta \beta_\mathrm{sym}$ &  $0.0148 \pm 0.0003$ &  $0.0075 \pm 0.0035$  \\
\bottomrule
\end{tabular}
\end{table}

Figures~\ref{alpha_beta_nutation}c and \ref{alpha_beta_nutation}d indicate modulations in the jet speed $\delta\beta_{\mathrm{sym}}$ on a timescale of $\sim$6.3~d, comparable to the nutation period. From table~\ref{beta_sym_model}, they fluctuate around $\sim0.26 \pm 0.01c$ in 2024 and $\sim0.30 \pm 0.01c$ in 2025, with a phase offset of about $-90^{\circ}$ relative to the nutation cycle. These results indicate day-scale modulations as well as a modest year-scale increase in the jet speed between the two campaigns. 

Jet speeds have been reported to vary on multiple timescales.
In the optical band, variations on timescales comparable to the orbital period ($\sim$13.1~d) with amplitudes of $\sim0.02c$ have been observed (e.g., \cite{Blundell_2007}).
At radio wavelengths, changes over several tens of days reaching up to $\sim0.3c$ have been reported (e.g., \cite{Blundell_2011, Jeffrey_2016}), sometimes accompanied by enhanced radio and optical emission. The variations observed with Resolve are comparable in amplitude to these earlier results but occur on shorter timescales of $\sim$6.3~d. The higher velocity detected in 2025, reaching $\sim0.3c$, may be related to the scenario suggested by \citet{Blundell_2011}, as our optical observations in 2025 showed relatively high fluxes (see subsection~\ref{sec:observation summary}). 
We note, however, that no contemporaneous radio monitoring was conducted during our campaign, and thus a direct correlation between optical brightness and radio activity could not be evaluated in this work. 

The new Resolve measurements show that the jet velocity varies on timescales comparable to nutation, highlighting that the physical mechanism driving these modulations is still not well understood.
The fundamental nature of SS~433's baryonic jets---whether they form a continuous stream or consist of discrete, blob-like ejecta---remains uncertain, and it is not yet clear whether the two jets are truly symmetric or share identical velocities.
In this analysis, our modeling assumes symmetric jets with equal speeds; however, this simplification may not fully represent the intrinsic dynamics of the system.
Interpreting the observed modulations may therefore require considering possible intrinsic asymmetries or time-variable jet speeds linked to the jet-launching physics.
Polarization studies in the radio (e.g., \cite{Miller_2008}) and X-ray bands (e.g., \cite{Kaaret_2024}) suggest that magnetic fields tend to align with the jet axis, although the symmetry between the jet and counter-jet may not be perfect.
In a super-Eddington accretion regime, radiation pressure and magnetic fields likely interact to shape both the stability and symmetry of the outflows.
Since SS~433's jets evolve under the combined influence of velocity and density contrasts, gravity, radiation, and magnetic stresses, long-term, multi-wavelength, and phase-resolved observations will be essential to determine whether these modulations are periodic and to clarify the physical processes that drive relativistic outflows.

\section{Conclusion} \label{sec:Conclusion}

We have presented results from XRISM's coordinated multiwavelength campaign of SS~433, conducted in April 2024 and March 2025, together with optical and near-infrared observations. 
The XRISM exposures were $\sim$200~ks in 2024 and $\sim$100~ks in 2025, covering distinct orbital and precessional configurations: in 2024 the approaching jet was inclined toward the observer, while in 2025 the jet axis was nearly perpendicular to the line of sight.

The key advancement from these campaigns comes from time-resolved Doppler-shift measurements with XRISM/Resolve. Resolve's high spectral resolution and effective area enabled precise separation of numerous emission lines and substantially improved Doppler-shift accuracy compared with earlier studies. Simultaneous optical spectroscopy suggested a tendency for the optical emission to lag slightly behind the X-rays. Resolve revealed an asymmetry between the two jet components, with velocities of $\sim0.26c$ in 2024 and $\sim0.30c$ in 2025, showing sinusoidal modulations with amplitudes $\sim\pm0.01c$. These variations follow the 6.3-d nutation period but are offset in phase by about $-90^{\circ}$, suggesting a possible connection between nutation and jet-velocity fluctuations.
These results likely reflect changes in viewing geometry between the two epochs, which affected how the inner jet regions were seen. 
The blue jet in 2024, observed during the out-of-eclipse phase, showed a larger velocity width than the red jet, whereas in 2025 both jets showed comparable widths.
The jet brightness also followed this geometrical trend, consistent with the interpretation by \citet{Shidatsu_2025} that the line widths increase toward the inner regions of the jet.

Optical photometry detected short flares of $\sim$400~s in 2024 and $\sim$1600~s in 2025, with amplitudes up to $\sim$15\% during out-of-eclipse intervals. X-ray variability was not detected within the statistical limits and within the limited duration of the simultaneous observations. Near-infrared observations with kSIRIUS in 2024 did not show significant variability, consistent with the X-ray data. The optical flux in 2025 was higher by $\sim$0.1 mag relative to 2024 at comparable orbital phases, while the color indices remained consistent across the two epochs. Continued long-term monitoring across a broader range of orbital, precessional, and nutation phases---with XRISM and complementary multiwavelength observations---will be essential for disentangling these effects and refining our understanding of the dynamical properties of the relativistic jets in SS~433.

\section*{Funding}
D.H. acknowledges support from NASA grant HST-GO-17770.002-A.  
This work was also supported by the U.S. National Science Foundation (NSF) under Cooperative Agreement PHY-2019786 (The NSF AI Institute for Artificial Intelligence and Fundamental Interactions, \url{http://iaifi.org/}).  
K.O. acknowledges support from the Korea Astronomy and Space Science Institute under the R\&D program (Project No. 2026-1-831-01), supervised by the Korea AeroSpace Administration, and from the National Research Foundation of Korea (NRF) grant funded by the Korea government (MSIT) (RS-2025-00553982).  
Additional support was provided by JSPS KAKENHI Grant Numbers 24KJ2067 (Y.S.), 19K14762, 23K03459, 24H01812 (M.S.), 20H01946 (Y.U.), 22H01272, 23K22543, 24K00672 (M.M.), 22K20386 (H.S.), 24H00248, 24K00680, and 25K01041 (K.N.).

\begin{ack}
We thank all colleagues involved in the coordinated multiwavelength campaign, including the teams of the XRISM, Seimei, LCO, Tomo-e Gozen, MITSuME, and kSIRIUS, for their complementary observations and support.  
M.S., T.T., T.U., M.U., and M.Y. acknowledge support by Ehime University Grant-in-Aid Research Empowerment Program.
The 3.8~m Seimei Telescope observations were carried out through programs 24A-N-CT07 (NAOJ open use) and 24A-K-0015 (Kyoto University time).  
We are grateful to the graduate and undergraduate students who conducted near-infrared observations of kSIRIUS using the 1-m Kagoshima telescope at Iriki Observatory.
This work was partially supported by the joint research program of the Institute for Cosmic Ray Research (ICRR), the University of Tokyo.
The Seimei telescope at the Okayama Observatory is jointly operated by Kyoto University and the National Astronomical Observatory of Japan (NAOJ), with assistance provided by the OISTER program.
The Kagoshima University 1 m telescope and the MITSuME 50 cm telescopes at Akeno and Okayama are members of the OISTER program, funded by MEXT of Japan.

We further acknowledge the use of public data and services from the High Energy Astrophysics Science Archive Research Center (HEASARC) at NASA/GSFC, and the astrometric calibration tools provided by WCS coordinate routines and the open-source project Astrometry.net (\url{https://astrometry.net/}) \citep{Lang_2010}. We thank the developers for making these resources freely available to the community.
We also thank the referee for constructive and helpful comments that improved the clarity of the manuscript.
\end{ack}


\appendix

\section{Tomo-e Gozen field of view and photometric aperture}\label{appendix:tomoe_fov}

\begin{figure}[ht!]
 \includegraphics[width=1\linewidth]{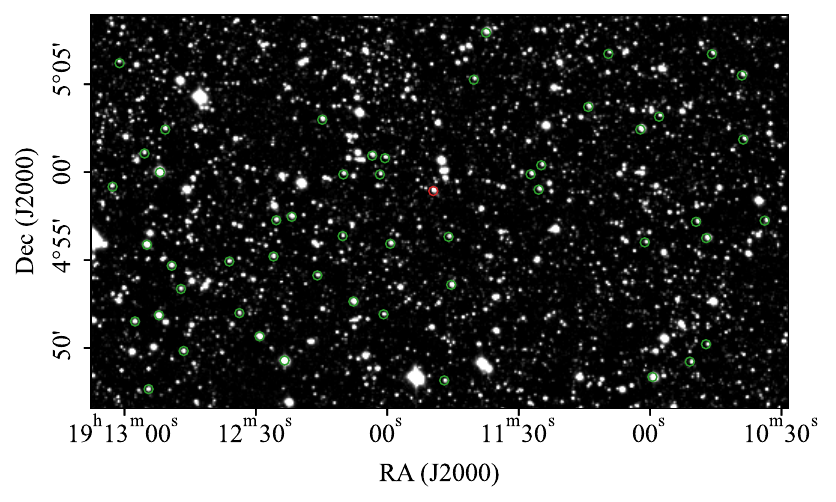}
\caption{
Full-field optical image of the SS~433 region obtained with Tomo-e Gozen. The image corresponds to the first data cube acquired on 2024 April 10 and is generated by stacking 10 consecutive frames with background subtraction applied. SS~433 and the photometric calibration stars are indicated by red and green circles, respectively. All markers have a radius of 13 pixels ($\sim 15\arcsec$), matching the photometric aperture used in the analysis. 
}
 \label{ss433_tomoe_fov}
\end{figure}

This appendix presents the field of view and photometric aperture used in the Tomo-e Gozen analysis. Figure~\ref{ss433_tomoe_fov} shows the observation from 2024 April 10, highlighting SS~433 and the calibration stars. This image illustrates the aperture placement and source isolation, supporting the calibration procedure described in subsection~\ref{Tomo-e Gozen}.

\section{Color index diagrams}\label{appendix:color_index}

Color--magnitude and color--color diagrams are commonly used to examine the variability of astrophysical sources (e.g., \cite{Bessell_1988}).  
Here we present such diagrams derived from MITSuME and kSIRIUS photometric data for reference.

\begin{figure*}[ht!]

 \includegraphics[width=0.9\linewidth]{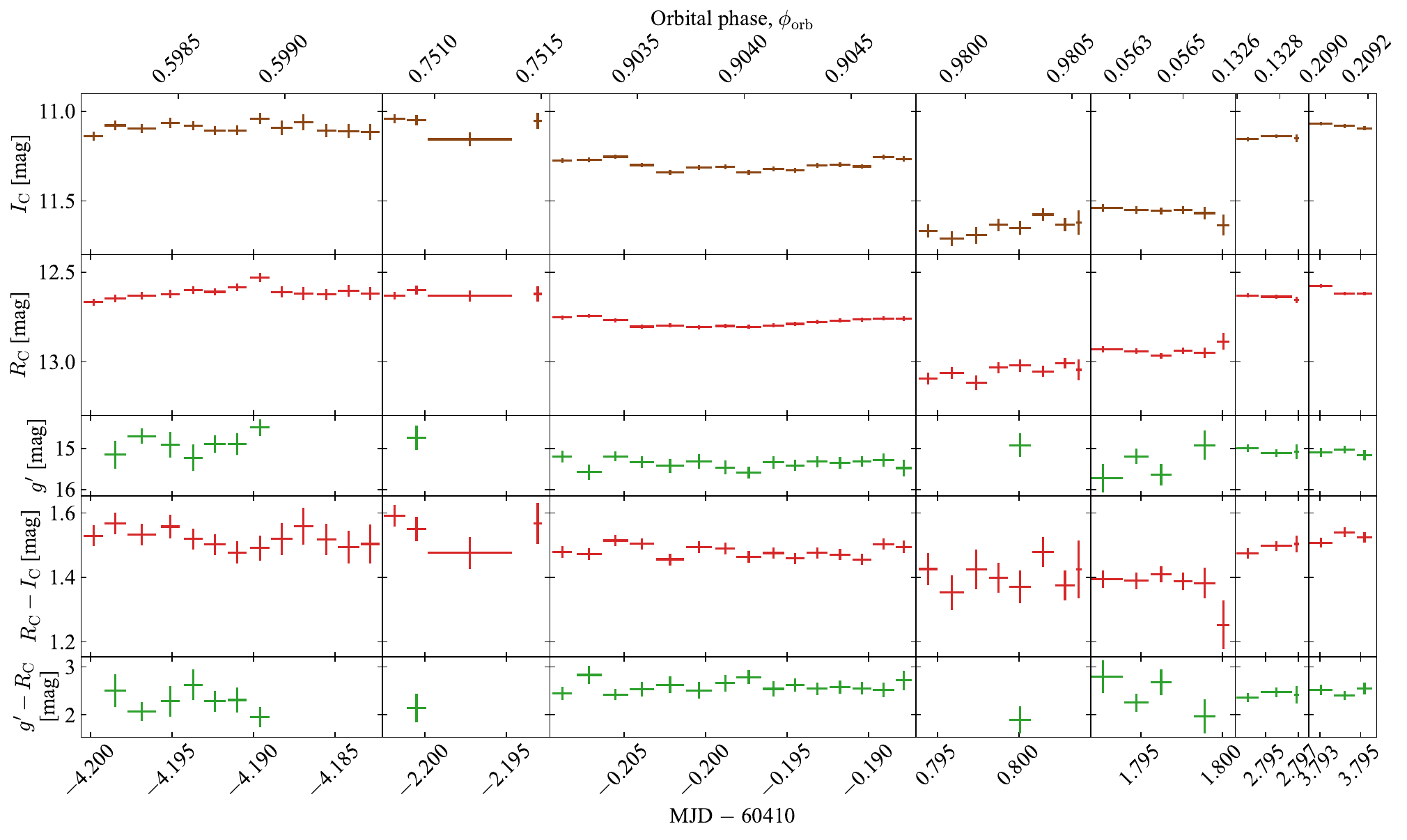}
\caption{
Light curves obtained with MITSuME during April 5--13, 2024.
From top to bottom, the panels show the $I_\mathrm{C}$, $R_\mathrm{C}$, $g'$ bands, followed by the color indices $R_\mathrm{C} - I_\mathrm{C}$ and $g' - R_\mathrm{C}$.
}
 \label{mitsume_light_curve_2024}
\end{figure*}

\begin{figure}[ht!]
 \includegraphics[width=1\linewidth]{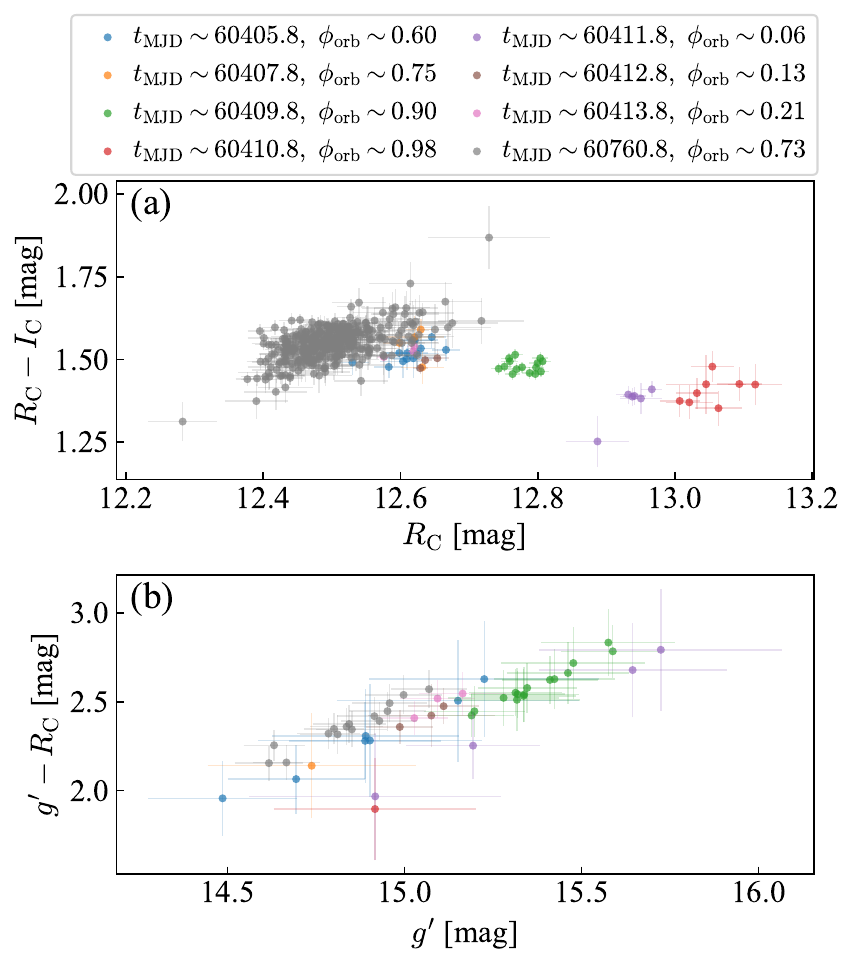}
\caption{
Scatter plots of magnitude versus color index based on the MITSuME light curve data shown in figure~\ref{mitsume_light_curve_2024} (2024) and figure~\ref{light_curve_2025} (2025).  
Panel (a) shows $R_\mathrm{C}$ versus $R_\mathrm{C} - I_\mathrm{C}$, and panel (b) shows $g'$ versus $g' - R_\mathrm{C}$.  
}
 \label{mitsume_color_index}
\end{figure}

\begin{figure}[ht!]
 \includegraphics[width=1\linewidth]{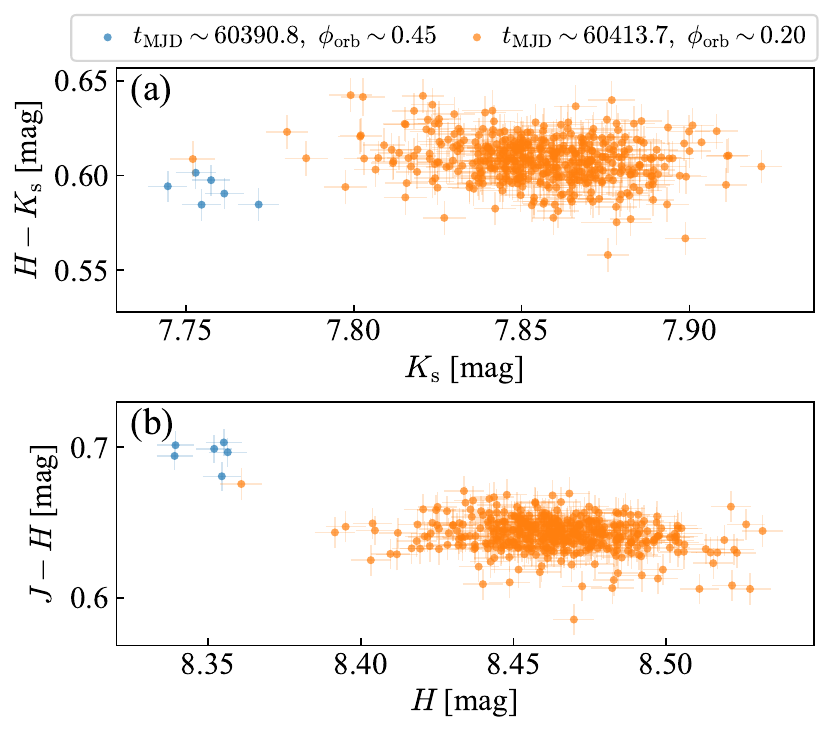}
\caption{
Scatter plots of magnitude versus color index based on the kSIRIUS light curve data shown in figure~\ref{ksirius_light_curve_2024}.
Panel (a) shows $K_\mathrm{s}$ versus $H - K_\mathrm{s}$, and panel (b) shows $H$ versus $J - H$.  
}
 \label{ksirius_color_index}
\end{figure}

Figure~\ref{mitsume_color_index} shows the MITSuME color--magnitude and color--color diagrams, derived from figures~\ref{mitsume_light_curve_2024} and \ref{light_curve_2025}.  
Figure~\ref{ksirius_color_index} presents the near-infrared diagrams from the kSIRIUS data shown in figure~\ref{ksirius_light_curve_2024}.  
Some optical panels appear to show weak trends between color and magnitude, whereas the near-infrared diagrams do not exhibit any obvious pattern.  

\section{Full-band spectrum of XRISM/Resolve} \label{appendix:full_resolve_spectrum}

\begin{figure}[ht!]
 \includegraphics[width=1.0\linewidth]{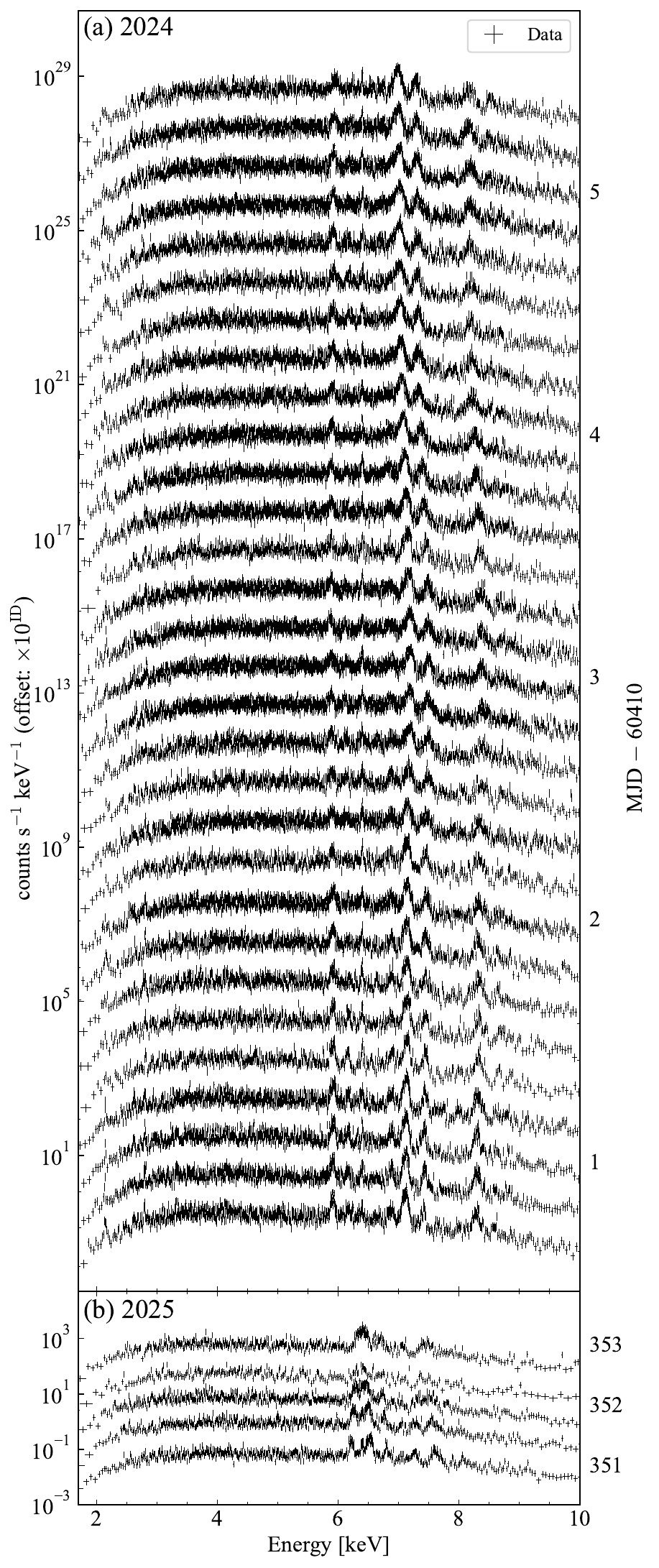}
\caption{Time-resolved XRISM/Resolve spectra in the 1.7--10~keV band from 2024 (a) and 2025 (b).  
Black points represent the observed data, shown with vertical offsets for clarity.  
The 2024 data are divided into 30 time segments, and the 2025 data into 5 segments.
}
 \label{resolve_all_spectra_full}
\end{figure}

For reference, we show the full-band (1.7--10 keV) XRISM/Resolve spectra.
Figure~\ref{resolve_all_spectra_full} displays the data for the same time intervals as in subsection~\ref{sec:spectral_analysis}, showing emission lines from lighter elements such as Si and S in the soft X-ray band (see also \cite{Shidatsu_2025}).

\begin{table*}[ht!]
\caption{Doppler shifts from the time-resolved Resolve spectra, fitted with a thermal model (see figure~\ref{resolve_all_spectra}).}
\label{tab:xray_doppler}
\begin{tabular}{r|rrlllllll}
\toprule
\textbf{ID} & \textbf{MJD$_{\mathrm{mid}}$} & \textbf{Exposure (s)} &
$\boldsymbol{z_\mathrm{b}}$ &
$\boldsymbol{W_{v_\mathrm{b}}}$\footnotemark[$*$] $\boldsymbol{(10^3)}$ &
$\boldsymbol{N_\mathrm{b}}$\footnotemark[$\dag$] $\boldsymbol{(10^1)}$ &
$\boldsymbol{z_\mathrm{r}}$ &
$\boldsymbol{W_{v_\mathrm{r}}}$\footnotemark[$*$] $\boldsymbol{(10^3)}$ &
$\boldsymbol{N_\mathrm{r}}$\footnotemark[$\dag$] $\boldsymbol{(10^1)}$ \\
\midrule
\multicolumn{9}{c}{\textbf{2024}} \\
\midrule
0 & $60410.684$ & $7995$ & $-0.0603 \pm 0.0003$ & $2.9^{+0.4}_{-0.3}$ & $0.87 \pm 0.07$ & $0.1308 \pm 0.0006$ & $2.2^{+0.5}_{-0.4}$ & $0.48 \pm 0.05$ \\
1 & $60410.843$ & $8130$ & $-0.0614^{+0.0004}_{-0.0002}$ & $2.6^{+0.3}_{-0.2}$ & $0.85 \pm 0.07$ & $0.1315^{+0.0008}_{-0.0007}$ & $2.4^{+0.5}_{-0.4}$ & $0.45 \pm 0.04$ \\
2 & $60411.002$ & $7921$ & $-0.0612 \pm 0.0003$ & $2.5 \pm 0.3$ & $0.78^{+0.07}_{-0.06}$ & $0.1298 \pm 0.0007$ & $2.3^{+0.5}_{-0.4}$ & $0.44 \pm 0.04$ \\
3 & $60411.161$ & $9192$ & $-0.0623 \pm 0.0003$ & $2.6 \pm 0.3$ & $0.86 \pm 0.06$ & $0.1295^{+0.0007}_{-0.0006}$ & $2.0^{+0.5}_{-0.4}$ & $0.43 \pm 0.04$ \\
4 & $60411.320$ & $4684$ & $-0.0638^{+0.0006}_{-0.0005}$ & $2.9 \pm 0.5$ & $0.77^{+0.08}_{-0.07}$ & $0.1301^{+0.0007}_{-0.0006}$ & $1.7^{+0.6}_{-0.4}$ & $0.43 \pm 0.05$ \\
5 & $60411.479$ & $5600$ & $-0.0641 \pm 0.0004$ & $2.9 \pm 0.4$ & $1.00^{+0.09}_{-0.08}$ & $0.1299^{+0.0007}_{-0.0006}$ & $1.7^{+0.5}_{-0.4}$ & $0.42^{+0.06}_{-0.05}$ \\
6 & $60411.638$ & $6230$ & $-0.0641 \pm 0.0005$ & $3.3 \pm 0.5$ & $0.98^{+0.13}_{-0.09}$ & $0.1304^{+0.0008}_{-0.0009}$ & $3.6^{+1.2}_{-1.3}$ & $0.46^{+0.07}_{-0.06}$ \\
7 & $60411.797$ & $7449$ & $-0.0633 \pm 0.0004$ & $3.2 \pm 0.4$ & $0.98^{+0.07}_{-0.06}$ & $0.1315 \pm 0.0009$ & $2.7^{+0.7}_{-0.5}$ & $0.42 \pm 0.05$ \\
8 & $60411.956$ & $9239$ & $-0.0645^{+0.0004}_{-0.0005}$ & $3.8 \pm 0.3$ & $1.10 \pm 0.07$ & $0.1304^{+0.0008}_{-0.0009}$ & $2.9^{+1.0}_{-0.8}$ & $0.45 \pm 0.05$ \\
9 & $60412.115$ & $4097$ & $-0.0641 \pm 0.0005$ & $3.6 \pm 0.5$ & $1.47 \pm 0.12$ & $0.1328^{+0.0016}_{-0.0017}$ & $3.0^{+1.2}_{-0.8}$ & $0.47 \pm 0.09$ \\
10 & $60412.274$ & $6982$ & $-0.0649 \pm 0.0005$ & $4.6 \pm 0.4$ & $1.57 \pm 0.11$ & $0.1338 \pm 0.0011$ & $3.1^{+0.8}_{-0.7}$ & $0.58 \pm 0.08$ \\
11 & $60412.433$ & $4455$ & $-0.0688^{+0.0006}_{-0.0005}$ & $4.6 \pm 0.6$ & $1.43^{+0.13}_{-0.12}$ & $0.1333^{+0.0012}_{-0.0011}$ & $2.4^{+0.9}_{-0.6}$ & $0.62 \pm 0.10$ \\
12 & $60412.592$ & $5285$ & $-0.0695^{+0.0009}_{-0.0008}$ & $5.0 \pm 0.6$ & $1.60^{+0.12}_{-0.11}$ & $0.1350^{+0.0012}_{-0.0013}$ & $2.8^{+1.0}_{-0.8}$ & $0.56 \pm 0.09$ \\
13 & $60412.751$ & $8682$ & $-0.0697^{+0.0003}_{-0.0005}$ & $4.1 \pm 0.4$ & $1.72 \pm 0.09$ & $0.1380 \pm 0.0007$ & $1.4^{+0.8}_{-0.5}$ & $0.46 \pm 0.07$ \\
14 & $60412.910$ & $8121$ & $-0.0700^{+0.0003}_{-0.0004}$ & $4.1 \pm 0.3$ & $1.55^{+0.09}_{-0.08}$ & $0.1357^{+0.0008}_{-0.0013}$ & $4.7^{+0.8}_{-0.9}$ & $0.65 \pm 0.07$ \\
15 & $60413.069$ & $8179$ & $-0.0695^{+0.0004}_{-0.0003}$ & $4.8^{+0.5}_{-0.4}$ & $1.54 \pm 0.09$ & $0.1399 \pm 0.0008$ & $2.5^{+0.6}_{-0.5}$ & $0.58 \pm 0.07$ \\
16 & $60413.228$ & $7401$ & $-0.0679^{+0.0003}_{-0.0004}$ & $3.7 \pm 0.4$ & $1.69^{+0.11}_{-0.10}$ & $0.1377^{+0.0007}_{-0.0008}$ & $2.0^{+0.9}_{-0.6}$ & $0.54 \pm 0.08$ \\
17 & $60413.387$ & $3884$ & $-0.0652 \pm 0.0005$ & $3.6 \pm 0.5$ & $1.72^{+0.14}_{-0.12}$ & $0.1378^{+0.0007}_{-0.0006}$ & $1.1^{+0.7}_{-0.4}$ & $0.48 \pm 0.09$ \\
18 & $60413.546$ & $6527$ & $-0.0617^{+0.0003}_{-0.0004}$ & $4.2 \pm 0.3$ & $1.65^{+0.12}_{-0.11}$ & $0.1355^{+0.0013}_{-0.0012}$ & $2.8 \pm 1.0$ & $0.51 \pm 0.08$ \\
19 & $60413.705$ & $7132$ & $-0.0585^{+0.0003}_{-0.0004}$ & $3.9 \pm 0.4$ & $1.75^{+0.10}_{-0.09}$ & $0.1360 \pm 0.0009$ & $2.1^{+0.6}_{-0.5}$ & $0.50 \pm 0.07$ \\
20 & $60413.864$ & $7067$ & $-0.0551 \pm 0.0005$ & $4.5 \pm 0.4$ & $1.69^{+0.11}_{-0.09}$ & $0.1345^{+0.0011}_{-0.0010}$ & $3.0 \pm 0.8$ & $0.51 \pm 0.07$ \\
21 & $60414.023$ & $6388$ & $-0.0516^{+0.0003}_{-0.0004}$ & $4.6 \pm 0.4$ & $1.66^{+0.12}_{-0.11}$ & $0.1336 \pm 0.0014$ & $3.6^{+1.1}_{-1.0}$ & $0.53^{+0.09}_{-0.08}$ \\
22 & $60414.182$ & $6528$ & $-0.0504 \pm 0.0005$ & $4.0 \pm 0.4$ & $1.52 \pm 0.10$ & $0.1312^{+0.0010}_{-0.0009}$ & $2.6^{+0.7}_{-0.6}$ & $0.59 \pm 0.08$ \\
23 & $60414.341$ & $5523$ & $-0.0487 \pm 0.0005$ & $4.2^{+0.5}_{-0.6}$ & $1.66^{+0.12}_{-0.11}$ & $0.1301^{+0.0011}_{-0.0012}$ & $2.2^{+0.9}_{-0.7}$ & $0.48 \pm 0.08$ \\
24 & $60414.500$ & $4872$ & $-0.0478 \pm 0.0006$ & $4.5 \pm 0.5$ & $1.47 \pm 0.12$ & $0.1295^{+0.0018}_{-0.0016}$ & $3.3^{+1.2}_{-0.9}$ & $0.50 \pm 0.09$ \\
25 & $60414.659$ & $6164$ & $-0.0474 \pm 0.0004$ & $4.2 \pm 0.4$ & $1.64^{+0.14}_{-0.11}$ & $0.1304^{+0.0012}_{-0.0011}$ & $2.6^{+0.8}_{-0.6}$ & $0.58 \pm 0.08$ \\
26 & $60414.818$ & $9222$ & $-0.0472^{+0.0004}_{-0.0005}$ & $4.6 \pm 0.4$ & $1.67 \pm 0.09$ & $0.1299 \pm 0.0007$ & $2.3^{+0.6}_{-0.5}$ & $0.49 \pm 0.06$ \\
27 & $60414.977$ & $8022$ & $-0.0464^{+0.0004}_{-0.0005}$ & $4.3 \pm 0.4$ & $1.48^{+0.09}_{-0.08}$ & $0.1291^{+0.0007}_{-0.0006}$ & $2.7^{+0.6}_{-0.5}$ & $0.57 \pm 0.07$ \\
28 & $60415.136$ & $8045$ & $-0.0454 \pm 0.0005$ & $3.8 \pm 0.4$ & $1.69^{+0.12}_{-0.11}$ & $0.1268^{+0.0012}_{-0.0013}$ & $3.3^{+0.8}_{-0.9}$ & $0.55^{+0.08}_{-0.09}$ \\
29 & $60415.295$ & $5312$ & $-0.0433 \pm 0.0004$ & $4.3 \pm 0.4$ & $1.72^{+0.14}_{-0.15}$ & $0.1209^{+0.0026}_{-0.0016}$ & $4.4 \pm 1.3$ & $0.44^{+0.10}_{-0.09}$ \\
\midrule
\multicolumn{9}{c}{\textbf{2025}} \\
\midrule
0 & $60760.840$ & $23163$ & $0.0242 \pm 0.0005$ & $2.7^{+0.5}_{-0.4}$ & $0.19^{+0.03}_{-0.02}$ & $0.0745 \pm 0.0007$ & $2.5^{+0.7}_{-0.5}$ & $0.13^{+0.02}_{-0.01}$ \\
1 & $60761.300$ & $16193$ & $0.0308 \pm 0.0006$ & $3.0^{+0.8}_{-0.6}$ & $0.23^{+0.04}_{-0.03}$ & $0.0681^{+0.0010}_{-0.0011}$ & $4.0^{+0.9}_{-0.8}$ & $0.20 \pm 0.02$ \\
2 & $60761.760$ & $22468$ & $0.0355^{+0.0008}_{-0.0007}$ & $3.2^{+0.7}_{-0.5}$ & $0.17^{+0.03}_{-0.02}$ & $0.0630^{+0.0013}_{-0.0014}$ & $4.1^{+1.0}_{-0.9}$ & $0.14 \pm 0.02$ \\
3 & $60762.220$ & $15005$ & $0.0376^{+0.0021}_{-0.0020}$ & $3.8^{+1.6}_{-1.5}$ & $0.11 \pm 0.02$ & $0.0607^{+0.0016}_{-0.0015}$ & $1.7^{+1.2}_{-1.7}$ & $0.07 \pm 0.02$ \\
4 & $60762.680$ & $21543$ & $0.0388^{+0.0018}_{-0.0016}$ & $4.2^{+1.2}_{-0.9}$ & $0.17 \pm 0.03$ & $0.0555^{+0.0016}_{-0.0017}$ & $2.6^{+1.5}_{-0.7}$ & $0.10 \pm 0.03$ \\
\bottomrule
\end{tabular}

\begin{tabnote}
Subscripts $\mathrm{b}$ and $\mathrm{r}$ denote the blue and red jet components, respectively. Uncertainties are at the 90\% confidence level.\\
\footnotemark[$*$] $W_v$ denotes the velocity width (FWHM) in km~s$^{-1}$.\\
\footnotemark[$\dag$] $N$ is the normalization as defined in XSPEC.\\
\footnotemark[$\ddag$] $kT$ is in keV.\\
\footnotemark[$\S$] Abundances relative to the solar value \citep{Lodders_2009}.\\
\footnotemark[$\|$] $N_\mathrm{gau}$ is in units of photons~cm$^{-2}$~s$^{-1}$.\\
\footnotemark[$\#$] $F$ is in erg~cm$^{-2}$~s$^{-1}$.\\
\end{tabnote}

\end{table*}

\begin{table*}[ht!]
\caption[\ref{tab:xray_doppler} (continued)]{Global parameters, flux, and fit quality.}

\begin{tabular}{r|llllllllll}
\toprule
\textbf{ID} &
$\boldsymbol{kT}$\footnotemark[$\ddag$] &
$\boldsymbol{A_\mathrm{Fe}}$\footnotemark[$\S$] &
$\boldsymbol{A_\mathrm{Ni}}$\footnotemark[$\S$] &
$\boldsymbol{N_\mathrm{gau}}$\footnotemark[$\|$] &
$\boldsymbol{F_\mathrm{b}}$\footnotemark[$\#$] $\boldsymbol{(10^{-11})}$ &
$\boldsymbol{F_\mathrm{r}}$\footnotemark[$\#$] $\boldsymbol{(10^{-11})}$ &
$\boldsymbol{F_\mathrm{all}^{(\mathrm{abs})}}$\footnotemark[$\#$] $\boldsymbol{(10^{-11})}$ &
$\boldsymbol{\chi^2}$/d.o.f. \\
\midrule
\multicolumn{9}{c}{\textbf{2024}} \\
\midrule
0 & $5.9 \pm 0.3$ & $1.01^{+0.11}_{-0.10}$ & $8.4^{+1.2}_{-1.1}$ & $3.9 \pm 1.3$ & $4.17 \pm 0.17$ & $1.52 \pm 0.14$ & $5.56 \pm 0.12$ & $398/402$ \\
1 & $6.3^{+0.4}_{-0.3}$ & $1.06^{+0.12}_{-0.10}$ & $8.9^{+1.3}_{-1.1}$ & $4.4 \pm 1.4$ & $4.29 \pm 0.17$ & $1.49^{+0.15}_{-0.14}$ & $5.65 \pm 0.12$ & $449/411$ \\
2 & $6.5^{+0.3}_{-0.4}$ & $1.12^{+0.13}_{-0.12}$ & $9.8^{+1.4}_{-1.3}$ & $5.1 \pm 1.5$ & $4.10 \pm 0.17$ & $1.55 \pm 0.14$ & $5.54 \pm 0.12$ & $387/397$ \\
3 & $6.3 \pm 0.3$ & $1.03^{+0.10}_{-0.09}$ & $8.7^{+1.2}_{-1.1}$ & $5.0 \pm 1.3$ & $4.29 \pm 0.16$ & $1.42 \pm 0.13$ & $5.59 \pm 0.11$ & $449/455$ \\
4 & $7.0^{+0.4}_{-0.5}$ & $1.20^{+0.17}_{-0.16}$ & $10.5^{+2.0}_{-1.8}$ & $6.0 \pm 2.0$ & $4.34 \pm 0.23$ & $1.61 \pm 0.19$ & $5.83 \pm 0.16$ & $269/258$ \\
5 & $6.4 \pm 0.4$ & $1.01^{+0.12}_{-0.11}$ & $6.9^{+1.3}_{-1.2}$ & $6.9 \pm 1.9$ & $4.98 \pm 0.22$ & $1.39 \pm 0.18$ & $6.24 \pm 0.15$ & $336/326$ \\
6 & $6.5^{+0.4}_{-0.5}$ & $0.93^{+0.13}_{-0.14}$ & $8.6 \pm 1.6$ & $7.0 \pm 1.9$ & $4.96^{+0.26}_{-0.28}$ & $1.49^{+0.25}_{-0.23}$ & $6.34^{+0.14}_{-0.15}$ & $430/369$ \\
7 & $6.8 \pm 0.3$ & $1.07 \pm 0.11$ & $8.0 \pm 1.2$ & $6.1 \pm 1.6$ & $5.15 \pm 0.21$ & $1.48 \pm 0.17$ & $6.49 \pm 0.13$ & $470/439$ \\
8 & $7.1 \pm 0.3$ & $1.05 \pm 0.10$ & $7.8^{+1.2}_{-1.1}$ & $7.1 \pm 1.6$ & $5.85 \pm 0.23$ & $1.59 \pm 0.20$ & $7.29 \pm 0.12$ & $660/583$ \\
9 & $7.2 \pm 0.4$ & $0.93^{+0.12}_{-0.11}$ & $8.1^{+1.6}_{-1.4}$ & $11.1 \pm 2.8$ & $7.70^{+0.37}_{-0.38}$ & $1.64^{+0.33}_{-0.32}$ & $9.17 \pm 0.21$ & $406/355$ \\
10 & $7.5 \pm 0.4$ & $0.91^{+0.10}_{-0.09}$ & $6.3^{+1.1}_{-1.0}$ & $11.0 \pm 2.2$ & $8.18^{+0.33}_{-0.34}$ & $1.98^{+0.30}_{-0.29}$ & $9.97 \pm 0.17$ & $642/608$ \\
11 & $8.5 \pm 0.6$ & $0.89^{+0.13}_{-0.11}$ & $8.0^{+1.7}_{-1.6}$ & $9.7 \pm 2.7$ & $7.96 \pm 0.45$ & $2.30^{+0.42}_{-0.41}$ & $10.04 \pm 0.21$ & $481/423$ \\
12 & $7.9 \pm 0.5$ & $1.01 \pm 0.12$ & $5.6^{+1.3}_{-1.2}$ & $10.8^{+2.5}_{-2.6}$ & $8.70^{+0.39}_{-0.37}$ & $2.03^{+0.32}_{-0.34}$ & $10.51 \pm 0.20$ & $561/511$ \\
13 & $8.1 \pm 0.4$ & $0.87 \pm 0.08$ & $4.5^{+0.9}_{-0.8}$ & $10.4 \pm 2.0$ & $9.04 \pm 0.30$ & $1.59^{+0.27}_{-0.26}$ & $10.41 \pm 0.15$ & $821/732$ \\
14 & $7.9 \pm 0.3$ & $0.94 \pm 0.07$ & $7.5 \pm 0.9$ & $11.3^{+2.0}_{-2.1}$ & $8.47^{+0.37}_{-0.38}$ & $2.31 \pm 0.34$ & $10.56 \pm 0.16$ & $743/708$ \\
15 & $7.5^{+0.4}_{-0.3}$ & $0.89^{+0.09}_{-0.08}$ & $6.2^{+1.0}_{-0.9}$ & $7.8 \pm 1.8$ & $8.05 \pm 0.29$ & $1.94 \pm 0.25$ & $9.77 \pm 0.15$ & $728/676$ \\
16 & $7.1^{+0.3}_{-0.4}$ & $0.78 \pm 0.08$ & $4.8 \pm 0.8$ & $11.8 \pm 2.2$ & $8.25^{+0.31}_{-0.29}$ & $1.71^{+0.25}_{-0.27}$ & $9.77 \pm 0.16$ & $719/627$ \\
17 & $7.1 \pm 0.5$ & $0.86^{+0.11}_{-0.10}$ & $5.6^{+1.2}_{-1.1}$ & $9.6 \pm 2.8$ & $8.56^{+0.37}_{-0.36}$ & $1.57 \pm 0.30$ & $9.91 \pm 0.22$ & $383/363$ \\
18 & $7.6 \pm 0.4$ & $1.06 \pm 0.11$ & $6.7^{+1.2}_{-1.1}$ & $10.5 \pm 2.3$ & $8.89^{+0.37}_{-0.34}$ & $1.82^{+0.30}_{-0.32}$ & $10.49 \pm 0.18$ & $705/596$ \\
19 & $7.2 \pm 0.3$ & $0.96^{+0.09}_{-0.08}$ & $6.0^{+1.0}_{-0.9}$ & $15.1 \pm 2.4$ & $8.87^{+0.30}_{-0.27}$ & $1.71^{+0.22}_{-0.26}$ & $10.42 \pm 0.17$ & $708/635$ \\
20 & $7.2 \pm 0.3$ & $1.02^{+0.09}_{-0.10}$ & $7.0 \pm 1.0$ & $12.2 \pm 2.3$ & $8.71^{+0.31}_{-0.30}$ & $1.79^{+0.26}_{-0.27}$ & $10.32 \pm 0.17$ & $658/622$ \\
21 & $7.1^{+0.3}_{-0.4}$ & $0.95 \pm 0.10$ & $6.0^{+1.1}_{-1.0}$ & $8.9 \pm 2.1$ & $8.23 \pm 0.35$ & $1.80 \pm 0.31$ & $9.82 \pm 0.17$ & $560/556$ \\
22 & $7.5 \pm 0.4$ & $0.97 \pm 0.10$ & $6.2^{+1.1}_{-0.9}$ & $9.6 \pm 2.1$ & $7.74^{+0.31}_{-0.33}$ & $2.08^{+0.29}_{-0.27}$ & $9.63 \pm 0.17$ & $616/560$ \\
23 & $7.4 \pm 0.4$ & $1.00 \pm 0.11$ & $5.1 \pm 1.0$ & $12.1 \pm 2.5$ & $8.31^{+0.51}_{-0.35}$ & $1.67 \pm 0.30$ & $9.81^{+0.18}_{-0.19}$ & $550/492$ \\
24 & $7.6^{+0.5}_{-0.4}$ & $1.07^{+0.14}_{-0.12}$ & $5.4 \pm 1.2$ & $9.4 \pm 2.4$ & $7.59^{+0.37}_{-0.40}$ & $1.79^{+0.36}_{-0.33}$ & $9.20 \pm 0.19$ & $416/419$ \\
25 & $6.6^{+0.3}_{-0.4}$ & $0.89 \pm 0.10$ & $4.9^{+1.0}_{-0.9}$ & $10.0 \pm 2.2$ & $7.54^{+0.30}_{-0.32}$ & $1.81^{+0.28}_{-0.26}$ & $9.18 \pm 0.17$ & $542/510$ \\
26 & $7.1 \pm 0.3$ & $1.02^{+0.09}_{-0.08}$ & $6.0 \pm 0.8$ & $9.7 \pm 1.8$ & $8.29 \pm 0.26$ & $1.67 \pm 0.23$ & $9.77 \pm 0.14$ & $838/718$ \\
27 & $7.6 \pm 0.3$ & $1.08 \pm 0.10$ & $6.6^{+1.0}_{-0.9}$ & $11.4 \pm 2.0$ & $7.73^{+0.30}_{-0.29}$ & $2.10^{+0.25}_{-0.26}$ & $9.65 \pm 0.15$ & $718/645$ \\
28 & $7.1 \pm 0.3$ & $0.90^{+0.09}_{-0.08}$ & $6.1 \pm 0.9$ & $11.5 \pm 2.1$ & $8.18^{+0.36}_{-0.33}$ & $1.87^{+0.30}_{-0.33}$ & $9.87 \pm 0.15$ & $761/658$ \\
29 & $7.1 \pm 0.4$ & $0.99^{+0.13}_{-0.10}$ & $4.4^{+1.1}_{-0.9}$ & $8.7 \pm 2.4$ & $8.23 \pm 0.40$ & $1.52 \pm 0.36$ & $9.56 \pm 0.19$ & $482/464$ \\
\midrule
\multicolumn{9}{c}{\textbf{2025}} \\
\midrule
0 & $6.9^{+0.5}_{-0.6}$ & $0.99 \pm 0.14$ & $6.6^{+1.3}_{-1.2}$ & $2.2 \pm 0.5$ & $0.80 \pm 0.06$ & $0.51 \pm 0.06$ & $1.30 \pm 0.03$ & $280/287$ \\
1 & $6.8^{+0.7}_{-0.5}$ & $1.20^{+0.20}_{-0.17}$ & $6.2^{+1.4}_{-1.3}$ & $2.5^{+0.7}_{-0.8}$ & $0.99 \pm 0.09$ & $0.80 \pm 0.09$ & $1.76 \pm 0.04$ & $263/274$ \\
2 & $6.7 \pm 0.5$ & $1.17^{+0.17}_{-0.16}$ & $6.9^{+1.5}_{-1.4}$ & $1.7^{+0.6}_{-0.7}$ & $0.72^{+0.06}_{-0.07}$ & $0.57 \pm 0.06$ & $1.27 \pm 0.03$ & $251/274$ \\
3 & $8.5^{+1.8}_{-1.3}$ & $0.62^{+0.27}_{-0.19}$ & $4.8^{+2.5}_{-2.0}$ & $0.6 \pm 0.5$ & $0.44 \pm 0.08$ & $0.27^{+0.08}_{-0.07}$ & $0.70 \pm 0.03$ & $116/98$ \\
4 & $7.1 \pm 0.5$ & $0.99^{+0.16}_{-0.15}$ & $5.1^{+1.4}_{-1.3}$ & $1.4^{+0.7}_{-0.8}$ & $0.39^{+0.13}_{-0.12}$ & $0.70^{+0.12}_{-0.13}$ & $1.07 \pm 0.03$ & $224/225$ \\
\bottomrule
\end{tabular}

\begin{tabnote}
Subscripts $\mathrm{b}$ and $\mathrm{r}$ denote the blue and red jet components, respectively. Uncertainties are at the 90\% confidence level.\\
\footnotemark[$*$] $W_v$ denotes the velocity width (FWHM) in km~s$^{-1}$.\\
\footnotemark[$\dag$] $N$ is the normalization as defined in XSPEC.\\
\footnotemark[$\ddag$] $kT$ is in keV.\\
\footnotemark[$\S$] Abundances relative to the solar value \citep{Lodders_2009}.\\
\footnotemark[$\|$] $N_\mathrm{gau}$ is in units of photons~cm$^{-2}$~s$^{-1}$.\\
\footnotemark[$\#$] $F$ is in erg~cm$^{-2}$~s$^{-1}$.\\
\end{tabnote}

\end{table*}

\begin{table}[ht!]
 \caption{Doppler shifts of the H$\alpha$ line from Seimei and LCO observations (see figure~\ref{optical_spectra_all}).}

 \label{tab:optical_doppler}
\begin{tabular}{r|rrrrr}
\toprule
\textbf{ID} & \textbf{MJD} &
$\boldsymbol{H\alpha_\mathrm{b}}$ [\AA] &
$\boldsymbol{H\alpha_\mathrm{r}}$ [\AA] &
$\boldsymbol{z_\mathrm{b}}$ &
$\boldsymbol{z_\mathrm{r}}$ \\
\midrule
\multicolumn{6}{c}{\textbf{Seimei}} \\
\midrule
$0$ & $60405.8042$ & $6067.50$ & $7525.00$ & $-0.0755$ & $0.1466$ \\
$1$ & $60406.7875$ & $6017.50$ & $7530.00$ & $-0.0831$ & $0.1474$ \\
$2$ & $60409.7653$ & $6102.50$ & $7440.00$ & $-0.0701$ & $0.1337$ \\
$3$ & $60410.7757$ & $6165.00$ & $7407.50$ & $-0.0606$ & $0.1287$ \\
$4$ & $60411.7812$ & $6142.50$ & $7407.50$ & $-0.0640$ & $0.1287$ \\
$5$ & $60412.7743$ & $6100.00$ & $7412.50$ & $-0.0705$ & $0.1295$ \\
$6$ & $60413.7715$ & $6142.50$ & $7455.00$ & $-0.0640$ & $0.1359$ \\
$7$ & $60416.7715$ & $6307.50$ & $7305.00$ & $-0.0389$ & $0.1131$ \\
$8$ & $60418.7799$ & $6262.50$ & $7292.50$ & $-0.0458$ & $0.1112$ \\
$9$ & $60424.7715$ & $6400.00$ & $7165.00$ & $-0.0248$ & $0.0918$ \\
$10$ & $60427.7799$ & $\mathrm{n/a}$\footnotemark[$*$] & $7050.00$ & $\mathrm{n/a}$\footnotemark[$*$] & $0.0742$ \\
$11$ & $60555.4750$ & $5907.50$ & $7577.50$ & $-0.0999$ & $0.1546$ \\
$12$ & $60556.4549$ & $5905.00$ & $7577.50$ & $-0.1002$ & $0.1546$ \\
$13$ & $60764.7910$ & $\mathrm{n/a}$\footnotemark[$*$] & $\mathrm{n/a}$\footnotemark[$*$] & $\mathrm{n/a}$\footnotemark[$*$] & $\mathrm{n/a}$\footnotemark[$*$] \\
\midrule
\multicolumn{6}{c}{\textbf{LCO}} \\
\midrule
$0$ & $60423.5962$ & $6408.88$ & $7161.14$ & $-0.0235$ & $0.0912$ \\
$1$ & $60430.6106$ & $\mathrm{n/a}$\footnotemark[$*$] & $7019.40$ & $\mathrm{n/a}$\footnotemark[$*$] & $0.0696$ \\
$2$ & $60437.5689$ & $6696.40$ & $6907.24$ & $0.0204$ & $0.0525$ \\
$3$ & $60451.5521$ & $6976.72$ & $6718.67$ & $0.0631$ & $0.0238$ \\
$4$ & $60465.6034$ & $7036.38$ & $6513.59$ & $0.0722$ & $-0.0075$ \\
$5$ & $60472.5187$ & $7073.92$ & $6481.86$ & $0.0779$ & $-0.0123$ \\
$6$ & $60479.5025$ & $7123.15$ & $6450.13$ & $0.0854$ & $-0.0172$ \\
$7$ & $60493.5267$ & $6890.02$ & $6758.46$ & $0.0499$ & $0.0298$ \\
$8$ & $60500.5151$ & $6653.91$ & $6842.83$ & $0.0139$ & $0.0427$ \\
\bottomrule
\end{tabular}

\begin{tabnote}
\footnotemark[$*$] $\mathrm{n/a}$ indicates values that are not available due to line blending with stationary components or insufficient signal-to-noise ratio, preventing reliable peak identification (see text for details).
\end{tabnote}

\end{table}

\bibliographystyle{apj}
\bibliography{main}

@article{Shidatsu_2025,
	adsurl = {https://ui.adsabs.harvard.edu/abs/2025PASJ.....Pf110S},
	archiveprefix = {arXiv},
	author = {{Shidatsu}, Megumi and {Kobayashi}, Shogo and {Sakai}, Yusuke and {Takagi}, Toshihiro and {Okada}, Yuta and {Yamada}, Shinya and {Ueda}, Yoshihiro and {Uchiyama}, Hideki and {Petre}, Robert},
	doi = {10.1093/pasj/psaf110},
	eid = {psaf110},
	eprint = {2510.24341},
	journal = {\pasj},
	month = oct,
	pages = {psaf110},
	primaryclass = {astro-ph.HE},
	title = {{XRISM high-resolution spectroscopy of SS 433: Evidence of decreasing line-of-sight velocity dispersion along the jet}},
	year = 2025}

@article{Takagi_2025,
	author = {{Takagi}, Toshihiro and {Shidatsu}, Megumi and {Okada}, Yuta and {Ueda}, Yoshihiro and {Sakai}, Yusuke and {Kobayashi}, Shogo and {Yamada}, Shinya and {Uchiyama}, Hideki and {Yoshimoto}, Marina and {Uenishi}, Miyu and {Usuki}, Tomoya and {Kobayashi}, Soma and {Petre}, Robert},
	journal = {PASJ},
	title = {{XRISM Observation of the Neutral Iron and Nickel Emission Lines in the Microquasar SS 433}},
	volume = {accepted},
	year = 2025}

@article{Sikora_1997,
	adsurl = {https://ui.adsabs.harvard.edu/abs/1997ApJ...484..108S},
	author = {{Sikora}, Marek and {Madejski}, Greg and {Moderski}, Rafa{\L} and {Poutanen}, Juri},
	doi = {10.1086/304305},
	journal = {\apj},
	month = jul,
	number = {1},
	pages = {108-117},
	title = {{Learning about Active Galactic Nucleus Jets from Spectral Properties of Blazars}},
	volume = {484},
	year = 1997}

@article{Sakai_2025,
	adsurl = {https://ui.adsabs.harvard.edu/abs/2025PASJ...77.1113S},
	author = {{Sakai}, Yusuke and {Yamada}, Shinya and {Sakemi}, Haruka and {Machida}, Mami and {Igarashi}, Taichi and {Hayakawa}, Ryota and {Tan}, Miho and {Furuyama}, Taisei},
	doi = {10.1093/pasj/psaf088},
	journal = {\pasj},
	month = oct,
	number = {5},
	pages = {1113-1125},
	title = {{Arcsecond-scale X-ray imaging and spectroscopy of SS 433 with the Chandra High-Energy Transmission Grating}},
	volume = {77},
	year = 2025}

@article{Blundell_2007,
	adsurl = {https://ui.adsabs.harvard.edu/abs/2007A&A...474..903B},
	archiveprefix = {arXiv},
	author = {{Blundell}, K.~M. and {Bowler}, M.~G. and {Schmidtobreick}, L.},
	doi = {10.1051/0004-6361:20077924},
	eprint = {0708.2930},
	journal = {\aap},
	month = nov,
	number = {3},
	pages = {903-910},
	primaryclass = {astro-ph},
	title = {{Fluctuations and symmetry in the speed and direction of the jets of SS 433 on different timescales}},
	volume = {474},
	year = 2007}

@article{Kaaret_2024,
	adsurl = {https://ui.adsabs.harvard.edu/abs/2024ApJ...961L..12K},
	archiveprefix = {arXiv},
	author = {{Kaaret}, Philip and {Ferrazzoli}, Riccardo and {Silvestri}, Stefano and {Negro}, Michela and {Manfreda}, Alberto and {Wu}, Kinwah and {Costa}, Enrico and {Soffitta}, Paolo and {Safi-Harb}, Samar and {Poutanen}, Juri and {Veledina}, Alexandra and {Di Marco}, Alessandro and {Slane}, Patrick and {Bianchi}, Stefano and {Ingram}, Adam and {Romani}, Roger W. and {Cibrario}, Nicol{\`o} and {Mac Intyre}, Brydyn and {Mikus̆incov{\'a}}, Romana and {Ratheesh}, Ajay and {Steiner}, James F. and {Svoboda}, Jiri and {Tugliani}, Stefano and {Agudo}, Iv{\'a}n and {Antonelli}, Lucio A. and {Bachetti}, Matteo and {Baldini}, Luca and {Baumgartner}, Wayne H. and {Bellazzini}, Ronaldo and {Bongiorno}, Stephen D. and {Bonino}, Raffaella and {Brez}, Alessandro and {Bucciantini}, Niccol{\`o} and {Capitanio}, Fiamma and {Castellano}, Simone and {Cavazzuti}, Elisabetta and {Chen}, Chien-Ting and {Ciprini}, Stefano and {De Rosa}, Alessandra and {Del Monte}, Ettore and {Di Gesu}, Laura and {Di Lalla}, Niccol{\`o} and {Donnarumma}, Immacolata and {Doroshenko}, Victor and {Dov{\v{c}}iak}, Michal and {Ehlert}, Steven R. and {Enoto}, Teruaki and {Evangelista}, Yuri and {Fabiani}, Sergio and {Garc{\'\i}a}, Javier A. and {Gunji}, Shuichi and {Hayashida}, Kiyoshi and {Heyl}, Jeremy and {Iwakiri}, Wataru and {Jorstad}, Svetlana G. and {Karas}, Vladimir and {Kislat}, Fabian and {Kitaguchi}, Takao and {Kolodziejczak}, Jeffery J. and {Krawczynski}, Henric and {La Monaca}, Fabio and {Latronico}, Luca and {Liodakis}, Ioannis and {Maldera}, Simone and {Marin}, Fr{\'e}d{\'e}ric and {Marinucci}, Andrea and {Marscher}, Alan P. and {Marshall}, Herman L. and {Massaro}, Francesco and {Matt}, Giorgio and {Mitsuishi}, Ikuyuki and {Mizuno}, Tsunefumi and {Muleri}, Fabio and {Ng}, Chi-Yung and {O'Dell}, Stephen L. and {Omodei}, Nicola and {Oppedisano}, Chiara and {Papitto}, Alessandro and {Pavlov}, George G. and {Peirson}, Abel L. and {Perri}, Matteo and {Pesce-Rollins}, Melissa and {Petrucci}, Pierre-Olivier and {Pilia}, Maura and {Possenti}, Andrea and {Puccetti}, Simonetta and {Ramsey}, Brian D. and {Rankin}, John and {Roberts}, Oliver J. and {Sgr{\`o}}, Carmelo and {Spandre}, Gloria and {Swartz}, Douglas A. and {Tamagawa}, Toru and {Tavecchio}, Fabrizio and {Taverna}, Roberto and {Tawara}, Yuzuru and {Tennant}, Allyn F. and {Thomas}, Nicholas E. and {Tombesi}, Francesco and {Trois}, Alessio and {Tsygankov}, Sergey S. and {Turolla}, Roberto and {Vink}, Jacco and {Weisskopf}, Martin C. and {Xie}, Fei and {Zane}, Silvia},
	doi = {10.3847/2041-8213/ad103b},
	eid = {L12},
	eprint = {2311.16313},
	journal = {\apjl},
	month = jan,
	number = {1},
	pages = {L12},
	primaryclass = {astro-ph.HE},
	title = {{X-Ray Polarization of the Eastern Lobe of SS 433}},
	volume = {961},
	year = 2024}

@article{Middleton_2021,
	adsurl = {https://ui.adsabs.harvard.edu/abs/2021MNRAS.506.1045M},
	archiveprefix = {arXiv},
	author = {{Middleton}, M.~J. and {Walton}, D.~J. and {Alston}, W. and {Dauser}, T. and {Eikenberry}, S. and {Jiang}, Y. -F. and {Fabian}, A.~C. and {Fuerst}, F. and {Brightman}, M. and {Marshall}, H. and {Parker}, M. and {Pinto}, C. and {Harrison}, F.~A. and {Bachetti}, M. and {Altamirano}, D. and {Bird}, A.~J. and {Perez}, G. and {Miller-Jones}, J. and {Charles}, P. and {Boggs}, S. and {Christensen}, F. and {Craig}, W. and {Forster}, K. and {Grefenstette}, B. and {Hailey}, C. and {Madsen}, K. and {Stern}, D. and {Zhang}, W.},
	doi = {10.1093/mnras/stab1280},
	eprint = {1810.10518},
	journal = {\mnras},
	month = sep,
	number = {1},
	pages = {1045-1058},
	primaryclass = {astro-ph.HE},
	title = {{NuSTAR reveals the hidden nature of SS433}},
	volume = {506},
	year = 2021}

@dataset{Cutri_2003,
	adsurl = {https://ui.adsabs.harvard.edu/abs/2003yCat.2246....0C},
	author = {{Cutri}, R.~M. and {Skrutskie}, M.~F. and {van Dyk}, S. and {Beichman}, C.~A. and {Carpenter}, J.~M. and {Chester}, T. and {Cambresy}, L. and {Evans}, T. and {Fowler}, J. and {Gizis}, J. and {Howard}, E. and {Huchra}, J. and {Jarrett}, T. and {Kopan}, E.~L. and {Kirkpatrick}, J.~D. and {Light}, R.~M. and {Marsh}, K.~A. and {McCallon}, H. and {Schneider}, S. and {Stiening}, R. and {Sykes}, M. and {Weinberg}, M. and {Wheaton}, W.~A. and {Wheelock}, S. and {Zacarias}, N.},
	eid = {II/246},
	howpublished = {VizieR On-line Data Catalog: II/246. Originally published in: University of Massachusetts and Infrared Processing and Analysis Center, (IPAC/California Institute of Technology) (2003)},
	month = jun,
	title = {{VizieR Online Data Catalog: 2MASS All-Sky Catalog of Point Sources (Cutri+ 2003)}},
	year = 2003}

@article{Blundell_2011,
	author = {Blundell, Katherine M. and Schmidtobreick, Linda and Trushkin, Sergei},
	doi = {10.1111/j.1365-2966.2011.18785.x},
	eprint = {https://academic.oup.com/mnras/article-pdf/417/4/2401/3808559/mnras0417-2401.pdf},
	issn = {0035-8711},
	journal = {Monthly Notices of the Royal Astronomical Society},
	month = {11},
	number = {4},
	pages = {2401-2410},
	title = {SS433's accretion disc, wind and jets: before, during and after a major flare},
	url = {https://doi.org/10.1111/j.1365-2966.2011.18785.x},
	volume = {417},
	year = {2011}}

@article{Medvedev_2018,
	adsurl = {https://ui.adsabs.harvard.edu/abs/2018AstL...44..390M},
	archiveprefix = {arXiv},
	author = {{Medvedev}, P.~S. and {Khabibullin}, I.~I. and {Sazonov}, S. Yu. and {Churazov}, E.~M. and {Tsygankov}, S.~S.},
	doi = {10.1134/S1063773718060038},
	eprint = {1804.01828},
	journal = {Astronomy Letters},
	month = jun,
	number = {6},
	pages = {390-410},
	primaryclass = {astro-ph.HE},
	title = {{An Upper Limit on Nickel Overabundance in the Supercritical Accretion Disk Wind of SS 433 from X-ray Spectroscopy}},
	volume = {44},
	year = 2018}

@article{Goranskij_2011,
	adsurl = {https://ui.adsabs.harvard.edu/abs/2011PZ.....31....5G},
	archiveprefix = {arXiv},
	author = {{Goranskij}, V.},
	doi = {10.48550/arXiv.1110.5304},
	eprint = {1110.5304},
	journal = {Peremennye Zvezdy},
	month = oct,
	number = {5},
	pages = {5},
	primaryclass = {astro-ph.HE},
	title = {{Photometric Mass Estimate for the Compact Component of SS 433: And Yet It Is a Neutron Star}},
	volume = {31},
	year = 2011}

@article{Kodaira_1985,
	adsurl = {https://ui.adsabs.harvard.edu/abs/1985ApJ...296..232K},
	author = {{Kodaira}, K. and {Nakada}, Y. and {Backman}, D.~E.},
	doi = {10.1086/163441},
	journal = {\apj},
	month = sep,
	pages = {232-239},
	title = {{Infrared variability of SS 433.}},
	volume = {296},
	year = 1985}

@article{Cherepashchuk_2005,
	adsurl = {https://ui.adsabs.harvard.edu/abs/2005A&A...437..561C},
	archiveprefix = {arXiv},
	author = {{Cherepashchuk}, A.~M. and {Sunyaev}, R.~A. and {Fabrika}, S.~N. and {Postnov}, K.~A. and {Molkov}, S.~V. and {Barsukova}, E.~A. and {Antokhina}, E.~A. and {Irsmambetova}, T.~R. and {Panchenko}, I.~E. and {Seifina}, E.~V. and {Shakura}, N.~I. and {Timokhin}, A.~N. and {Bikmaev}, I.~F. and {Sakhibullin}, N.~A. and {Aslan}, Z. and {Khamitov}, I. and {Pramsky}, A.~G. and {Sholukhova}, O. and {Gnedin}, Yu. N. and {Arkharov}, A.~A. and {Larionov}, V.~M.},
	doi = {10.1051/0004-6361:20041563},
	eprint = {astro-ph/0503352},
	journal = {\aap},
	month = jul,
	number = {2},
	pages = {561-573},
	primaryclass = {astro-ph},
	title = {{INTEGRAL observations of SS433: Results of a coordinated campaign}},
	volume = {437},
	year = 2005}

@article{Blanton_2007,
	adsurl = {https://ui.adsabs.harvard.edu/abs/2007AJ....133..734B},
	archiveprefix = {arXiv},
	author = {{Blanton}, Michael R. and {Roweis}, Sam},
	doi = {10.1086/510127},
	eprint = {astro-ph/0606170},
	journal = {\aj},
	month = feb,
	number = {2},
	pages = {734-754},
	primaryclass = {astro-ph},
	title = {{K-Corrections and Filter Transformations in the Ultraviolet, Optical, and Near-Infrared}},
	volume = {133},
	year = 2007}

@article{Cherepashchuk_2021,
	adsurl = {https://ui.adsabs.harvard.edu/abs/2021MNRAS.507L..19C},
	archiveprefix = {arXiv},
	author = {{Cherepashchuk}, A.~M. and {Belinski}, A.~A. and {Dodin}, A.~V. and {Postnov}, K.~A.},
	doi = {10.1093/mnrasl/slab083},
	eprint = {2107.09005},
	journal = {\mnras},
	month = oct,
	number = {1},
	pages = {L19-L23},
	primaryclass = {astro-ph.SR},
	title = {{Discovery of orbital eccentricity and evidence for orbital period increase of SS433}},
	volume = {507},
	year = 2021}

@article{Verner_1996,
	adsurl = {https://ui.adsabs.harvard.edu/abs/1996ApJ...465..487V},
	archiveprefix = {arXiv},
	author = {{Verner}, D.~A. and {Ferland}, G.~J. and {Korista}, K.~T. and {Yakovlev}, D.~G.},
	doi = {10.1086/177435},
	eprint = {astro-ph/9601009},
	journal = {\apj},
	month = jul,
	pages = {487},
	primaryclass = {astro-ph},
	title = {{Atomic Data for Astrophysics. II. New Analytic Fits for Photoionization Cross Sections of Atoms and Ions}},
	volume = {465},
	year = 1996}

@article{Lang_2010,
	author = {Lang, Dustin and Hogg, David W. and Mierle, Keir and Blanton, Michael and Roweis, Sam},
	doi = {10.1088/0004-6256/139/5/1782},
	journal = {The Astronomical Journal},
	month = {mar},
	number = {5},
	pages = {1782},
	publisher = {The American Astronomical Society},
	title = {ASTROMETRY.NET: BLIND ASTROMETRIC CALIBRATION OF ARBITRARY ASTRONOMICAL IMAGES},
	url = {https://dx.doi.org/10.1088/0004-6256/139/5/1782},
	volume = {139},
	year = {2010}}

@article{Brinkmann_1991,
	adsurl = {https://ui.adsabs.harvard.edu/abs/1991A&A...241..112B},
	author = {{Brinkmann}, W. and {Kawai}, N. and {Matsuoka}, M. and {Fink}, H.~H.},
	journal = {\aap},
	month = jan,
	pages = {112},
	title = {{The X-ray emission of SS 433.}},
	volume = {241},
	year = 1991}

@article{Collins_1988,
	adsurl = {https://ui.adsabs.harvard.edu/abs/1988ApJ...331..486C},
	author = {{Collins}, II, George W. and {Newsom}, Gerald H.},
	doi = {10.1086/166574},
	journal = {\apj},
	month = aug,
	pages = {486},
	title = {{Transient and Secular Variations of the Moving-Line Spectra from SS 433}},
	volume = {331},
	year = 1988}

@article{Atapin_2015,
	adsurl = {https://ui.adsabs.harvard.edu/abs/2015MNRAS.446..893A},
	archiveprefix = {arXiv},
	author = {{Atapin}, Kirill and {Fabrika}, Sergei and {Medvedev}, Aleksei and {Vinokurov}, Alexander},
	doi = {10.1093/mnras/stu2134},
	eprint = {1410.8495},
	journal = {\mnras},
	month = jan,
	number = {1},
	pages = {893-910},
	primaryclass = {astro-ph.HE},
	title = {{X-ray variability of SS 433: effects of the supercritical accretion disc}},
	volume = {446},
	year = 2015}

@article{Cherepashchuk_2025,
	adsurl = {https://ui.adsabs.harvard.edu/abs/2025arXiv250601106C},
	archiveprefix = {arXiv},
	author = {{Cherepashchuk}, A.~M. and {Dodin}, A.~V. and {Postnov}, K.~A.},
	doi = {10.48550/arXiv.2506.01106},
	eid = {arXiv:2506.01106},
	eprint = {2506.01106},
	journal = {arXiv e-prints},
	month = jun,
	pages = {arXiv:2506.01106},
	primaryclass = {astro-ph.HE},
	title = {{Unique microquasar SS433: new results, new issues}},
	year = 2025}

@article{Fabrika_2004,
	adsurl = {https://ui.adsabs.harvard.edu/abs/2004ASPRv..12....1F},
	archiveprefix = {arXiv},
	author = {{Fabrika}, S.},
	doi = {10.48550/arXiv.astro-ph/0603390},
	eprint = {astro-ph/0603390},
	journal = {\apspr},
	month = jan,
	pages = {1-152},
	primaryclass = {astro-ph},
	title = {{The jets and supercritical accretion disk in SS433}},
	volume = {12},
	year = 2004}

@article{Bessell_1988,
	adsurl = {https://ui.adsabs.harvard.edu/abs/1988PASP..100.1134B},
	author = {{Bessell}, M.~S. and {Brett}, J.~M.},
	doi = {10.1086/132281},
	journal = {\pasp},
	month = sep,
	pages = {1134},
	title = {{JHKLM Photometry: Standard Systems, Passbands, and Intrinsic Colors}},
	volume = {100},
	year = 1988}

@article{Chambers_2016,
	adsurl = {https://ui.adsabs.harvard.edu/abs/2016arXiv161205560C},
	archiveprefix = {arXiv},
	author = {{Chambers}, K.~C. and {Magnier}, E.~A. and {Metcalfe}, N. and {Flewelling}, H.~A. and {Huber}, M.~E. and {Waters}, C.~Z. and {Denneau}, L. and {Draper}, P.~W. and {Farrow}, D. and {Finkbeiner}, D.~P. and {Holmberg}, C. and {Koppenhoefer}, J. and {Price}, P.~A. and {Rest}, A. and {Saglia}, R.~P. and {Schlafly}, E.~F. and {Smartt}, S.~J. and {Sweeney}, W. and {Wainscoat}, R.~J. and {Burgett}, W.~S. and {Chastel}, S. and {Grav}, T. and {Heasley}, J.~N. and {Hodapp}, K.~W. and {Jedicke}, R. and {Kaiser}, N. and {Kudritzki}, R. -P. and {Luppino}, G.~A. and {Lupton}, R.~H. and {Monet}, D.~G. and {Morgan}, J.~S. and {Onaka}, P.~M. and {Shiao}, B. and {Stubbs}, C.~W. and {Tonry}, J.~L. and {White}, R. and {Ba{\~n}ados}, E. and {Bell}, E.~F. and {Bender}, R. and {Bernard}, E.~J. and {Boegner}, M. and {Boffi}, F. and {Botticella}, M.~T. and {Calamida}, A. and {Casertano}, S. and {Chen}, W. -P. and {Chen}, X. and {Cole}, S. and {Deacon}, N. and {Frenk}, C. and {Fitzsimmons}, A. and {Gezari}, S. and {Gibbs}, V. and {Goessl}, C. and {Goggia}, T. and {Gourgue}, R. and {Goldman}, B. and {Grant}, P. and {Grebel}, E.~K. and {Hambly}, N.~C. and {Hasinger}, G. and {Heavens}, A.~F. and {Heckman}, T.~M. and {Henderson}, R. and {Henning}, T. and {Holman}, M. and {Hopp}, U. and {Ip}, W. -H. and {Isani}, S. and {Jackson}, M. and {Keyes}, C.~D. and {Koekemoer}, A.~M. and {Kotak}, R. and {Le}, D. and {Liska}, D. and {Long}, K.~S. and {Lucey}, J.~R. and {Liu}, M. and {Martin}, N.~F. and {Masci}, G. and {McLean}, B. and {Mindel}, E. and {Misra}, P. and {Morganson}, E. and {Murphy}, D.~N.~A. and {Obaika}, A. and {Narayan}, G. and {Nieto-Santisteban}, M.~A. and {Norberg}, P. and {Peacock}, J.~A. and {Pier}, E.~A. and {Postman}, M. and {Primak}, N. and {Rae}, C. and {Rai}, A. and {Riess}, A. and {Riffeser}, A. and {Rix}, H.~W. and {R{\"o}ser}, S. and {Russel}, R. and {Rutz}, L. and {Schilbach}, E. and {Schultz}, A.~S.~B. and {Scolnic}, D. and {Strolger}, L. and {Szalay}, A. and {Seitz}, S. and {Small}, E. and {Smith}, K.~W. and {Soderblom}, D.~R. and {Taylor}, P. and {Thomson}, R. and {Taylor}, A.~N. and {Thakar}, A.~R. and {Thiel}, J. and {Thilker}, D. and {Unger}, D. and {Urata}, Y. and {Valenti}, J. and {Wagner}, J. and {Walder}, T. and {Walter}, F. and {Watters}, S.~P. and {Werner}, S. and {Wood-Vasey}, W.~M. and {Wyse}, R.},
	doi = {10.48550/arXiv.1612.05560},
	eid = {arXiv:1612.05560},
	eprint = {1612.05560},
	journal = {arXiv e-prints},
	month = dec,
	pages = {arXiv:1612.05560},
	primaryclass = {astro-ph.IM},
	title = {{The Pan-STARRS1 Surveys}},
	year = 2016}

@article{Kotani_2005,
	adsurl = {https://ui.adsabs.harvard.edu/abs/2005NCimC..28..755K},
	archiveprefix = {arXiv},
	author = {{Kotani}, T. and {Kawai}, N. and {Yanagisawa}, K. and {Watanabe}, J. and {Arimoto}, M. and {Fukushima}, H. and {Hattori}, T. and {Inata}, M. and {Izumiura}, H. and {Kataoka}, J. and {Koyano}, H. and {Kubota}, K. and {Kuroda}, D. and {Mori}, M. and {Nagayama}, S. and {Ohta}, K. and {Okada}, T. and {Okita}, K. and {Sato}, R. and {Serino}, Y. and {Shimizu}, Y. and {Shimokawabe}, T. and {Suzuki}, M. and {Toda}, H. and {Ushiyama}, T. and {Yatsu}, Y. and {Yoshida}, A. and {Yoshida}, M.},
	doi = {10.1393/ncc/i2005-10190-5},
	eprint = {astro-ph/0702708},
	journal = {Nuovo Cimento C Geophysics Space Physics C},
	month = jul,
	number = {4},
	pages = {755},
	primaryclass = {astro-ph},
	title = {{MITSuME---Multicolor Imaging Telescopes for Survey and Monstrous Explosions}},
	volume = {28},
	year = 2005}

@article{Niwano_2021,
	adsurl = {https://ui.adsabs.harvard.edu/abs/2021PASJ...73...14N},
	archiveprefix = {arXiv},
	author = {{Niwano}, Masafumi and {Murata}, Katsuhiro L. and {Adachi}, Ryo and {Wang}, Sili and {Tachibana}, Yutaro and {Yatsu}, Yoichi and {Kawai}, Nobuyuki and {Shimokawabe}, Takashi and {Itoh}, Ryosuke},
	doi = {10.1093/pasj/psaa091},
	eprint = {2008.11486},
	journal = {\pasj},
	month = feb,
	number = {1},
	pages = {14-24},
	primaryclass = {astro-ph.IM},
	title = {{A GPU-accelerated image reduction pipeline}},
	volume = {73},
	year = 2021}

@inproceedings{Shimokawabe_2009,
	adsurl = {https://ui.adsabs.harvard.edu/abs/2009AIPC.1133...79S},
	author = {{Shimokawabe}, Takashi and {Kawai}, Nobuyuki and {Mori}, Yuki A. and {Kudo}, Yusuke and {Nakajima}, Hideya and {Yoshida}, Michitoshi and {Yanagisawa}, Kenshi and {Nagayama}, Shogo and {Toda}, Hiroyuki and {Shimizu}, Yasuhiro and {Kuroda}, Daisuke and {Watanabe}, Jun'ichi and {Fukushima}, Hideo and {Mori}, Masaki},
	booktitle = {Gamma-ray Burst: Sixth Huntsville Symposium},
	doi = {10.1063/1.3155974},
	editor = {{Meegan}, Charles and {Kouveliotou}, Chryssa and {Gehrels}, Neil},
	month = may,
	pages = {79-81},
	publisher = {AIP},
	series = {American Institute of Physics Conference Series},
	title = {{MITSuME-multicolor optical/NIR telescopes for GRB afterglows-}},
	volume = {1133},
	year = 2009}

@article{Tonry_2012,
	adsurl = {https://ui.adsabs.harvard.edu/abs/2012ApJ...750...99T},
	archiveprefix = {arXiv},
	author = {{Tonry}, J.~L. and {Stubbs}, C.~W. and {Lykke}, K.~R. and {Doherty}, P. and {Shivvers}, I.~S. and {Burgett}, W.~S. and {Chambers}, K.~C. and {Hodapp}, K.~W. and {Kaiser}, N. and {Kudritzki}, R. -P. and {Magnier}, E.~A. and {Morgan}, J.~S. and {Price}, P.~A. and {Wainscoat}, R.~J.},
	doi = {10.1088/0004-637X/750/2/99},
	eid = {99},
	eprint = {1203.0297},
	journal = {\apj},
	month = may,
	number = {2},
	pages = {99},
	primaryclass = {astro-ph.IM},
	title = {{The Pan-STARRS1 Photometric System}},
	volume = {750},
	year = 2012}

@article{Yatsu_2007,
	adsurl = {https://ui.adsabs.harvard.edu/abs/2007PhyE...40..434Y},
	author = {{Yatsu}, Yoichi and {Kawai}, Nobuyuki and {Shimokawabe}, Takashi and {Vasquez}, Nicolas and {Ishimura}, Takuto and {Kotani}, Taro and {Yanagisawa}, Kenshi and {Yoshida}, Michitoshi and {Nagayama}, Sinngo and {Shimizu}, Hiroyasu and {Toda}, Hiroyuki and {Kuroda}, Daisuke},
	doi = {10.1016/j.physe.2007.06.050},
	journal = {Physica E Low-Dimensional Systems and Nanostructures},
	month = dec,
	number = {2},
	pages = {434-437},
	title = {{Development of MITSuME{\textemdash}Multicolor imaging telescopes for survey and monstrous explosions}},
	volume = {40},
	year = 2007}

@article{Uchida_2025,
	adsurl = {https://ui.adsabs.harvard.edu/abs/2025PASJ..tmp...49U},
	archiveprefix = {arXiv},
	author = {{Uchida}, Hiroyuki and {Mori}, Koji and {Tomida}, Hiroshi and {Nakajima}, Hiroshi and {Noda}, Hirofumi and {Tanaka}, Takaaki and {Murakami}, Hiroshi and {Suzuki}, Hiromasa and {Kobayashi}, Shogo Benjamin and {Yoneyama}, Tomokage and {Hagino}, Kouichi and {Nobukawa}, Kumiko Kawabata and {Uchiyama}, Hideki and {Nobukawa}, Masayoshi and {Matsumoto}, Hironori and {Tsuru}, Takeshi Go and {Yamauchi}, Makoto and {Hatsukade}, Isamu and {Odaka}, Hirokazu and {Kohmura}, Takayoshi and {Yamaoka}, Kazutaka and {Yoshida}, Tessei and {Kanemaru}, Yoshiaki and {Ishi}, Daiki and {Dotani}, Tadayasu and {Ozaki}, Masanobu and {Tsunemi}, Hiroshi and {Miyazaki}, Keitaro and {Kusunoki}, Kohei and {Otsuka}, Yoshinori and {Yokosu}, Haruhiko and {Yonemaru}, Wakana and {Ichikawa}, Kazuhiro and {Nakano}, Hanako and {Takemoto}, Reo and {Matsushima}, Tsukasa and {Urase}, Reika and {Kurashima}, Jun and {Fuchi}, Kotomi and {Hayakawa}, Kaito and {Fukuda}, Masahiro and {Inoue}, Shun and {Aoki}, Yuma and {Takayama}, Kouta and {Sako}, Takashi and {Yoshimoto}, Marina and {Shima}, Kohei and {Higuchi}, Mayu and {Ninoyu}, Kaito and {Aoki}, Daiki and {Tsunomachi}, Shun and {Okajima}, Takashi and {Ishida}, Manabu and {Maeda}, Yoshitomo and {Hayashi}, Takayuki and {Tamura}, Keisuke and {Boissay-Malaquin}, Rozenn and {Sato}, Toshiki and {Takeo}, Mai and {Miyamoto}, Asca and {Matsumoto}, Gakuto and {Eckart}, Megan E. and {Hell}, Natalie and {Leutenegger}, Maurice A. and {Hayashida}, Kiyoshi},
	doi = {10.1093/pasj/psaf030},
	eprint = {2503.20180},
	journal = {\pasj},
	month = may,
	primaryclass = {astro-ph.IM},
	title = {{In-orbit performance of the soft X-ray imaging telescope Xtend aboard XRISM}},
	year = 2025}

@article{Valenti_2014,
	adsurl = {https://ui.adsabs.harvard.edu/abs/2014MNRAS.438L.101V},
	archiveprefix = {arXiv},
	author = {{Valenti}, S. and {Sand}, D. and {Pastorello}, A. and {Graham}, M.~L. and {Howell}, D.~A. and {Parrent}, J.~T. and {Tomasella}, L. and {Ochner}, P. and {Fraser}, M. and {Benetti}, S. and {Yuan}, F. and {Smartt}, S.~J. and {Maund}, J.~R. and {Arcavi}, I. and {Gal-Yam}, A. and {Inserra}, C. and {Young}, D.},
	doi = {10.1093/mnrasl/slt171},
	eprint = {1309.4269},
	journal = {\mnras},
	month = feb,
	number = {1},
	pages = {L101-L105},
	primaryclass = {astro-ph.CO},
	title = {{The first month of evolution of the slow-rising Type IIP SN 2013ej in M74$^{★}$}},
	volume = {438},
	year = 2014}

@inproceedings{Tody_1986,
	adsurl = {https://ui.adsabs.harvard.edu/abs/1986SPIE..627..733T},
	author = {{Tody}, Doug},
	booktitle = {Instrumentation in astronomy VI},
	doi = {10.1117/12.968154},
	editor = {{Crawford}, David L.},
	month = jan,
	pages = {733},
	series = {Society of Photo-Optical Instrumentation Engineers (SPIE) Conference Series},
	title = {{The IRAF Data Reduction and Analysis System}},
	volume = {627},
	year = 1986}

@article{Revnivtsev_2004,
	adsurl = {https://ui.adsabs.harvard.edu/abs/2004A&A...424L...5R},
	archiveprefix = {arXiv},
	author = {{Revnivtsev}, M. and {Burenin}, R. and {Fabrika}, S. and {Postnov}, K. and {Bikmaev}, I. and {Pavlinsky}, M. and {Sunyaev}, R. and {Khamitov}, I. and {Aslan}, Z.},
	doi = {10.1051/0004-6361:200400012},
	eprint = {astro-ph/0405388},
	journal = {\aap},
	month = sep,
	pages = {L5-L8},
	primaryclass = {astro-ph},
	title = {{First simultanous X-ray and optical observations of rapid variability of supercritical accretor SS433}},
	volume = {424},
	year = 2004}

@article{Collins_2002,
	adsurl = {https://ui.adsabs.harvard.edu/abs/2002MNRAS.336.1011C},
	author = {{Collins}, George W. and {Scher}, Robert W.},
	doi = {10.1046/j.1365-8711.2002.05844.x},
	journal = {\mnras},
	month = nov,
	number = {3},
	pages = {1011-1020},
	title = {{A revised dynamical model for SS433 and the nature of the system}},
	volume = {336},
	year = 2002}

@article{Kayama_2025,
	author = {Kayama, Kazuho and Tanaka, Takaaki and Uchida, Hiroyuki and Tsuru, Takeshi Go and Inoue, Yoshiyuki and Khangulyan, Dmitry and Tsuji, Naomi and Yamamoto, Hiroaki},
	doi = {10.1093/pasj/psaf059},
	eprint = {https://academic.oup.com/pasj/advance-article-pdf/doi/10.1093/pasj/psaf059/63489957/psaf059.pdf},
	issn = {2053-051X},
	journal = {Publications of the Astronomical Society of Japan},
	month = {06},
	pages = {psaf059},
	title = {X-ray study of the propagation of non-thermal particles in microquasar SS 433/W 50 extended jets},
	url = {https://doi.org/10.1093/pasj/psaf059},
	year = {2025}}

@inproceedings{Nagayama_2024,
	adsurl = {https://ui.adsabs.harvard.edu/abs/2024SPIE13096E..3IN},
	author = {{Nagayama}, Takahiro and {Nakaya}, Hidehiko},
	booktitle = {Ground-based and Airborne Instrumentation for Astronomy X},
	doi = {10.1117/12.3016593},
	editor = {{Bryant}, Julia J. and {Motohara}, Kentaro and {Vernet}, Jo{\"e}l. R.~D.},
	eid = {130963I},
	month = jul,
	pages = {130963I},
	series = {Society of Photo-Optical Instrumentation Engineers (SPIE) Conference Series},
	title = {{kSIRIUS: a simultaneous JHKs camera for the Kagoshima University 1m telescope using newly developed Japanese InGaAs array detectors}},
	volume = {13096},
	year = 2024}

@article{Cherepashchuk_2018,
	adsurl = {https://ui.adsabs.harvard.edu/abs/2018ARep...62..747C},
	author = {{Cherepashchuk}, A.~M. and {Esipov}, V.~F. and {Dodin}, A.~V. and {Davydov}, V.~V. and {Belinskii}, A.~A.},
	doi = {10.1134/S106377291811001X},
	journal = {Astronomy Reports},
	month = nov,
	number = {11},
	pages = {747-763},
	title = {{Spectroscopic Monitoring of SS 433. Stability of Parameters of the Kinematic Model over 40 Years}},
	volume = {62},
	year = 2018}

@article{Noda_2025,
	adsurl = {https://ui.adsabs.harvard.edu/abs/2025PASJ..tmp...16N},
	archiveprefix = {arXiv},
	author = {{Noda}, Hirofumi and {Mori}, Koji and {Tomida}, Hiroshi and {Nakajima}, Hiroshi and {Tanaka}, Takaaki and {Murakami}, Hiroshi and {Uchida}, Hiroyuki and {Suzuki}, Hiromasa and {Kobayashi}, Shogo Benjamin and {Yoneyama}, Tomokage and {Hagino}, Kouichi and {Nobukawa}, Kumiko and {Uchiyama}, Hideki and {Nobukawa}, Masayoshi and {Matsumoto}, Hironori and {Tsuru}, Takeshi Go and {Yamauchi}, Makoto and {Hatsukade}, Isamu and {Odaka}, Hirokazu and {Kohmura}, Takayoshi and {Yamaoka}, Kazutaka and {Yoshida}, Tessei and {Kanemaru}, Yoshiaki and {Hiraga}, Junko and {Dotani}, Tadayasu and {Ozaki}, Masanobu and {Tsunemi}, Hiroshi and {Sato}, Jin and {Takaki}, Toshiyuki and {Terada}, Yuta and {Miyazaki}, Keitaro and {Kusunoki}, Kohei and {Otsuka}, Yoshinori and {Yokosu}, Haruhiko and {Yonemaru}, Wakana and {Ichikawa}, Kazuhiro and {Nakano}, Hanako and {Takemoto}, Reo and {Matsushima}, Tsukasa and {Urase}, Reika and {Kurashima}, Jun and {Fuchi}, Kotomi and {Hayakawa}, Kaito and {Fukuda}, Masahiro and {Kamei}, Takamitsu and {Asahina}, Yoh and {Inoue}, Shun and {Amano}, Yuki and {Aoki}, Yuma and {Ito}, Yamato and {Kamatani}, Tomoya and {Takayama}, Kouta and {Sako}, Takashi and {Yoshimoto}, Marina and {Shima}, Kohei and {Higuchi}, Mayu and {Ninoyu}, Kaito and {Aoki}, Daiki and {Tsunomachi}, Shun and {Hayashida}, Kiyoshi},
	doi = {10.1093/pasj/psaf011},
	eprint = {2502.08030},
	journal = {\pasj},
	month = mar,
	primaryclass = {astro-ph.IM},
	title = {{Soft X-ray Imager of the Xtend system on board XRISM}},
	year = 2025}

@inproceedings{Mori_2022,
	adsurl = {https://ui.adsabs.harvard.edu/abs/2022SPIE12181E..1TM},
	archiveprefix = {arXiv},
	author = {{Mori}, Koji and {Tomida}, Hiroshi and {Nakajima}, Hiroshi and {Okajima}, Takashi and {Noda}, Hirofumi and {Tanaka}, Takaaki and {Uchida}, Hiroyuki and {Hagino}, Kouichi and {Kobayashi}, Shogo Benjamin and {Suzuki}, Hiromasa and {Yoshida}, Tessei and {Murakami}, Hiroshi and {Uchiyama}, Hideki and {Nobukawa}, Masayoshi and {Nobukawa}, Kumiko and {Yoneyama}, Tomokage and {Matsumoto}, Hironori and {Tsuru}, Takeshi and {Yamauchi}, Makoto and {Hatsukade}, Isamu and {Ishida}, Manabu and {Maeda}, Yoshitomo and {Hayashi}, Takayuki and {Tamura}, Keisuke and {Boissay-Malaquin}, Rozenn and {Sato}, Toshiki and {Hiraga}, Junko and {Kohmura}, Takayoshi and {Yamaoka}, Kazutaka and {Dotani}, Tadayasu and {Ozaki}, Masanobu and {Tsunemi}, Hiroshi and {Kanemaru}, Yoshiaki and {Sato}, Jin and {Takaki}, Toshiyuki and {Terada}, Yuta and {Miyazaki}, Keitaro and {Kusunoki}, Kohei and {Otsuka}, Yoshinori and {Yokosu}, Haruhiko and {Yonemaru}, Wakana and {Asahina}, Yoh and {Asakura}, Kazunori and {Yoshimoto}, Marina and {Ode}, Yuichi and {Sato}, Junya and {Hakamata}, Tomohiro and {Aoyagi}, Mio and {Aoki}, Yuma and {Tsunomachi}, Shun and {Doi}, Toshiki and {Aoki}, Daiki and {Fujisawa}, Kaito and {Kitajima}, Masatoshi and {Hayashida}, Kiyoshi},
	booktitle = {Space Telescopes and Instrumentation 2022: Ultraviolet to Gamma Ray},
	doi = {10.1117/12.2626894},
	editor = {{den Herder}, Jan-Willem A. and {Nikzad}, Shouleh and {Nakazawa}, Kazuhiro},
	eid = {121811T},
	eprint = {2303.07575},
	month = aug,
	pages = {121811T},
	primaryclass = {astro-ph.IM},
	series = {Society of Photo-Optical Instrumentation Engineers (SPIE) Conference Series},
	title = {{Xtend, the soft x-ray imaging telescope for the X-Ray Imaging and Spectroscopy Mission (XRISM)}},
	volume = {12181},
	year = 2022}

@inproceedings{Ishisaki_2022,
	adsurl = {https://ui.adsabs.harvard.edu/abs/2022SPIE12181E..1SI},
	author = {{Ishisaki}, Yoshitaka and {Kelley}, Richard L. and {Awaki}, Hisamitsu and {Balleza}, Jesus C. and {Barnstable}, Kim R. and {Bialas}, Thomas G. and {Boissay-Malaquin}, Rozenn and {Brown}, Gregory V. and {Canavan}, Edgar R. and {Cumbee}, Renata S. and {Carnahan}, Timothy M. and {Chiao}, Meng P. and {Comber}, Brian J. and {Costantini}, Elisa and {den Herder}, Jan-Willem and {Dercksen}, Johannes and {de Vries}, Cor P. and {DiPirro}, Michael J. and {Eckart}, Megan E. and {Ezoe}, Yuichiro and {Ferrigno}, Carlo and {Fujimoto}, Ryuichi and {Gorter}, Nathalie and {Graham}, Steven M. and {Grim}, Martin and {Hartz}, Leslie S. and {Hayakawa}, Ryota and {Hayashi}, Takayuki and {Hell}, Natalie and {Hoshino}, Akio and {Ichinohe}, Yuto and {Ishida}, Manabu and {Ishikawa}, Kumi and {James}, Bryan L. and {Kenyon}, Steven J. and {Kilbourne}, Caroline A. and {Kimball}, Mark O. and {Kitamoto}, Shunji and {Leutenegger}, Maurice A. and {Maeda}, Yoshitomo and {McCammon}, Dan and {Miko}, Joseph J. and {Mizumoto}, Misaki and {Okajima}, Takashi and {Okamoto}, Atsushi and {Paltani}, Stephane and {Porter}, Frederick S. and {Sato}, Kosuke and {Sato}, Toshiki and {Sawada}, Makoto and {Shinozaki}, Keisuke and {Shipman}, Russell and {Shirron}, Peter J. and {Sneiderman}, Gary A. and {Soong}, Yang and {Szymkiewicz}, Richard and {Szymkowiak}, Andrew E. and {Takei}, Yoh and {Tamura}, Keisuke and {Tsujimoto}, Masahiro and {Uchida}, Yuusuke and {Wasserzug}, Stephen and {Witthoeft}, Michael C. and {Wolfs}, Rob and {Yamada}, Shinya and {Yasuda}, Susumu},
	booktitle = {Space Telescopes and Instrumentation 2022: Ultraviolet to Gamma Ray},
	doi = {10.1117/12.2630654},
	editor = {{den Herder}, Jan-Willem A. and {Nikzad}, Shouleh and {Nakazawa}, Kazuhiro},
	eid = {121811S},
	month = aug,
	pages = {121811S},
	series = {Society of Photo-Optical Instrumentation Engineers (SPIE) Conference Series},
	title = {{Status of resolve instrument onboard X-Ray Imaging and Spectroscopy Mission (XRISM)}},
	volume = {12181},
	year = 2022}

@article{Matsubayashi_2019,
	adsurl = {https://ui.adsabs.harvard.edu/abs/2019PASJ...71..102M},
	archiveprefix = {arXiv},
	author = {{Matsubayashi}, Kazuya and {Ohta}, Kouji and {Iwamuro}, Fumihide and {Iwata}, Ikuru and {Kambe}, Eiji and {Tsutsui}, Hironori and {Izumiura}, Hideyuki and {Yoshida}, Michitoshi and {Hattori}, Takashi},
	doi = {10.1093/pasj/psz087},
	eid = {102},
	eprint = {1905.05430},
	journal = {\pasj},
	month = oct,
	number = {5},
	pages = {102},
	primaryclass = {astro-ph.IM},
	title = {{KOOLS-IFU: Kyoto Okayama Optical Low-dispersion Spectrograph with optical-fiber Integral Field Unit}},
	volume = {71},
	year = 2019}

@article{Yoshida_2005,
	adsurl = {https://ui.adsabs.harvard.edu/abs/2005JKAS...38..117Y},
	author = {{Yoshida}, Michitoshi},
	doi = {10.5303/JKAS.2005.38.2.117},
	journal = {Journal of Korean Astronomical Society},
	month = jun,
	number = {2},
	pages = {117-120},
	title = {{Current Status of the Instruments, Instrumentation and Open Use of Okayama Astrophysical Observatory}},
	volume = {38},
	year = 2005}

@article{Kurita_2020,
	adsurl = {https://ui.adsabs.harvard.edu/abs/2020PASJ...72...48K},
	author = {{Kurita}, Mikio and {Kino}, Masaru and {Iwamuro}, Fumihide and {Ohta}, Kouji and {Nogami}, Daisaku and {Izumiura}, Hideyuki and {Yoshida}, Michitoshi and {Matsubayashi}, Kazuya and {Kuroda}, Daisuke and {Nakatani}, Yoshikazu and {Yamamoto}, Kodai and {Tsutsui}, Hironori and {Iribe}, Masatsugu and {Jikuya}, Ichiro and {Ohtani}, Hiroshi and {Shibata}, Kazunari and {Takahashi}, Keisuke and {Tokoro}, Hitoshi and {Maihara}, Toshinori and {Nagata}, Tetsuya},
	doi = {10.1093/pasj/psaa036},
	eid = {48},
	journal = {\pasj},
	month = jun,
	number = {3},
	pages = {48},
	title = {{The Seimei telescope project and technical developments}},
	volume = {72},
	year = 2020}

@article{Brown_2013,
	adsurl = {https://ui.adsabs.harvard.edu/abs/2013PASP..125.1031B},
	archiveprefix = {arXiv},
	author = {{Brown}, T.~M. and {Baliber}, N. and {Bianco}, F.~B. and {Bowman}, M. and {Burleson}, B. and {Conway}, P. and {Crellin}, M. and {Depagne}, {\'E}. and {De Vera}, J. and {Dilday}, B. and {Dragomir}, D. and {Dubberley}, M. and {Eastman}, J.~D. and {Elphick}, M. and {Falarski}, M. and {Foale}, S. and {Ford}, M. and {Fulton}, B.~J. and {Garza}, J. and {Gomez}, E.~L. and {Graham}, M. and {Greene}, R. and {Haldeman}, B. and {Hawkins}, E. and {Haworth}, B. and {Haynes}, R. and {Hidas}, M. and {Hjelstrom}, A.~E. and {Howell}, D.~A. and {Hygelund}, J. and {Lister}, T.~A. and {Lobdill}, R. and {Martinez}, J. and {Mullins}, D.~S. and {Norbury}, M. and {Parrent}, J. and {Paulson}, R. and {Petry}, D.~L. and {Pickles}, A. and {Posner}, V. and {Rosing}, W.~E. and {Ross}, R. and {Sand}, D.~J. and {Saunders}, E.~S. and {Shobbrook}, J. and {Shporer}, A. and {Street}, R.~A. and {Thomas}, D. and {Tsapras}, Y. and {Tufts}, J.~R. and {Valenti}, S. and {Vander Horst}, K. and {Walker}, Z. and {White}, G. and {Willis}, M.},
	doi = {10.1086/673168},
	eprint = {1305.2437},
	journal = {\pasp},
	month = sep,
	number = {931},
	pages = {1031},
	primaryclass = {astro-ph.IM},
	title = {{Las Cumbres Observatory Global Telescope Network}},
	volume = {125},
	year = 2013}

@article{Lodders_2009,
	adsurl = {https://ui.adsabs.harvard.edu/abs/2009LanB...4B..712L},
	archiveprefix = {arXiv},
	author = {{Lodders}, K. and {Palme}, H. and {Gail}, H. -P.},
	doi = {10.1007/978-3-540-88055-4_34},
	eprint = {0901.1149},
	journal = {Landolt B{\"o}rnstein},
	month = jan,
	pages = {712},
	primaryclass = {astro-ph.EP},
	title = {{Abundances of the Elements in the Solar System}},
	volume = {4B},
	year = 2009}

@article{Lopez_2006,
	adsurl = {https://ui.adsabs.harvard.edu/abs/2006ApJ...650..338L},
	archiveprefix = {arXiv},
	author = {{Lopez}, Laura A. and {Marshall}, Herman L. and {Canizares}, Claude R. and {Schulz}, Norbert S. and {Kane}, Julie F.},
	doi = {10.1086/506174},
	eprint = {astro-ph/0605574},
	journal = {\apj},
	month = oct,
	number = {1},
	pages = {338-349},
	primaryclass = {astro-ph},
	title = {{Determining the Nature of the SS 433 Binary from an X-Ray Spectrum during Eclipse}},
	volume = {650},
	year = 2006}

@article{Smith_2001,
	adsurl = {https://ui.adsabs.harvard.edu/abs/2001ApJ...556L..91S},
	archiveprefix = {arXiv},
	author = {{Smith}, Randall K. and {Brickhouse}, Nancy S. and {Liedahl}, Duane A. and {Raymond}, John C.},
	doi = {10.1086/322992},
	eprint = {astro-ph/0106478},
	journal = {\apjl},
	month = aug,
	number = {2},
	pages = {L91-L95},
	primaryclass = {astro-ph},
	title = {{Collisional Plasma Models with APEC/APED: Emission-Line Diagnostics of Hydrogen-like and Helium-like Ions}},
	volume = {556},
	year = 2001}

@article{Marshall_2013,
	adsurl = {https://ui.adsabs.harvard.edu/abs/2013ApJ...775...75M},
	archiveprefix = {arXiv},
	author = {{Marshall}, Herman L. and {Canizares}, Claude R. and {Hillwig}, Todd and {Mioduszewski}, Amy and {Rupen}, Michael and {Schulz}, Norbert S. and {Nowak}, Michael and {Heinz}, Sebastian},
	doi = {10.1088/0004-637X/775/1/75},
	eid = {75},
	eprint = {1307.8427},
	journal = {\apj},
	month = sep,
	number = {1},
	pages = {75},
	primaryclass = {astro-ph.HE},
	title = {{Multiwavelength Observations of the SS 433 Jets}},
	volume = {775},
	year = 2013}

@article{Burenin_2011,
	adsurl = {https://ui.adsabs.harvard.edu/abs/2011AstL...37..100B},
	archiveprefix = {arXiv},
	author = {{Burenin}, R.~A. and {Revnivtsev}, M.~G. and {Khamitov}, I.~M. and {Bikmaev}, I.~F. and {Nosov}, A.~S. and {Pavlinsky}, M.~N. and {Sunyaev}, R.~A.},
	doi = {10.1134/S1063773711010026},
	eprint = {1101.2134},
	journal = {Astronomy Letters},
	month = feb,
	number = {2},
	pages = {100-112},
	primaryclass = {astro-ph.HE},
	title = {{Fast optical variability of SS 433}},
	volume = {37},
	year = 2011}

@article{Nishino_2022,
	adsurl = {https://ui.adsabs.harvard.edu/abs/2022PASJ...74L..17N},
	archiveprefix = {arXiv},
	author = {{Nishino}, Yohei and {Kimura}, Mariko and {Sako}, Shigeyuki and {Beniyama}, Jin and {Enoto}, Teruaki and {Minezaki}, Takeo and {Nakaniwa}, Nozomi and {Ohsawa}, Ryou and {Takita}, Satoshi and {Yamada}, Shinya and {Gendreau}, Keith C.},
	doi = {10.1093/pasj/psac027},
	eprint = {2205.08721},
	journal = {\pasj},
	month = jun,
	number = {3},
	pages = {L17-L22},
	primaryclass = {astro-ph.CO},
	title = {{Detection of highly correlated optical and X-ray variations in SS Cygni with Tomo-e Gozen and NICER}},
	volume = {74},
	year = 2022}

@article{Beniyama_2022,
	adsurl = {https://ui.adsabs.harvard.edu/abs/2022PASJ...74..877B},
	archiveprefix = {arXiv},
	author = {{Beniyama}, Jin and {Sako}, Shigeyuki and {Ohsawa}, Ryou and {Takita}, Satoshi and {Kobayashi}, Naoto and {Okumura}, Shin-ichiro and {Urakawa}, Seitaro and {Yoshikawa}, Makoto and {Usui}, Fumihiko and {Yoshida}, Fumi and {Doi}, Mamoru and {Niino}, Yuu and {Shigeyama}, Toshikazu and {Tanaka}, Masaomi and {Tominaga}, Nozomu and {Aoki}, Tsutomu and {Arima}, Noriaki and {Arimatsu}, Ko and {Kasuga}, Toshihiro and {Kondo}, Sohei and {Mori}, Yuki and {Takahashi}, Hidenori and {Watanabe}, Jun-ichi},
	doi = {10.1093/pasj/psac043},
	eprint = {2207.07071},
	journal = {\pasj},
	month = aug,
	number = {4},
	pages = {877-903},
	primaryclass = {astro-ph.EP},
	title = {{Video observations of tiny near-Earth objects with Tomo-e Gozen}},
	volume = {74},
	year = 2022}

@inproceedings{Kojima_2018,
	author = {Yuto Kojima and Shigeyuki Sako and Ryou Ohsawa and Hidenori Takahashi and Mamoru Doi and Naoto Kobayashi and Tsutomu Aoki and Noriaki Arima and Ko Arimatsu and Makoto Ichiki and Shiro Ikeda and Kota Inooka and Yoshifusa Ita and Toshihiro Kasuga and Mitsuru Kokubo and Masahiro Konishi and Hiroyuki Maehara and Noriyuki Matsunaga and Kazuma Mitsuda and Takashi Miyata and Yuki Mori and Mikio Morii and Tomoki Morokuma and Kentaro Motohara and Yoshikazu Nakada and Shin-Ichiro Okumura and Yuki Sarugaku and Mikiya Sato and Toshikazu Shigeyama and Takao Soyano and Masaomi Tanaka and Ken'ichi Tarusawa and Nozomu Tominaga and Tomonori Totani and Seitaro Urakawa and Fumihiko Usui and Junichi Watanabe and Takuya Yamashita and Makoto Yoshikawa},
	booktitle = {High Energy, Optical, and Infrared Detectors for Astronomy VIII},
	doi = {10.1117/12.2311301},
	editor = {Andrew D. Holland and James Beletic},
	organization = {International Society for Optics and Photonics},
	pages = {107091T},
	publisher = {SPIE},
	title = {{Evaluation of large pixel CMOS image sensors for the Tomo-e Gozen wide field camera}},
	url = {https://doi.org/10.1117/12.2311301},
	volume = {10709},
	year = {2018}}

@inproceedings{Sako_2018,
	author = {Shigeyuki Sako and Ryou Ohsawa and Hidenori Takahashi and Yuto Kojima and Mamoru Doi and Naoto Kobayashi and Tsutomu Aoki and Noriaki Arima and Ko Arimatsu and Makoto Ichiki and Shiro Ikeda and Kota Inooka and Yoshifusa Ita and Toshihiro Kasuga and Mitsuru Kokubo and Masahiro Konishi and Hiroyuki Maehara and Noriyuki Matsunaga and Kazuma Mitsuda and Takashi Miyata and Yuki Mori and Mikio Morii and Tomoki Morokuma and Kentaro Motohara and Yoshikazu Nakada and Shin-Ichiro Okumura and Yuki Sarugaku and Mikiya Sato and Toshikazu Shigeyama and Takao Soyano and Masaomi Tanaka and Ken'ichi Tarusawa and Nozomu Tominaga and Tomonori Totani and Seitaro Urakawa and Fumihiko Usui and Junichi Watanabe and Takuya Yamashita and Makoto Yoshikawa},
	booktitle = {Ground-based and Airborne Instrumentation for Astronomy VII},
	doi = {10.1117/12.2310049},
	editor = {Christopher J. Evans and Luc Simard and Hideki Takami},
	organization = {International Society for Optics and Photonics},
	pages = {107020J},
	publisher = {SPIE},
	title = {{The Tomo-e Gozen wide field CMOS camera for the Kiso Schmidt telescope}},
	url = {https://doi.org/10.1117/12.2310049},
	volume = {10702},
	year = {2018}}

@article{Gaia_2023,
	adsurl = {https://ui.adsabs.harvard.edu/abs/2023A&A...674A...1G},
	archiveprefix = {arXiv},
	author = {{Gaia Collaboration} and {Vallenari}, A. and {Brown}, A.~G.~A. and {Prusti}, T. and {de Bruijne}, J.~H.~J. and {Arenou}, F. and {Babusiaux}, C. and {Biermann}, M. and {Creevey}, O.~L. and {Ducourant}, C. and {Evans}, D.~W. and {Eyer}, L. and {Guerra}, R. and {Hutton}, A. and {Jordi}, C. and {Klioner}, S.~A. and {Lammers}, U.~L. and {Lindegren}, L. and {Luri}, X. and {Mignard}, F. and {Panem}, C. and {Pourbaix}, D. and {Randich}, S. and {Sartoretti}, P. and {Soubiran}, C. and {Tanga}, P. and {Walton}, N.~A. and {Bailer-Jones}, C.~A.~L. and {Bastian}, U. and {Drimmel}, R. and {Jansen}, F. and {Katz}, D. and {Lattanzi}, M.~G. and {van Leeuwen}, F. and {Bakker}, J. and {Cacciari}, C. and {Casta{\~n}eda}, J. and {De Angeli}, F. and {Fabricius}, C. and {Fouesneau}, M. and {Fr{\'e}mat}, Y. and {Galluccio}, L. and {Guerrier}, A. and {Heiter}, U. and {Masana}, E. and {Messineo}, R. and {Mowlavi}, N. and {Nicolas}, C. and {Nienartowicz}, K. and {Pailler}, F. and {Panuzzo}, P. and {Riclet}, F. and {Roux}, W. and {Seabroke}, G.~M. and {Sordo}, R. and {Th{\'e}venin}, F. and {Gracia-Abril}, G. and {Portell}, J. and {Teyssier}, D. and {Altmann}, M. and {Andrae}, R. and {Audard}, M. and {Bellas-Velidis}, I. and {Benson}, K. and {Berthier}, J. and {Blomme}, R. and {Burgess}, P.~W. and {Busonero}, D. and {Busso}, G. and {C{\'a}novas}, H. and {Carry}, B. and {Cellino}, A. and {Cheek}, N. and {Clementini}, G. and {Damerdji}, Y. and {Davidson}, M. and {de Teodoro}, P. and {Nu{\~n}ez Campos}, M. and {Delchambre}, L. and {Dell'Oro}, A. and {Esquej}, P. and {Fern{\'a}ndez-Hern{\'a}ndez}, J. and {Fraile}, E. and {Garabato}, D. and {Garc{\'\i}a-Lario}, P. and {Gosset}, E. and {Haigron}, R. and {Halbwachs}, J. -L. and {Hambly}, N.~C. and {Harrison}, D.~L. and {Hern{\'a}ndez}, J. and {Hestroffer}, D. and {Hodgkin}, S.~T. and {Holl}, B. and {Jan{\ss}en}, K. and {Jevardat de Fombelle}, G. and {Jordan}, S. and {Krone-Martins}, A. and {Lanzafame}, A.~C. and {L{\"o}ffler}, W. and {Marchal}, O. and {Marrese}, P.~M. and {Moitinho}, A. and {Muinonen}, K. and {Osborne}, P. and {Pancino}, E. and {Pauwels}, T. and {Recio-Blanco}, A. and {Reyl{\'e}}, C. and {Riello}, M. and {Rimoldini}, L. and {Roegiers}, T. and {Rybizki}, J. and {Sarro}, L.~M. and {Siopis}, C. and {Smith}, M. and {Sozzetti}, A. and {Utrilla}, E. and {van Leeuwen}, M. and {Abbas}, U. and {{\'A}brah{\'a}m}, P. and {Abreu Aramburu}, A. and {Aerts}, C. and {Aguado}, J.~J. and {Ajaj}, M. and {Aldea-Montero}, F. and {Altavilla}, G. and {{\'A}lvarez}, M.~A. and {Alves}, J. and {Anders}, F. and {Anderson}, R.~I. and {Anglada Varela}, E. and {Antoja}, T. and {Baines}, D. and {Baker}, S.~G. and {Balaguer-N{\'u}{\~n}ez}, L. and {Balbinot}, E. and {Balog}, Z. and {Barache}, C. and {Barbato}, D. and {Barros}, M. and {Barstow}, M.~A. and {Bartolom{\'e}}, S. and {Bassilana}, J. -L. and {Bauchet}, N. and {Becciani}, U. and {Bellazzini}, M. and {Berihuete}, A. and {Bernet}, M. and {Bertone}, S. and {Bianchi}, L. and {Binnenfeld}, A. and {Blanco-Cuaresma}, S. and {Blazere}, A. and {Boch}, T. and {Bombrun}, A. and {Bossini}, D. and {Bouquillon}, S. and {Bragaglia}, A. and {Bramante}, L. and {Breedt}, E. and {Bressan}, A. and {Brouillet}, N. and {Brugaletta}, E. and {Bucciarelli}, B. and {Burlacu}, A. and {Butkevich}, A.~G. and {Buzzi}, R. and {Caffau}, E. and {Cancelliere}, R. and {Cantat-Gaudin}, T. and {Carballo}, R. and {Carlucci}, T. and {Carnerero}, M.~I. and {Carrasco}, J.~M. and {Casamiquela}, L. and {Castellani}, M. and {Castro-Ginard}, A. and {Chaoul}, L. and {Charlot}, P. and {Chemin}, L. and {Chiaramida}, V. and {Chiavassa}, A. and {Chornay}, N. and {Comoretto}, G. and {Contursi}, G. and {Cooper}, W.~J. and {Cornez}, T. and {Cowell}, S. and {Crifo}, F. and {Cropper}, M. and {Crosta}, M. and {Crowley}, C. and {Dafonte}, C. and {Dapergolas}, A. and {David}, M. and {David}, P. and {de Laverny}, P. and {De Luise}, F. and {De March}, R.},
	doi = {10.1051/0004-6361/202243940},
	eid = {A1},
	eprint = {2208.00211},
	journal = {\aap},
	month = jun,
	pages = {A1},
	primaryclass = {astro-ph.GA},
	title = {{Gaia Data Release 3. Summary of the content and survey properties}},
	volume = {674},
	year = 2023}

@article{Cherepashchuk_2022,
	adsurl = {https://ui.adsabs.harvard.edu/abs/2022ARep...66..451C},
	author = {{Cherepashchuk}, A.~M. and {Dodin}, A.~V. and {Postnov}, K.~A. and {Belinski}, A.~A. and {Burlak}, M.~A. and {Ikonnikova}, N.~P. and {Irsmambetova}, T.~R. and {Trushkin}, S.~A.},
	doi = {10.1134/S1063772922060026},
	journal = {Astronomy Reports},
	month = jun,
	number = {6},
	pages = {451-465},
	title = {{Optical Monitoring of SS 433 in 2017-2021}},
	volume = {66},
	year = 2022}

@article{Cherepashchuk_2023,
	adsurl = {https://ui.adsabs.harvard.edu/abs/2023NewA..10302060C},
	archiveprefix = {arXiv},
	author = {{Cherepashchuk}, Anatol and {Belinski}, Alexander and {Dodin}, Alexander and {Postnov}, Konstantin},
	doi = {10.1016/j.newast.2023.102060},
	eid = {102060},
	eprint = {2305.07093},
	journal = {New Astronomy},
	month = oct,
	pages = {102060},
	primaryclass = {astro-ph.HE},
	title = {{Evolutionary increase of the orbital separation and change of the Roche lobe size in SS433}},
	volume = {103},
	year = 2023}

@inproceedings{xspec_tool,
	adsurl = {https://ui.adsabs.harvard.edu/abs/1996ASPC..101...17A},
	author = {{Arnaud}, K.~A.},
	booktitle = {Astronomical Data Analysis Software and Systems V},
	editor = {{Jacoby}, George H. and {Barnes}, Jeannette},
	month = jan,
	pages = {17},
	series = {Astronomical Society of the Pacific Conference Series},
	title = {{XSPEC: The First Ten Years}},
	volume = {101},
	year = 1996}

@article{Dubner_1998,
	adsurl = {https://ui.adsabs.harvard.edu/abs/1998AJ....116.1842D},
	author = {{Dubner}, G.~M. and {Holdaway}, M. and {Goss}, W.~M. and {Mirabel}, I.~F.},
	doi = {10.1086/300537},
	journal = {\aj},
	month = oct,
	number = {4},
	pages = {1842-1855},
	title = {{A High-Resolution Radio Study of the W50-SS 433 System and the Surrounding Medium}},
	volume = {116},
	year = 1998}

@article{Medvedev_2019,
	adsurl = {https://ui.adsabs.harvard.edu/abs/2019AstL...45..299M},
	archiveprefix = {arXiv},
	author = {{Medvedev}, P.~S. and {Khabibullin}, I.~I. and {Sazonov}, S. Yu.},
	doi = {10.1134/S1063773719050049},
	eprint = {2005.12416},
	journal = {Astronomy Letters},
	month = may,
	number = {5},
	pages = {299-320},
	primaryclass = {astro-ph.HE},
	title = {{Diagnostics of Parameters for the X-ray Jets of SS 433 from High-Resolution Chandra Spectroscopy}},
	volume = {45},
	year = 2019}

@article{Tashiro_2022,
	adsurl = {https://ui.adsabs.harvard.edu/abs/2022IJMPD..3130001T},
	author = {{Tashiro}, Makoto S.},
	doi = {10.1142/S0218271822300014},
	eid = {2230001},
	journal = {International Journal of Modern Physics D},
	month = jan,
	number = {2},
	pages = {2230001},
	title = {{XRISM: X-ray imaging and spectroscopy mission}},
	volume = {31},
	year = 2022}

@article{Marshall_2002,
	adsurl = {https://ui.adsabs.harvard.edu/abs/2002ApJ...564..941M},
	archiveprefix = {arXiv},
	author = {{Marshall}, Herman L. and {Canizares}, Claude R. and {Schulz}, Norbert S.},
	doi = {10.1086/324398},
	eprint = {astro-ph/0108206},
	journal = {\apj},
	month = jan,
	number = {2},
	pages = {941-952},
	primaryclass = {astro-ph},
	title = {{The High-Resolution X-Ray Spectrum of SS 433 Using the Chandra HETGS}},
	volume = {564},
	year = 2002}

@article{Eikenberry_2001,
	adsurl = {https://ui.adsabs.harvard.edu/abs/2001ApJ...561.1027E},
	archiveprefix = {arXiv},
	author = {{Eikenberry}, S.~S. and {Cameron}, P.~B. and {Fierce}, B.~W. and {Kull}, D.~M. and {Dror}, D.~H. and {Houck}, J.~R. and {Margon}, B.},
	doi = {10.1086/323380},
	eprint = {astro-ph/0107296},
	journal = {\apj},
	month = nov,
	number = {2},
	pages = {1027-1033},
	primaryclass = {astro-ph},
	title = {{Twenty Years of Timing SS 433}},
	volume = {561},
	year = 2001}

@article{Katz_1982,
	adsurl = {https://ui.adsabs.harvard.edu/abs/1982ApJ...260..780K},
	author = {{Katz}, J.~I. and {Anderson}, S.~F. and {Margon}, B. and {Grandi}, S.~A.},
	doi = {10.1086/160297},
	journal = {\apj},
	month = sep,
	pages = {780-793},
	title = {{Nodding motions of accretion rings and disks : a short-term period inSS 433.}},
	volume = {260},
	year = 1982}

@article{Waisberg_2019,
	adsurl = {https://ui.adsabs.harvard.edu/abs/2019A&A...624A.127W},
	archiveprefix = {arXiv},
	author = {{Waisberg}, Idel and {Dexter}, Jason and {Olivier-Petrucci}, Pierre and {Dubus}, Guillaume and {Perraut}, Karine},
	doi = {10.1051/0004-6361/201834747},
	eid = {A127},
	eprint = {1811.12564},
	journal = {\aap},
	month = apr,
	pages = {A127},
	primaryclass = {astro-ph.HE},
	title = {{Collimated radiation in SS 433. Constraints from spatially resolved optical jets and Cloudy modeling of the optical bullets}},
	volume = {624},
	year = 2019}

@article{Davydov_2008,
	adsurl = {https://ui.adsabs.harvard.edu/abs/2008ARep...52..487D},
	author = {{Davydov}, V.~V. and {Esipov}, V.~F. and {Cherepashchuk}, A.~M.},
	doi = {10.1134/S1063772908060061},
	journal = {Astronomy Reports},
	month = jun,
	number = {6},
	pages = {487-506},
	title = {{Spectroscopic monitoring of SS 433: A search for long-term variations of kinematic model parameters}},
	volume = {52},
	year = 2008}

@article{Geldzahler_1980,
	adsurl = {https://ui.adsabs.harvard.edu/abs/1980A&A....84..237G},
	author = {{Geldzahler}, B.~J. and {Pauls}, T. and {Salter}, C.~J.},
	journal = {\aap},
	month = apr,
	pages = {237-244},
	title = {{Continuum observations of the SNR W50 and G 74.9+1.2 at 2695 MHz.}},
	volume = {84},
	year = 1980}

@article{Hillwig_2004,
	author = {T. C. Hillwig and D. R. Gies and W. Huang and M. V. McSwain and M. A. Stark and A. van der Meer and L. Kaper},
	doi = {10.1086/423927},
	journal = {The Astrophysical Journal},
	month = {nov},
	number = {1},
	pages = {422},
	title = {Identification of the Mass Donor Star's Spectrum in SS 433},
	url = {https://dx.doi.org/10.1086/423927},
	volume = {615},
	year = {2004}}

@article{Miller_2008,
	adsurl = {https://ui.adsabs.harvard.edu/abs/2008ApJ...682.1141M},
	archiveprefix = {arXiv},
	author = {{Miller-Jones}, J.~C.~A. and {Migliari}, S. and {Fender}, R.~P. and {Thompson}, T.~W.~J. and {van der Klis}, M. and {M{\'e}ndez}, M.},
	doi = {10.1086/589144},
	eprint = {0804.1337},
	journal = {\apj},
	month = aug,
	number = {2},
	pages = {1141-1151},
	primaryclass = {astro-ph},
	title = {{Coupled Radio and X-Ray Emission and Evidence for Discrete Ejecta in the Jets of SS 433}},
	volume = {682},
	year = 2008}

@article{Sakemi_2023,
	adsurl = {https://ui.adsabs.harvard.edu/abs/2023PASJ...75..338S},
	archiveprefix = {arXiv},
	author = {{Sakemi}, Haruka and {Machida}, Mami and {Yamamoto}, Hiroaki and {Tachihara}, Kengo},
	doi = {10.1093/pasj/psad001},
	eprint = {2301.13333},
	journal = {\pasj},
	month = apr,
	number = {2},
	pages = {338-350},
	primaryclass = {astro-ph.GA},
	title = {{Molecular clouds at the eastern edge of radio nebula W 50}},
	volume = {75},
	year = 2023}

@article{Jeffrey_2016,
	adsurl = {https://ui.adsabs.harvard.edu/abs/2016MNRAS.461..312J},
	archiveprefix = {arXiv},
	author = {{Jeffrey}, Robert M. and {Blundell}, Katherine M. and {Trushkin}, Sergei A. and {Mioduszewski}, Amy J.},
	doi = {10.1093/mnras/stw1322},
	eprint = {1606.01240},
	journal = {\mnras},
	month = sep,
	number = {1},
	pages = {312-320},
	primaryclass = {astro-ph.HE},
	title = {{Fast launch speeds in radio flares, from a new determination of the intrinsic motions of SS 433's jet bolides}},
	volume = {461},
	year = 2016}

@article{Brinkmann_1996,
	adsurl = {https://ui.adsabs.harvard.edu/abs/1996A&A...312..306B},
	author = {{Brinkmann}, W. and {Aschenbach}, B. and {Kawai}, N.},
	journal = {\aap},
	month = aug,
	pages = {306-316},
	title = {{ROSAT observations of the W 50/SS 433 system.}},
	volume = {312},
	year = 1996}

@article{Kotani_1996,
	author = {Kotani, Taro and Kawai, Nobuyuki and Matsuoka, Masaru and Brinkmann, Wolfgang},
	doi = {10.1093/pasj/48.4.619},
	eprint = {https://academic.oup.com/pasj/article-pdf/48/4/619/9713443/pasj48-0619.pdf},
	issn = {0004-6264},
	journal = {Publications of the Astronomical Society of Japan},
	month = {08},
	number = {4},
	pages = {619-629},
	title = {{Iron-Line Diagnostics of the Jets of SS 433}},
	url = {https://doi.org/10.1093/pasj/48.4.619},
	volume = {48},
	year = {1996}}

@article{Crampton_1980,
	adsurl = {https://ui.adsabs.harvard.edu/abs/1981ApJ...251..604C},
	author = {{Crampton}, D. and {Hutchings}, J.~B.},
	doi = {10.1086/159505},
	journal = {\apj},
	month = dec,
	pages = {604-610},
	title = {{The SS 433 binary system.}},
	volume = {251},
	year = 1981}

@article{Margon_1984,
	adsurl = {https://ui.adsabs.harvard.edu/abs/1984ARA&A..22..507M},
	author = {{Margon}, Bruce},
	doi = {10.1146/annurev.aa.22.090184.002451},
	journal = {\araa},
	month = jan,
	pages = {507-536},
	title = {{Observations of SS 433}},
	volume = {22},
	year = 1984}

@article{Abell_1979,
	author = {Abell, George O. and Margon, Bruce},
	date = {1979/06/01},
	doi = {10.1038/279701a0},
	id = {ABELL1979},
	isbn = {1476-4687},
	journal = {Nature},
	number = {5715},
	pages = {701--703},
	title = {A kinematic model for SS433},
	url = {https://doi.org/10.1038/279701a0},
	volume = {279},
	year = {1979}}

@article{Gies_2002,
	author = {D. R. Gies and M. V. McSwain and R. L. Riddle and Z. Wang and P. J. Wiita and D. W. Wingert},
	doi = {10.1086/338335},
	journal = {The Astrophysical Journal},
	month = {feb},
	number = {2},
	pages = {1069},
	title = {The Spectral Components of SS 433},
	url = {https://dx.doi.org/10.1086/338335},
	volume = {566},
	year = {2002}}

@article{Hjellming_1981,
	adsurl = {https://ui.adsabs.harvard.edu/abs/1981ApJ...246L.141H},
	author = {{Hjellming}, R.~M. and {Johnston}, K.~J.},
	doi = {10.1086/183571},
	journal = {\apjl},
	month = jun,
	pages = {L141-L145},
	title = {{An analysis of the proper motions of SS 433 radio jets.}},
	volume = {246},
	year = 1981}

@article{Kubota_2010b,
	author = {Kubota, Kaori and Ueda, Yoshihiro and Kawai, Nobuyuki and Kotani, Taro and Namiki, Masaaki and Kinugasa, Kenzo and Ozaki, Shinobu and Iijima, Takashi and Fabrika, Sergei and Yuasa, Takayuki and Yamada, Shin'ya and Makishima, Kazuo},
	doi = {10.1093/pasj/62.2.323},
	eprint = {https://academic.oup.com/pasj/article-pdf/62/2/323/17440568/pasj62-0323.pdf},
	issn = {0004-6264},
	journal = {Publications of the Astronomical Society of Japan},
	month = {04},
	number = {2},
	pages = {323-333},
	title = {{Suzaku X-Ray and Optical Spectroscopic Observations of SS 433 in the 2006 April Multiwavelength Campaign}},
	url = {https://doi.org/10.1093/pasj/62.2.323},
	volume = {62},
	year = {2010}}

@article{Blundell_2004,
	author = {Katherine M. Blundell and Michael G. Bowler},
	doi = {10.1086/426542},
	journal = {The Astrophysical Journal},
	month = {oct},
	number = {2},
	pages = {L159},
	title = {Symmetry in the Changing Jets of SS 433 and Its True Distance from Us},
	url = {https://dx.doi.org/10.1086/426542},
	volume = {616},
	year = {2004}}






\end{document}